\documentclass[reqno,12pt]{article}
\usepackage[T1]{fontenc}
\usepackage[ansinew]{inputenc}
\usepackage{amsmath}
\usepackage{amssymb}
\usepackage{graphicx}
\usepackage{xcolor}
\usepackage[hidelinks]{hyperref}
\usepackage{dsdshorthand}
\usepackage{color}
\definecolor{darkblue}{cmyk}{0.9,0.9,0,0}
\usepackage{epsfig}
\usepackage{mdframed}
\usepackage{cite}
\usepackage{floatpag}


\usepackage{varioref}
\usepackage{makeidx}
\makeindex

\usepackage{mathrsfs}

\usepackage[english]{babel}
\begin{document}


\thispagestyle{empty}

\renewcommand{\thefootnote}{\fnsymbol{footnote}}
\setcounter{page}{1}
\setcounter{footnote}{0}
\setcounter{figure}{0}

\vspace{-0.4in}

\hfill CALT-TH 033

\begin{center}
$$$$
{\Large\textbf{\mathversion{bold}
Chaos on the mass shell
}\par}
\vspace{1.0cm}

\textrm{Alexandre Homrich, David Simmons-Duffin and George Yuzhe Zhou}
\\ \vspace{1cm}
\footnotesize{\textit{Walter Burke Institute for Theoretical Physics, Caltech, Pasadena, CA 91125, USA}}
\end{center}

\vspace{2mm}
\begin{abstract}
We study particle production in high-energy states for theories given by an RG flow between an asymptotically free or conformal theory in the UV and a massive theory in the IR. Using the framework of detector matching, we argue that distributions of produced particles satisfy universal evolution equations with energy, determined by the spectrum of light-ray operators of the UV theory. In particular, we interpret the UV Chew-Frautschi plot as a ``wave equation'' for particle densities on the mass shell, evolving with energy. (When the UV theory is a CFT, this wave equation is the same one that controls the spread of chaos in hyperbolic space at a particular temperature.) We test these ideas for integrable flows in 2d CFTs, finding an independent derivation of the evolution equations in terms of the thermodynamics of a ``Feynman gas.'' We also discuss consequences for 4d gauge theories, including the presence of a ``BFKL contrail'' produced by an outgoing wave of hadron number density.
\end{abstract}

\setcounter{page}{1}
\renewcommand{\thefootnote}{\arabic{footnote}}
\setcounter{footnote}{0}

\newpage


{
\tableofcontents
}


\newpage

\section{Introduction} 

Many collider observables can be modeled as matrix elements that mix UV and IR descriptions of the same theory:
\be
\label{eq:mixedmatrixelement}
\<\Psi^\textrm{UV}|\mathcal{D}^\mathrm{IR}|\Psi^\textrm{UV}\>.
\ee
Here, $|\Psi^\textrm{UV}\>$ is a state expressed in terms of UV degrees of freedom, while $\mathcal{D}^\mathrm{IR}$ is a ``detector'' expressed in terms of IR degrees of freedom. For example, in the QCD contribution to the process $e^+e^- \to \g^* \to \mathrm{hadrons}$, we can take $|\Psi^\textrm{UV}\> = J^{\mu}_\mathrm{EM}(p)|\Omega\>$, where $J^\mu_\mathrm{EM}(p)$ is the electromagnetic current expressed in terms of quark fields, while the detector $\mathcal{D}^\mathrm{IR}$ counts outgoing hadrons (or their decay products).

When an observable involves different effective descriptions of the same theory, we can evaluate it using {\it matching}: we take objects in one description and ``match'' them onto linear combinations of objects in the other description. Once everything is encoded in terms of the same degrees of freedom, the resulting quantities are (hopefully) calculable. In the case of the matrix element (\ref{eq:mixedmatrixelement}), there are two ways we could imagine performing matching:
\begin{enumerate}
\item ``State matching'': match the state $|\Psi^\textrm{UV}\>$ onto a linear combination of states expressed in terms of IR degrees of freedom.
\item ``Detector matching'': match the detector $\mathcal{D}^\mathrm{IR}$ onto a linear combination of operators expressed in terms of UV degrees of freedom.
\end{enumerate}

We will be particularly interested in collider observables at high energy. In this case, state matching is a potentially daunting and complicated task. The typical number of IR particles may be large, so state matching could require specifying a tremendous amount of data. Relatedly, the resulting sums over IR states may not converge well in the high-energy limit. (We will see examples shortly.) By contrast, detector matching can be much simpler.

Reference \cite{Chang:2025zib} argued that detector matching is the correct choice for understanding the high-energy scaling of observables like (\ref{eq:mixedmatrixelement}), and it is natural to match $\mathcal{D}^\mathrm{IR}$ onto light-ray operators $\cD^\mathrm{UV}_i$ of the UV theory:
\be
\label{eq:detectormatchingequation}
\cD^{\mathrm{IR}} = \sum_i c_i \cD^\mathrm{UV}_i.
\ee
This leads to predictions for the high-energy scaling in QCD of particle number and various generalizations \cite{Chang:2025zib,Gonzalez:2026aha}, charge correlations \cite{Chen:2026charge}, energy correlators in the post-confinement region \cite{Chang:2025kgq}, and more. For simplicity, \cite{Chang:2025zib} mostly adopted an approximation where the IR degrees of freedom were treated as massless. One goal of this work is to explore in detail how detector matching works when the IR particles are massive.

\begin{figure}[t!]
  \centering
  \includegraphics{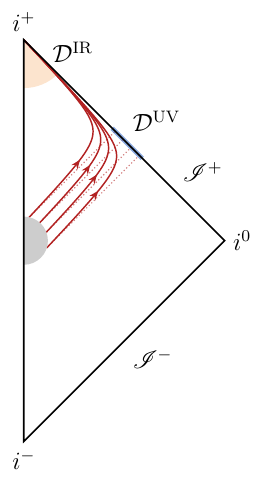}
  \caption{A high-energy insertion creates radiation that emanates from a region near the origin (gray) and initially propagates outward at nearly the speed of light. Massive particles form, and their trajectories bend upwards towards future infinity $i^+$, where they are measured by $\mathcal{D}^\mathrm{IR}$. To a UV observer, the initial radiation appears to head towards future null infinity $\mathscr{I}^+$, where it is measured by $\mathcal{D}^\mathrm{UV}$.}
  \label{fig:penrose}
\end{figure}

We can understand why UV light-ray operators appear in (\ref{eq:detectormatchingequation}) using a simple picture (figure~\ref{fig:penrose}). Consider a state created by a high-energy insertion in the middle of Minkowski space. The energy from this insertion initially propagates outward at the speed of light, in a way dictated by the dynamics of the UV theory. Eventually, massive particles form, and their trajectories ``bend'' upwards on the Penrose diagram towards future infinity $i^+$, where they are measured by $\mathcal{D}^\mathrm{IR}$. Detector matching expresses measurements in a neighborhood of $i^+$ in terms of measurements of the outgoing radiation in the UV region. Because the outgoing radiation travels at the speed of light, it is naturally measured by operators that appear to live at future null infinity $\mathscr{I}^+$, relative to the UV region. These are ``light-ray'' operators at infinity~\cite{Hofman:2008ar,Kravchuk:2018htv}. Said differently, although degrees of freedom do not eventually end up at $\mathscr{I}^+$, they appear to head towards $\mathscr{I}^+$ from the perspective of a UV observer.

In this work, we start by exploring detector matching in theories that are under better theoretical control than QCD. Our general setting will be a UV CFT, deformed so that it undergoes an RG flow to a massive theory with mass gap $m$. Let $|\Psi^\mathrm{UV}\>$ be given by the action of a local CFT operator $\cO$ on the vacuum $|\Omega\>$:
\be
|\Psi^\mathrm{UV}\>=\cO(\tau_E=-\tfrac r 2)|\Omega\>.
\ee
To make a normalizable state, we place $\cO$ at Euclidean time $\tau_E=-r/2$. As an example of $\cD^\mathrm{IR}$, we might ask: what is the density of asymptotic particles in this state? In formulas, let $a^\dag(p),a(p)$ be creation and annihilation operators for a stable particle with mass $m$, with on-shell momentum $p$, normalized so they transform like Lorentz scalars. We would like to measure the expectation value of the number density operator $n(p)=a^\dag(p)a(p)$:
\be
\<\Psi^\textrm{UV}|\mathcal{D}^\mathrm{IR}|\Psi^\textrm{UV}\> = \<\Omega|\cO(x)n(p)\cO(0)|\Omega\>,
\ee
where $x$ is at Euclidean time $r \ll 1/m$. Note that ``state matching'' in this context is just the form-factor expansion:
\be
\label{eq:formfactorexpansion}
&\<\Omega|\cO(x)n(p)\cO(0)|\Omega\> 
\nn\\
&= \sum_N \int d\mathrm{LIPS}_N\, |\<p_1,\dots,p_N|\cO(0)\>|^2 \p{\sum_{i=1}^N (2\pi)^{d-1}2E_p\,\de^{d-1}(\vec p-\vec p_i)}\, e^{-r\sum_i E_i}.
\ee
This expansion, while correct, converges more and more slowly as $r\to 0$. We would like to understand how to resum it in the $r\to 0$ limit.

Our first key observation is that detector matching predicts a ``flow'' equation for the number density $\<n(p)\>$ as a function of radial time $t=-\log r$ and momentum $p$. The large-$t$ dynamics of the flow are determined by the Chew-Frautschi plot of the UV CFT, and are thus universal for all RG flows starting from the same UV fixed point.\footnote{Subleading corrections coming from conformal perturbation theory will depend on the RG flow.}

To explain how this works, let us simplify the discussion by specializing to 2d. We parametrize on-shell momenta as $p=m(\cosh \th,\sinh \th)$, and write $n(\theta)$ as a function of the rapidity $\theta$. According to \cite{Chang:2025zib}, the first step in detector matching is to decompose $n(\theta)$ into Lorentz irreps. In 2d, this is accomplished by Fourier transforming in $\theta$:
\be
\label{eq:fouriertransform}
\widehat n(J_L) &= \int_{-\oo}^\oo d\theta\, e^{-J_L \theta} n(\theta),
\ee
where $J_L\in i \R$ is a boost weight. By Lorentz symmetry, the IR detector $\widehat n(J_L)$ must match onto a linear combination of UV detectors with the same boost weight $J_L$:
\be
\label{eq:matchingfixedboostweight}
\widehat n(J_L) &= \sum_i c_i(J_L) \cD^{\mathrm{UV}}_{J_L,i}.
\ee
Each UV detector has mass dimension $-\De_{L,i}(J_L)$,\footnote{The minus sign is because $\cD^{\mathrm{UV}}_{J_L,i}$ transforms like a primary at spatial infinity.} so plugging (\ref{eq:matchingfixedboostweight}) into our matrix element gives
\be
\label{eq:therexpansion}
\<\widehat n(J_L)\>_{\cO(r)} &= \sum_i c_i(J_L) \<\cD^{\mathrm{UV}}_{J_L,i}\>_{\cO(r)} = \sum_i d_i(J_L) r^{\De_{L,i}(J_L)}(1+ \mathrm{CPT}),
\ee
where we have adopted the notation $\<\cD\>_{\cO(r)}= \<\cO(x) \cD \cO(0)\>/\<\cO(x)\cO(0)\>$. Here, $d_i(J_L)$ are (dimensionful) products of matching coefficients $c_i(J_L)$ and one-point coefficients for $\cD^{\mathrm{UV}}_{J_L,i}$ in the given state. The leading $r$-dependence of (\ref{eq:therexpansion}) is fixed by dimensional analysis, and ``CPT'' denotes corrections from conformal perturbation theory in the UV deformation.\footnote{It is interesting to ask whether (\ref{eq:therexpansion}) is a convergent or asymptotic expansion. We comment on this further in the discussion, section~\ref{sec:discussion}.}

To go further, we must understand which $\De_{L,i}$ appear in (\ref{eq:therexpansion}). Let us assume for now that the light-ray operators $\cD^\textrm{UV}_{J_L,i}$ appearing in the matching equation (\ref{eq:matchingfixedboostweight}) are the same as those appearing in the Regge limit of local four-point correlators in the UV CFT 
\cite{Kravchuk:2018htv}.\footnote{As explained below, we will find some examples of IR detectors that {\it violate\/} this assumption.} The spectrum of these operators is described by the CFT's Chew-Frautschi plot, which can be computed, for example, using the Lorentzian inversion formula \cite{Caron-Huot:2017vep}.

\begin{figure}[t!]
  \centering
  \includegraphics{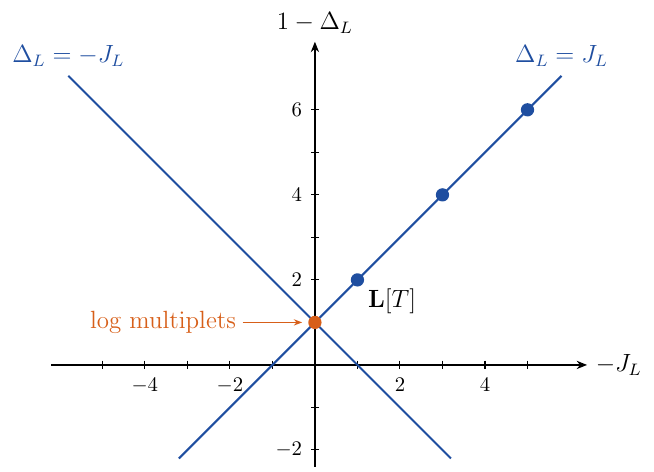}
  \caption{Chew-Frautschi plot of the leading Regge trajectories in a 2d CFT, plotted as $1-\Delta_L$ versus $-J_L$. Dots are light transforms of Virasoro descendants of the identity with spin $J=2,4,6,\dots$, starting with the light transform $\mathbf{L}[T]$ of the stress tensor. The trajectory $\Delta_L=J_L$ and its shadow $\Delta_L=-J_L$ intersect at $J_L=\Delta_L=0$, where they can combine into logarithmic multiplets.}
  \label{fig:chewfrautschi}
\end{figure}

In a 2d CFT, the light-ray operators with minimal $\De_{L,i}$ for each $J_L$ are ``$45^\circ$ trajectories'' that analytically continue between Virasoro descendants of the identity operator and their shadows, see figure~\ref{fig:chewfrautschi}. (Note that there are an infinite number of such trajectories, but they have degenerate quantum numbers.) These leading trajectories satisfy $\De_L^2 - J_L^2=0$. Writing $t=-\log r$, this leads to an approximate differential equation for the density
\be
\p{\frac{\ptl^2}{\ptl t^2} - J_L^2} \<\widehat n(J_L)\>_{\cO(r)} = 0 + \textrm{subleading} + \textrm{CPT} \qquad (r\to 0),
\ee
where ``subleading'' represents contributions from trajectories with larger $\De_{L,i}$, and ``CPT'' represents conformal perturbation theory corrections, both of which are suppressed as $r\to 0$. Transforming back to rapidity space, we obtain an approximate wave equation for the particle number density:
\be
\label{eq:thewaveequation}
\p{\frac{\ptl^2}{\ptl t^2} - \frac{\ptl^2}{\ptl \theta^2}} \<n(\theta)\>_{\cO(r)} &= 0 + \textrm{subleading} + \textrm{CPT} \qquad (r\to 0).
\ee
This is a prediction of the UV CFT --- it should hold for any particle density in any RG flow from the given 2d CFT. We can capture subleading terms by including additional Regge trajectories, yielding flow equations with higher degree in $\ptl_t$, with smaller subleading corrections.

To summarize, the Chew-Frautschi plot is a wave equation for the evolution of conformal matter in rapidity space. As the conformal matter expands, it eventually converts into particles (which is how we observe it). The details of this conversion depend on the RG flow, but the way particle properties change with energy depends only on the UV CFT. This is reminiscent of the seeding of structure by primordial fluctuations in inflation \cite{Mukhanov:1981xt,Hawking:1982cz,Guth:1982ec}.

Integrable flows from 2d minimal models \cite{Zamolodchikov:1978xm,Zamolodchikov:1989hfa} provide a natural playground for testing these ideas. In this setting, we can compute form factors and sum the form factor expansion (\ref{eq:formfactorexpansion}) to check predictions like (\ref{eq:therexpansion}) and (\ref{eq:thewaveequation}). In sections~\ref{sec:ising101} and \ref{sec:isingdetails}, we explore perhaps the simplest such flow: the thermal deformation of the 2d Ising CFT. Although this is a free theory, the $\sigma$ form factor has support on arbitrary numbers of particles, so densities like $\<n(\theta)\>_{\sigma(r)}$ are nontrivial to compute. Fortunately, $\<n(\theta)\>_{\sigma(r)}$ was computed recently in \cite{Liu:2026asy} using powerful integrability methods, and we can analyze and interpret those results. In section~\ref{sec:isingdetails}, we also give a simple derivation of some of the main features of $\<n(\theta)\>_{\sigma(r)}$ using elementary methods, and in section~\ref{sec:ising101} we reproduce these features with numerics. We find that the flow equation (\ref{eq:thewaveequation}) indeed holds.

From our exploration of $\<n(\theta)\>$ in the thermally-deformed 2d Ising CFT, we identify a general physical mechanism that explains the flow equation (\ref{eq:thewaveequation}) in a broad class of integrable flows. The key insight is that many integrable form factors factorize when subsets of particles become far-separated in rapidity space. Consequently, the squared form factor $|\<\theta_1,\dots,\theta_n|\cO\>|^2$ can be treated as the exponentiated potential $e^{-V(\theta_1,\dots,\theta_n)}$ of a classical ``gas'' of particles with short-range interactions. This gas is sometimes called a ``Feynman gas'' \cite{Feynman:1969PRL,Wilson:1970zzb,Cardy:1990pc}. The form factor expansion computes the partition function of the Feynman gas in a box of size $2\log(1/mr)$. As $r\to 0$, the box expands, and the density of the gas moves like a wave in rapidity space.

In section~\ref{sec:magnetic}, we study a different deformation of the 2d Ising CFT: the magnetic deformation, which flows to Zamolodchikov's $E_8$ theory \cite{Zamolodchikov:1989hfa,Zamolodchikov:1989fp}. We compute form factors up to 8 particles, and integrate them numerically to estimate thermodynamic properties of the Feynman gas in this theory. Our numerics support the above picture. 

Altogether, the Feynman gas picture gives an independent physical argument for (\ref{eq:thewaveequation}), providing support for the detector matching idea in a nonperturbative setting. Turning the discussion around, the flow equation (\ref{eq:thewaveequation}) predicts that a local Feynman gas appears in any form factor in any (not necessarily integrable) massive RG flow from a 2d CFT. It would be interesting to test this prediction in the future.

In our investigation of thermally-deformed 2d Ising in section~\ref{sec:multiscale}, we do find a puzzle that challenges our current understanding of detector matching. We find that it is possible to construct IR observables that involve large rapidity separations that do not appear to match into individual light-ray operators on the Chew-Frautschi plot of the UV theory. We refer to such observables as ``multiscale detectors.'' Multiscale detectors appear to match onto products of UV light-ray operators --- i.e. {\it composite\/} UV detectors. We do not have a systematic understanding of when this should happen: which IR observables match onto composite UV detectors? In fact, we do not know what a ``composite'' detector means in a general theory. Clearly, we need to study more examples and gather more data.

In higher-dimensional CFTs, detector matching leads to approximate flow equations that are polynomial in $\ptl_t$ with coefficients given by nontrivial integration kernels on the mass shell. We discuss the higher-dimensional case in section~\ref{sec:higherDsec}. Similar logic applies in theories with a running coupling, where the flow equations have explicit $t$-dependence through the running coupling. We discuss flow equations for QCD in sections~\ref{sec:PYM} to \ref{sec:generalevol}.

\subsection{Relation to chaos in hyperbolic space}

In a CFT, detectors at future null infinity are related by conformal symmetry to the light-ray operators that appear in the Regge limit of four-point functions of local operators. This is the so-called ``spacelike-timelike'' correspondence of \cite{Caron-Huot:2013fea,Caron-Huot:2015bja}. The dictionary relating their quantum numbers is \cite{Kravchuk:2018htv} $\De_L=1-J$, where $\Delta_L$ is the scaling dimension of a detector at future null infinity, and $J$ is the spin of a light-ray operator appearing in a four-point correlation function. Under a boost $e^{K t}$, CFT correlators in the Regge limit evolve as $e^{(J-1) t}=e^{-\Delta_L t} = r^{\Delta_L}$ with $t=-\log r$. Thus, the dynamics of detectors under dilatations and four-point functions under boosts are controlled by the same quantum numbers. 

There is also a correspondence between Minkowski correlators in the Regge limit and out-of-time-order correlators (OTOCs) in hyperbolic space $H^{d-1}$ \cite{Maldacena:2015waa}. Stringing together these correspondences, we conclude that late-time OTOCs in hyperbolic space are controlled by the same ``flow'' equations as IR detector observables, where the hyperbolic space becomes the mass shell. More concretely, we can consider the number density $\<n(m y)\>_{\cO(r)}$ as a function of a point $y\in H^{d-1}$ and a ``time'' $t=-\log r$. We claim that, up to conformal perturbation theory corrections, this observable satisfies the same approximate flow equations as the OTOC
\be
\Tr(V(e,0)W(y,t)e^{-\pi H} V(e,0) W(y,t) e^{-\pi H}),
\ee
where $V(y,t)$ and $W(y,t)$ are primaries in the UV theory on $H^{d-1} \x \R$ with Hamiltonian $H$. The OTOC describes the spread of chaos. Thus, the particle number density evolves on the mass shell in (approximately) the same way that chaos evolves in hyperbolic space, hence our title ``Chaos on the mass shell.''\footnote{Turtle power!} We give more technical details in appendix~\ref{app:conformalreggedetails}.

\section{Example:  $\sigma$ evolution in the 2d Ising model at $T>T_c$}
\label{sec:ising101}
As our first example, let us consider the computation of $\langle \widehat{n}(J_L) \rangle_{\sigma(r)}$ in the 2d Ising field theory at $T>T_c$:
\begin{equation}
        S = S_\text{Ising CFT} + \tau \int d^2x\, \epsilon(x). \label{eq:thermalising}
\end{equation}
This is the theory of a free massive fermion. Nevertheless, the operator $\sigma$ sources an arbitrary odd number of fermions and thus its two-point function 
admits a nontrivial form factor expansion,
\begin{equation}
\langle \sigma(r)\sigma(0)\rangle = \sum_{N \text{ odd}} \frac{1}{N!} \int \frac{d^N\theta}{(2\pi)^N}\,
\exp\left({-mr\sum_i \cosh\theta_i}\right) 
|\langle \sigma(0)|\theta_1,\dots,\theta_N\rangle|^2, \label{eq:sigmaffexpasion}
\end{equation}
where we parametrize the on-shell momenta through the rapidities $p(\theta)=m( \cosh\theta, \sinh\theta)$. The form factors are given by \cite{McCoy:1977er,Berg:1978sw}
\begin{equation}
\langle \sigma(0)|\theta_1,\dots,\theta_N\rangle= \mathcal{N}\prod_{i<j} \tanh\!\left(\frac{\theta_i-\theta_j}{2}\right),\label{eq:ffsigma}
\end{equation}
where $\mathcal{N} = A^3 2^{1/3} e^{-1/4}$ is a normalization constant, with $A$ the Glaisher constant.

The series (\ref{eq:sigmaffexpasion}) can be thought of as the grand-canonical partition
 function $Z$ for a 1d classical gas whose particles sit at positions $\theta_i$ \cite{Cardy:1990pc,2020sfta.book.....M}. The matrix elements
  $|\langle \sigma|\theta_1,\dots,\theta_N\rangle|^2$ function as a repulsive, short-range,
   and pairwise potential for the particles
\begin{equation}
   |\langle \sigma|\theta_1,\dots,\theta_N\rangle|^2/\mathcal{N}^2 \equiv  \exp{\left(-\sum_{i<j} V_{ij}\right)} = \exp\left(\sum_{i<j}  \log \tanh^2\left(\frac{\theta_i-\theta_j}{2}\right)\right),
\end{equation}
while the factor $z(\theta) = \frac{1}{2\pi}\exp(-mr\cosh \theta)$ acts as a position-dependent fugacity.\footnote{\label{ft:sigmamu}The restriction to odd $N$ is immaterial in the thermodynamic limit. It corresponds to a subextensive shift in the entropy. The sum over even $N$ computes the two-point function of the disorder operator  $\mu$.}

The high-energy limit $r\,\to\,0$ is, from this perspective, a thermodynamic limit.
 At small $r$, the doubly exponential fugacity $z(\theta)$ confines the particles to
  a box of size ${2 L \simeq 2 \log(1/mr)}$, and the partition function is dominated by 
  thermodynamic states with $N = O(L)$ particles. The partition function of a classical gas in a box of size $V=2L$ is given by
  \be
  Z &= e^{V p(z)},
  \ee
where $p(z)$ is the pressure of the gas as a function of fugacity $z$. The short-distance conformal behavior $\<\s(r)\s(0)\> \sim r^{-2\De_\s}$ is reproduced because $p(z_c)=\De_\s$, where $z_c$ is the ``critical'' value of the fugacity in the bulk of the box:
  \begin{equation}
    z(\theta) \to z_c = 1/2\pi \qquad \text{for } |\theta| = O(L^0)   .
 \end{equation} 
  We will solve this thermodynamic system explicitly in section~\ref{sec:fredholm}. 

 A number of features of $\langle n(\theta)\rangle_{\sigma(r)}$ at small $r$ can be understood from thermodynamics. 
  Firstly, since the fugacity is approximately constant in the bulk of the box, 
the equilibrium configuration is homogeneous in $\theta$ in this region. Thus, to leading order in $r$, we expect (and will soon confirm) a constant number density 
\be
n_c &=z_c p'(z_c)
\ee
inside the box. Near the edge of the box, the particle density develops a nontrivial `kink' profile. As $r\to0$, this kink density $n_{\text{kink}}(\pm \theta-L)$ converges to an $r$-independent shape satisfying
\begin{equation}
n_{\text{kink}}(x)\to
\begin{cases}
n_{c}, & x\to -\infty,\\
0, & x\to +\infty.
\end{cases}
\end{equation}
To leading order in $r$, we can approximate
\begin{equation}
       \langle n(\theta)\rangle_{\sigma(r)} \sim  n_{\text{kink}}(\theta - L) +  n_{\text{kink}}(-\theta - L) - n_c,
       \label{eq:kinkansatz}
\end{equation}
 which satisfies the wave equation (\ref{eq:thewaveequation}).
 
 We can also understand the CFT scaling (\ref{eq:therexpansion}) from this thermodynamic picture.
 The $\widehat{n}(J_L)$ detector matrix element can be obtained through an infinitesimal variation of the fugacity,
 \begin{equation}
        z(\theta)\to z(\theta)\left(1 + \lambda e^{-J_L \theta}\right)
 \end{equation}
 so that 
 \begin{equation}
        \label{eq:laplacetransformKink}
\langle \widehat{n}(J_L)\rangle_{\sigma(r)} =  {\partial_\lambda \log Z }\vert_{\lambda=0}= \left \langle e^{-J_L \theta} \right \rangle_\text{gas}=\int_{-\infty}^\infty d\theta e^{-J_L \theta} \langle n(\theta)\rangle_{\sigma(r)}.
 \end{equation}
At high energies ($r\to0$), this can be computed with the kink approximation (\ref{eq:kinkansatz}). It is enough to integrate over positive $\theta$ and symmetrize the result in $\pm J_L $. We write
\begin{equation}
       \int_{0}^\infty d\theta e^{-J_L \theta} \langle n(\theta)\rangle_{\sigma(r)} \overset{r\to0}{=}\int_{-\infty}^\infty d\theta e^{-J_L \theta} \left(n_\text{kink}(\theta-L) - \Theta(-\theta) n_c \right)
        = \frac{n_c}{J_L} + (m r)^{J_L} \widehat{n}_\text{kink}(J_L), \label{eq:usingkink}
\end{equation}      
where $\widehat{n}_\text{kink}(J_L)=\int_{-\infty}^\infty dx\,e^{-J_L x}n_\text{kink}(x)$ is the bilateral Laplace transform of $n_\text{kink}$. Symmetrizing under $\pm J_L$  we obtain
\begin{equation}
        \langle \widehat{n}(J_L)\rangle_{\sigma(r)} =  (m r)^{J_L} \widehat{n}_\text{kink}(J_L) +  (m r)^{-J_L} \widehat{n}_\text{kink}(-J_L). \label{eq:scalingSigmanHat}
\end{equation}
This result is valid at leading order in $m r$. We discuss subleading corrections to this result in section~\ref{sec:subleadingsigma}.

We thus see that, even though the overall coefficient $\widehat{n}_\text{kink}(\pm{J_L})$ depends on the precise shape of the kink solution (which we will compute soon), 
the $r\to 0$ scaling is entirely controlled by the position of the edges. As argued in (\ref{eq:therexpansion}),
 the $r^{\pm J_L}$ factor follows from the matching of the detector $\widehat{n}(J_L)$ with the leading CFT light-ray operators corresponding to the trajectories $\Delta_L({J_L}) = \pm J_L$.
 
 For $J_L=0$ the integral (\ref{eq:laplacetransformKink}) receives extensive contributions from the bulk of the box. Of course, in this case we are simply computing the total number of particles
 as $r\to0$,
 \begin{equation}
   \label{eq:logscale}     \langle \widehat{n}(J_L=0)\rangle_{\sigma(r)} \to \langle N \rangle_\text{gas} = 2 \log(1/mr) n_c + O(L^0).
 \end{equation}
 This logarithmic scaling can also be understood through detector matching. For $J_L=0$ the $\Delta(J_L) = \pm J_L$ trajectories intersect and degenerate, see figure~\ref{fig:chewfrautschi}.
 At this point, the UV dilatation operator ceases to be diagonalizable and develops nontrivial Jordan blocks. This corresponds to the possibility of combining the $\pm J_L$ trajectories into
 logarithmic multiplets at $J_L=0$.\footnote{For example, introducing an arbitrary reference scale $\mu$, we can define \cite{Chang:2025zib}
 \begin{equation}
        \mathcal{D}^\text{log} = \underset{J_L \to 0}{\lim} \frac{\mathcal{D}_{J_L} -\mu^{-2J_L} \mathcal{D}_{-J_L}}{J_L}
 \end{equation}
 where here $\mathcal{D}_{-J_L}$ denotes the shadow transform of the $\mathcal{D}_{J_L}$ trajectory. $ \mathcal{D}^\text{log}$ is linearly independent from $\mathcal{D}_0$ and transforms under dilatations as 
 \begin{equation}
        D\cdot    \mathcal{D}^\text{log} =  2 \mathcal{D}_{0},
 \end{equation} 
 i.e.\ $\mathcal{D}_0$ and $ \mathcal{D}^\text{log}$ form a logarithmic multiplet.%
 \label{fn:log_multiplet}} 
This unusual transformation under dilatations produces $\log r$ factors in correlation functions such as (\ref{eq:logscale}).

\begin{figure}[t!]
\vspace{-1cm}
  \centering
  \includegraphics[width=\linewidth]{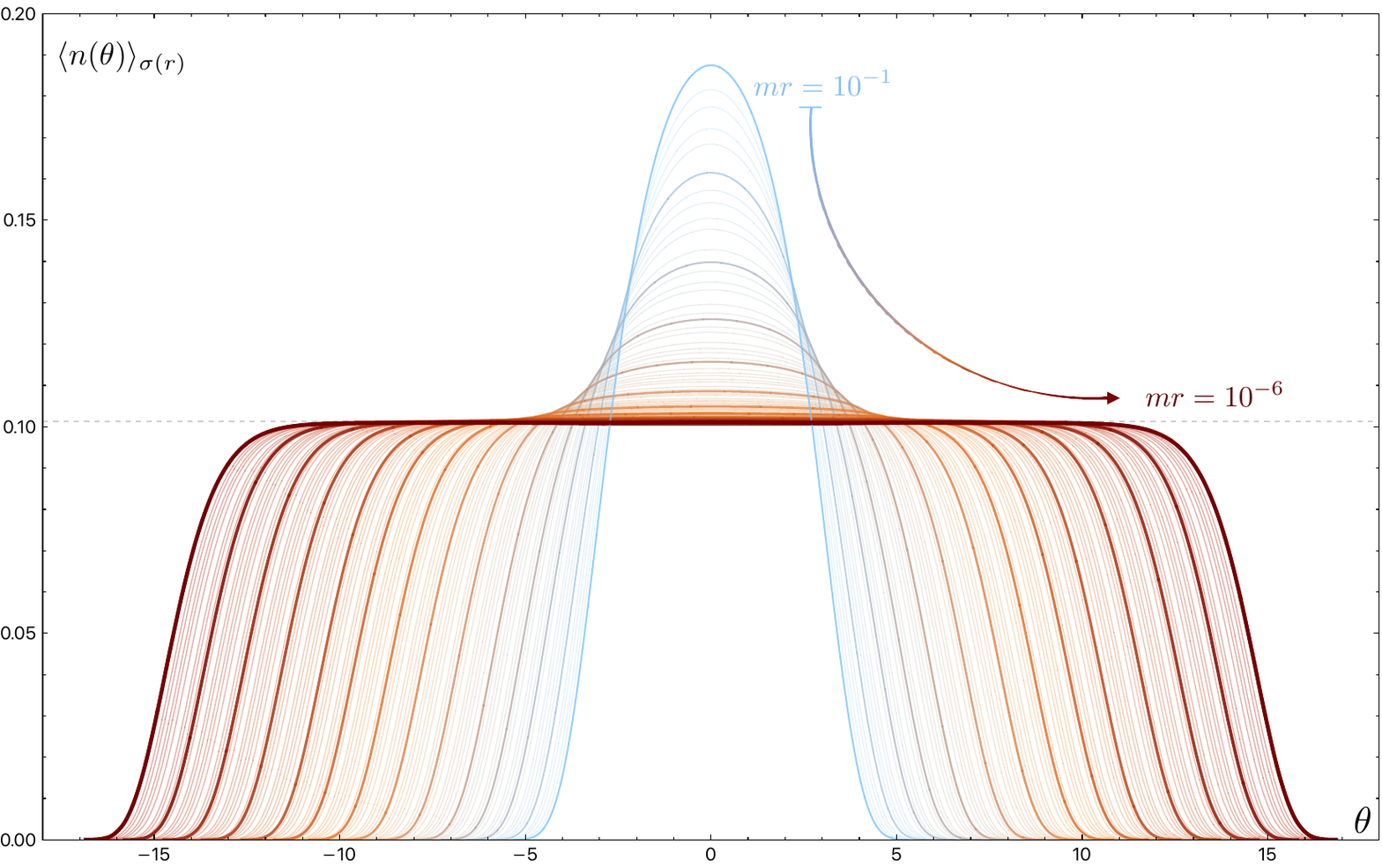}
  \vspace{-1.5cm}
  \caption{The particle number density $\langle n(\theta)\rangle_{\sigma(r)}$ as $mr$ is varied from $10^{-1}$ (light blue) to $10^{-6}$ (brown). We interpret this as a solution to the flow equations (\ref{eq:thewaveequation}) defined by the Chew-Frautschi plot. At sufficiently small $mr$, this evolution equation reduces to a massless wave equation corresponding to the contribution from the leading Regge trajectories. Its solution is a plateau with boundaries which propagate outwards with unit velocity. Subleading trajectories correspond to massive waves which dissipate as $mr \to 0$, see section~\ref{sec:isingdetails}.}
  \label{fig:sigmadensity}
\end{figure}

 A nice feature of the form factor expansion (\ref{eq:sigmaffexpasion}) is its fast convergence rate. 
 In fact, it is sufficient to truncate the sum in (\ref{eq:sigmaffexpasion}) to $N=7$ in order to compute $\langle \sigma \sigma\rangle$ and $ \langle n(\theta)\rangle_{\sigma(r)}$ down to $mr = 10^{-10}$ with several digits of relative precision,
 thus allowing us to numerically observe the UV scaling laws with exponents $\Delta_\sigma$, $\Delta_{L}(J_L)$ over several decades in $mr$. 
 While expected, it is remarkable how efficient the elementary sum over a few IR asymptotic particles is in extracting  UV data. 
 In figure~\ref{fig:sigmadensity}, we use this method to display the evolution of 
 $ \langle n(\theta)\rangle_{\sigma(r)}$ in $r$. At small $r$,
  the wave-like evolution (\ref{eq:thewaveequation}) is clearly visible.

 Although in the thermal deformation of the Ising CFT we can resum the form factor expansion exactly, see section~\ref{sec:fredholm}, this method provides a practical strategy to study detector matching in more
 nontrivial strongly interacting integrable flows for which the form factors can be computed but the series cannot be resummed. We apply this strategy to the magnetic 
 deformation of the 2d Ising model in section~\ref{sec:magnetic}.

 We have so far given an alternative thermodynamic derivation of detector matching to leading order 
 for the $\widehat{n}(J_L)$ operators in the $\sigma$ state of the thermal Ising model. 
 How general is this argument? The crucial step was the identification of the form factors  $|\langle \sigma|\theta_1,\dots,\theta_N\rangle|^2$ 
 as a short-range, factorized potential for the various particles contributing to the correlator, 
 from which thermodynamics is expected to follow in the large-volume limit $r\to0$. This rapidity-space factorization property of form factors can be argued for any relevant operator in a broad class of integrable flows \cite{Delfino:1996nf}. 
 For example, it applies equally well to the $\sigma$ operator in the magnetic deformation of the Ising model, 
 and so we obtain a thermodynamic gas in that case as well.%

 An interesting case to consider is that of Virasoro descendants. Consider $\mathcal{O}_\sigma^{(\ell,\ell)}$, a scalar Virasoro descendant of $\sigma$ at level $(\ell,\ell)$.
 As shown in \cite{Cardy:1990pc}, these operators have form factors of the form
 \begin{equation}
        \langle \mathcal{O}_\sigma^{(\ell,\ell)} |\theta_1 \dots \theta_N\rangle =  P_\ell(\{e^{\theta_i}\}, \{e^{-\theta_i}\})  \langle \sigma |\theta_1 \dots \theta_N\rangle
 \end{equation}
 where $P_\ell$ is a symmetric polynomial of homogeneous degree $\ell$ in each set of variables. 

 The effect of the additional factor $|P_\ell|^2$ in the thermodynamic picture can be understood by rewriting
 \begin{equation}
        P_\ell(\{e^{\theta_i}\}, \{e^{-\theta_i}\}) = e^{2 \ell L} P_\ell(\{e^{\theta_i -L}\}, \{e^{-\theta_i - L}\}).
 \end{equation} 
We see that, after extracting the overall homogeneity factor, the effect of the polynomial term is concentrated at the boundaries of the box. Importantly, at fixed descendant level, 
it has the effect of marking up to $4\ell$ particles at the boundaries, regardless of $N$. 
These marked particles do affect the precise shape of the boundary profiles, but they do
 not affect the bulk density and its volume growth in the thermodynamic limit $r\to0$. We thus expect the same leading scaling $(mr)^{\pm J_L}$ expected from detector matching  (\ref{eq:therexpansion}) to also hold for $\langle \widehat{n}(J_L)\rangle_{\mathcal{O}_\sigma^{(\ell,\ell)}}$.
 
Of course, detector matching is expected to hold universally, with the thermodynamic argument serving here as a robust check. 
It is tempting to reverse the logic: detector matching suggests that the gas in a box picture could hold in general.
It is natural to conjecture that any generic\footnote{A non-generic example is the $\epsilon$ state in the thermal Ising model. This is a two-particle state, and so there is no thermodynamic interpretation. See section~\ref{sec:epsilonsection} for further discussion.} scaling operator in any massive flow has form factors which, 
as potentials, behave thermodynamically. It would be fantastic to establish so. Evidence could be gathered via the MPS method \cite{Jha:2024jan}.

\section{More detectors, more states, and more trajectories in the 2d Ising model at $T>T_c$}
\label{sec:isingdetails} 
In this section we analyze in more detail various aspects of detector matching in the 2d Ising model at $T>T_c$. In section~\ref{sec:fredholm} we provide further details on the thermodynamic limit of the $\sigma$ state. In section~\ref{sec:subleadingsigma} we discuss the evolution of detector correlators to subleading order. Section~\ref{sec:epsilonsection} discusses detector matching in the $\epsilon$ state. Section~\ref{sec:multiscale} discusses the matching of more general ``multiparticle'' detectors.
\subsection{Fredholm solution}
\label{sec:fredholm}

Let us now describe the exact solution of the Feynman gas introduced in section~\ref{sec:ising101}. Here we provide a pedagogical review of the results in 
\cite{Wu:1975mw, Tracy:1990tn, Babelon:1992sn, Bernard:1994re,  Liu:2026asy}.

The main observation is that the form factor expansion of the correlator $\langle \sigma \sigma \rangle$ can be resummed into a Fredholm determinant. Noting that 
\begin{equation}
        \prod_{i=1}^N z(\theta_i)  \prod_{j>i} \tanh^2\left(\frac{\theta_i-\theta_j}{2}\right) =
         \underset{N\times N}{\det}\left(H_{ij} \equiv \frac{ z(\theta_i)^{\frac{1}{2}}z(\theta_j)^{\frac{1}{2}}}{\cosh\left(\frac{\theta_i - \theta_j}{2}\right)}\right)
\end{equation} 
we find
\begin{equation}
\langle \sigma(r) \sigma(0)\rangle = \sum_{N \text{ odd}} \frac{\mathcal{N}^2 }{ N!} \int d^N \theta \underset{N \times N }{\det} H_{ij} = \mathcal{N}^2   \left(\tfrac{1}{2} \text{Det}(1 + H) -  \tfrac{1}{2}\text{Det}(1- H)\right) \label{eq:FredholmEq}
\end{equation}
where $\text{Det}$ denotes the Fredholm determinant and $H \equiv H(\theta, \theta')$ is understood as a kernel. 
In the thermodynamic limit it is sufficient to keep the $\text{Det}(1 + H)$ term up to subextensive corrections. This is equivalent to summing over both even and odd $N$, see footnote \ref{ft:sigmamu}. 

As a first exercise, let us compute the thermodynamic bulk pressure $p(z_c) = \Delta_\sigma$. We start with 
\begin{equation}
        \log Z = \sum_{n=1}^\infty \frac{(-1)^{n+1}}{n} \Tr  H^n  = \sum_{n=1}^\infty \frac{(-1)^{n+1}}{n} \int d^n\theta \prod_{i=1}^n H(\theta_i, \theta_{i+1}),
\end{equation}
where $\theta_{n+1}\equiv \theta_1$. In the thermodynamic limit, we can separate the integral over the center-of-mass coordinate,  
which runs over the width of the box and produces a trivial volume factor, and the relative coordinates $\eta_i = \theta_i - \theta_{i+1}$, 
whose integral can be extended to $\mathbb{R}$ due to the exponential decay of the kernel with separation. 
In the relative coordinate integrals we can set $z(\theta) \to z_c$  and the kernel $H$ becomes
 translation invariant. 
It can therefore be diagonalized in Fourier space,
\begin{equation}
 \widehat{H}(q) \to \int_{-\infty}^\infty {d\theta} \frac{\,  e^{i q \theta}\ z_c}{\cosh \frac{\theta}{2}} = \frac{1}{\cosh \pi q}.
\end{equation}
Since
\begin{equation}
\int d^{n} \eta \, \delta\left(\sum_i \eta_i\right)\prod_{i=1}^n H(\eta_i)  = \int \frac{dq}{2\pi} \widehat{H}(q)^n,
\end{equation}
we  obtain, up to subextensive contributions,
\begin{equation}
        \log Z = V \sum_{n=1}^\infty \frac{(-1)^{n+1}}{n} \int \frac{dq}{2\pi} \widehat{H}(q)^n = V \int \frac{dq}{2\pi} \log (1 + \widehat{H} ) = V/8 
\end{equation}
from which we read  $p(z_c)= \Delta_\sigma = 1/8 $.

 The Fredholm resummation (\ref{eq:FredholmEq}) can also be used to compute the particle number density $\langle n(\theta) \rangle_{\sigma(r)}$. Using the Jacobi identity,
\begin{equation}
        \delta \log \text{Det} (1 + H) = \text{Tr}( \delta H ( 1 + H )^{-1})
\end{equation}
we find
\begin{equation}
\langle n(\theta) \rangle_{\sigma(r)} = \frac{\delta\log Z}{\delta \log z(\theta)} = \int d\theta' H(\theta,\theta')(1 + H)^{-1}(\theta', \theta).
\end{equation}
It is useful to define the resolvent
\begin{equation}
         R = H(1+H)^{-1},
\end{equation}
 so that $\langle n(\theta) \rangle_{\sigma(r)} = R(\theta, \theta)$.  Thus, to compute the density, it is enough to compute the resolvent $R(\theta, \theta')$.
  The resolvent satisfies a beautiful integral equation:
 \begin{equation}
        R(\theta, \theta') + \int d\theta'' R(\theta,\theta'') H(\theta'',\theta') = H(\theta,\theta'), \label{eq:resolventintegraleq}
 \end{equation}
 which can be efficiently solved numerically at finite $r$ through discretization in $\theta$. This solution can be compared with the results depicted in figure~\ref{fig:sigmadensity}, with perfect agreement.

 Instead, let us solve for the resolvent analytically in the thermodynamic limit. 
It is sufficient to zoom in around the right kink region. Defining $y = \theta - \log(1/mr)$ and similarly for $y'$, we can approximate
\begin{equation}
        H(\theta,\theta')\to H_\text{kink}(y,y')\equiv\frac{\exp\left(-\frac{1}{4}{(e^y + e^{y'})}\right)}{2\pi \cosh\left(\frac{ y-y'}{2}\right)}.
\end{equation}
The kernel $H_\text{kink}$ can be diagonalized explicitly (as was done in \cite{Tracy:1990tn,Liu:2026asy}). 
The eigenfunctions are
\begin{equation}
        \psi_p(y) =     \frac{\sqrt{2 p \sinh(\pi p)}}{\pi} K_{i p}\left(\frac{e^{y}}{4}\right),
\end{equation}
 with $K_\nu(x)$ the modified Bessel function of the second kind and $p>0$. They satisfy completeness and orthonormality:
\begin{equation}
\int_{\mathbb{R}^+}dp \psi_p(y) \psi_p(y') = \delta(y-y'), \qquad \int_\mathbb{R} dy \psi_p(y) \psi_{p'}(y) = \delta(p-p').
\end{equation}
 The corresponding eigenvalues are
 \begin{equation}
      \int dy'  H_\text{kink}(y,y')  \psi_p(y') = \frac{\psi_p(y)}{\cosh(\pi p)}.
 \end{equation}
 The resolvent is then 
 \begin{equation}
        R_\text{kink}(y,y') = \int_0^\infty dp  \frac{\psi_p(y) \psi_p(y')}{1 + \cosh \pi p}.
 \end{equation}
Finally, we obtain the particle number density in the thermodynamic limit,
\begin{equation}
      n_\text{kink}(y) =   R_\text{kink}(y,y) = \frac{e^{2 y}}{16 \pi^2} \left(K^2_1\left(\frac{e^y}{4}\right)-K^2_0\left(\frac{e^y}{4}\right) \right).
\end{equation}
We will explore this result further in section~\ref{sec:subleadingsigma}.

A useful consequence of having introduced the resolvent technology is that we can now straightforwardly compute correlators with multiple detectors in the $\sigma$ state! 
Using that
\begin{equation}
        \frac{\delta R(\theta_i,\theta_j)}{\delta \log z(\theta_k)} = - R(\theta_i,\theta_k)R(\theta_k,\theta_j)
\end{equation}
away from coincident points, we obtain, by
taking repeated variations,
\begin{equation}
        \langle n(\theta_1) \dots n(\theta_n)\rangle_{\sigma(r)} = \sum_{\pi \in S_n} (-1)^{|\pi|}\prod_i R(\theta_{i},\theta_{\pi_i}) = \underset{i,j}{\det} R(\theta_i, \theta_j). \label{eq:densitydensitycorrelator}
\end{equation}
It follows from (\ref{eq:resolventintegraleq}) that the resolvent decays exponentially with rapidity separation. We therefore see from (\ref{eq:densitydensitycorrelator}) that the density correlators have a finite correlation length 
in the thermodynamic state created by $\sigma$. %

\subsection{Multiscale detectors}
\label{sec:multiscale}

The detectors $\widehat{n}(\theta)$ are local on the mass shell.  
It is interesting to consider a more general class of multiparticle detectors, such as 
\begin{equation}
        \widehat{n}_{\psi}(\theta) =    \int_{\mathbb{R}^+} d\chi \psi(\chi) a^\dagger(\theta + \chi)a^\dagger(\theta - \chi)a(\theta - \chi)a(\theta+\chi). \label{eq:multiscale}
\end{equation}
Here we restrict the detector to carry no momentum, $[ P^\mu,\widehat{n}_{\psi}(\theta)]=0$, which forces the initial and final rapidities to be the same. We call detectors such as (\ref{eq:multiscale}) multiscale detectors. 

In this section we will study the matching of $\widehat{n}_{\psi}(J_L) $ into UV detectors by studying the matrix element $\langle  \widehat{n}_{\psi}(J_L)  \rangle_{\sigma(r)}$ in the $r\to0$ limit. 
If the relative rapidity smearing $\psi(\chi)$ is of compact support on the mass shell, we will find that  $\widehat{n}_{\psi}(J_L) $ matches to $45^\circ$ identity trajectories. On the other hand, for exponential $\psi(\chi) = \exp(\chi p)$ with $p>0$, we find that $\widehat{n}_{\psi}(J_L)$  matches onto composite horizontal trajectories built out of pairing Virasoro trajectories. This requires broadening the class of detectors allowed in the detector matching equations (\ref{eq:detectormatchingequation}) in the case of multiscale detectors. 

The correlator $\langle \widehat{n}_{\psi}(J_L) \rangle_{\sigma(r)}$ can be computed with (\ref{eq:densitydensitycorrelator}). Defining
\begin{equation}
        g(\theta,\chi) = n(\theta + \chi)n(\theta - \chi) - R(\theta + \chi,\theta - \chi)^2, \label{eq:gdef}
\end{equation}
        we have 
\begin{equation}
        \langle \widehat{n}_{\psi}(J_L) \rangle_{\sigma(r)} = \int_\mathbb{R} d\theta e^{-J_L \theta}\int_{\mathbb{R}^+} d\chi \psi(\chi) g(\theta,\chi). \label{eq:twoparticles}
\end{equation}
Let us consider first the case in which $\psi(\chi)$ is of compact support. In this case, we can approximate the resolvents by their kink solutions,  
\begin{equation}
          \langle \widehat{n}_{\psi}(J_L) \rangle_{\sigma(r)}  \overset{r\to0}{=}  \int_\mathbb{R} d\theta e^{-J_L \theta}\int_{\mathbb{R}^+} d\chi \psi(\chi) \left(g_\text{kink}(\theta-L,\chi )+ g_\text{kink}(-\theta-L,\chi) - c(\chi)\right),
\end{equation}
where $c(\chi) = \underset{r \rightarrow 0}{\lim}\, g(\theta,\chi)$ with $\chi, \theta$ fixed. We can then use the same steps discussed in equations (\ref{eq:usingkink}) and (\ref{eq:scalingSigmanHat}) to conclude that
\begin{equation}
         \langle \widehat{n}_{\psi}(J_L) \rangle_{\sigma(r)} = (mr)^{J_L} \int d\theta  d\chi e^{-J_L \theta} \psi(\chi )g_{\text{kink}}(\theta,\chi) + (J_L \leftrightarrow -J_L). \label{eq:edge}
\end{equation}
This result is valid at leading order as $r\to 0$ and for $\psi(\chi)$ of compact support.

\begin{figure}[t!]
\vspace{-1cm}
  \centering
  \includegraphics[width=0.7\linewidth]{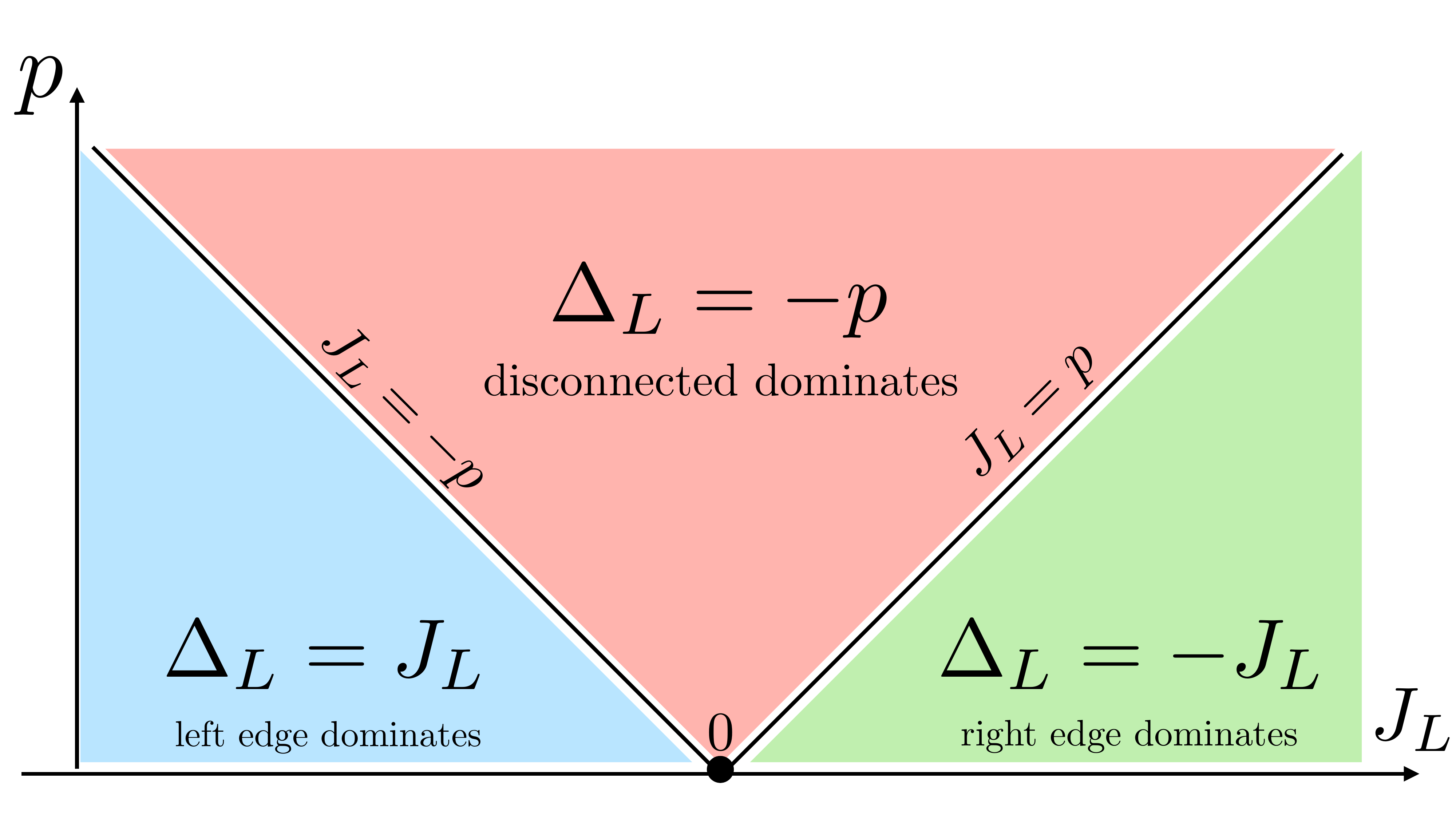}
  \caption{Leading contribution to detector matching for the multiscale detectors (\ref{eq:multiscale}) when $\psi(\chi) = \exp(\chi p)$. For $|\Re(J_L)|>p$, the two-particle detectors match onto the identity trajectory. However, for $|\Re(J_L)|<p$ they match onto the composite detectors (\ref{eq:composite}).} \label{fig:multiscale}
\end{figure}

Next, consider the case in which $\psi(\chi) = \exp(p \chi)$. The region of finite $\chi$ gives the edge contribution (\ref{eq:edge}). However, in this case, the correlator also receives important contributions from
 pairs of particles at opposite edges of the gas distribution, where $\chi = L + y$ with $y$ finite as $r\to0$. 
 Due to the exponential decay of the resolvent at large rapidity separations, in this regime we can drop the connected term in (\ref{eq:gdef}) and approximate 
 \begin{equation}
g(\theta,\chi)  \overset{\substack{r\to0}}{=} n_\text{kink}(y + \theta )n_\text{kink}(y-\theta ). \label{eq:disconnected}
 \end{equation} 
 Integrating (\ref{eq:disconnected}) in (\ref{eq:twoparticles}) produces
 \begin{equation}
        \langle \widehat{n}_{\psi}(J_L) \rangle_{\sigma(r)} \supset (mr)^{-p} \left( C(J_L, p) \equiv \int d\theta \int dy e^{-J_L \theta + y p} n_\text{kink}(y+ \theta )n_\text{kink}(y-\theta) \right) \label{eq:pscale}
 \end{equation}
which dominates the small-$r$ asymptotics relative to the contributions from individual edges (\ref{eq:edge}) if $p>|\Re{J_L}|$. 

The scaling (\ref{eq:pscale}) can be understood from detector matching.  Since the contribution comes from
 detecting particles which are far separated on the mass shell, 
the finite correlation length of density correlators implies that $\langle  \widehat{n}_{\psi}(J_L)  \rangle_{\sigma(r)}$ factorizes into one-particle detector
 contributions in the $r\to0$ limit.
 To leading order, each one-particle detector matches onto a $45^\circ$ detector in the identity trajectory. This supports the matching of the
 two-particle detector (\ref{eq:multiscale}) onto a composite UV detector \cite{Caron-Huot:2022eqs}
\begin{equation}
        \widehat{n}_{e^{\chi p}}(J_L)  \to \mathcal{H}_{J_L,p} \equiv \mathcal{D}^\text{UV}_{\frac{1}{2}(J_L + p), -}
        \mathcal{D}^\text{UV}_{\frac{1}{2}(J_L - p), +}, \label{eq:composite}
\end{equation}
where the UV detector $\mathcal{D}^\text{UV}_{s, \pm}$ has spin $J_L = s$ and dimension $\Delta_L = \pm s$. 
The horizontal detector $\mathcal{H}_{J_L,p}$ has quantum numbers $(\Delta_L, J_L) = (-p, J_L)$, which is
  compatible with the scaling (\ref{eq:pscale}). We summarize the dominant contributions in the detector matching in the $J_L \times p$ plane in figure~\ref{fig:multiscale}.

\subsection{Subleading matching}
\label{sec:subleadingsigma}
Next, we would like to explore subleading matching. In the rest of this section, we set $m=1$. The subleading behavior of $\<n(\theta)\>_{\sigma(r)}$ in $r$ was computed in~\cite{Liu:2026asy} using the Fredholm methods described earlier in this section. Taking the 
(two-sided) Laplace transform of their result yields
\begin{equation} \label{sigma rhoJ}
    \begin{aligned}
        \<\widehat{n}(J_L)\>_{\s(r),\text{kink}}&=-\frac{4^{-J_L}J_Lr^{J_L}\Gamma\left(\frac{-J_L}{2}\right)^3\left(2(J_L+r-J_Lr)+(J_L-2)r\log\left(\frac{8}{r e^{\g_E}}\right)\right)}{16\pi^{3/2}\Gamma\left(\frac{3-J_L}{2}\right)\left(-2+r\log\left(\frac{8}{re^{\g_E}}\right)\right)} \\
        &+(J_L\rightarrow -J_L)+O(r^{2\pm J_L}).
    \end{aligned}
\end{equation}
The number operator is obtained from the $J_L\rightarrow 0$ limit, which is given by
\begin{equation}
    \<N\>_{\s(r)}=\frac{1}{\pi^2}\left(-2\log{r}+4\log{2}-3\gamma_E-\psi^{(0)}\left(\frac{3}{2}\right)\right)+O(r).
\end{equation}

We see the $\log{r}$ scaling obtained in the previous section, along with additional $O(r^0)$ terms coming from the details of the wave profile. The first subleading terms are of order $r^{\pm J_L+1}$. This scaling is consistent with the fact that the first subleading
$\mathbb{Z}_2$-even Regge trajectory in the 2d Ising CFT is the $\epsilon$ trajectory given by $\Delta_L=\pm J_L+1$, and consistent with conformal perturbation theory. In $\th$ space, we can interpret these subleading contributions as massive waves that decay exponentially,
which we can see as the relaxation of the bulk density to the plateau value in figure~\ref{fig:sigmadensity}. 

One particular aspect about this example is that because the $\epsilon$ trajectory and the first conformal perturbation
theory corrections contribute to the same order in $r$, we are unable to distinguish their contributions. Nonetheless, the subleading scaling and evolution of the density at this order are both predictions of detector matching and conformal perturbation theory.

\subsection{Evolution in the $\epsilon$ state}
\label{sec:epsilonsection}
We also study the density in the $\epsilon$ state, to check to what extent matching and evolution are universal. The $\epsilon$ density is much simpler than the $\sigma$ density, since $\epsilon$ only creates two particles. Already, this tells us we cannot interpret the $\epsilon$
density as that of a gas. Additionally, the $\epsilon$ form factor, given by
\begin{equation}
        F_2^{\epsilon}(\th_1,\th_2)=-im\sinh{\frac{\th_1-\th_2}{2}},
\end{equation}
does not factorize. Nonetheless, our statements of evolution and matching still hold. We can exactly compute the density, which is given by
\begin{equation}
        \<n(\th)\>_{\epsilon(r)}=\frac{e^{-r\cosh{\th}}(\cosh{\th}K_1(r)-K_0(r))}{K_1^2(r)-K_0^2(r)}.
\end{equation}
As expected, this looks like two particles localized in rapidity space with rapidity $\pm\log{\frac{2}{r}}$, and to leading order in $r$ obeys the wave equation. As is the case with $\s$, we can get the leading behavior kink function:
\begin{equation}
        \<n(y)\>_{\e(r),\text{kink}}=e^{y-e^{y}}.
\end{equation}
We can also write the harmonic components exactly in terms of Bessel functions:
\begin{equation}
        \<\widehat{n}(J_L)\>_{\e(r)}=\frac{2K_0(r)K_{J_L}(r)-K_1(r)(K_{J_L-1}(r)+K_{J_L+1}(r))}{K_0^2(r)-K_1^2(r)}.
\end{equation}
Expanding in powers of $r$, we find $r^{\pm J_L}$ terms and $r^{\pm J_L+2}$ terms, but no $r^{\pm J_L+1}$ terms. This is consistent with the fact that the three-point function of three $\epsilon$s vanishes, and hence receives no contributions from
the $\epsilon$ trajectory, nor first-order corrections from conformal perturbation theory.

Here, we should comment on the $r$ scaling of the total particle number, which corresponds to $J_L=0$ of $\<\widehat{n}(J_L)\>$. As we have seen, the total particle number for $\sigma$ scales as $\log{r}$. As described in footnote~\ref{fn:log_multiplet},
this is because the leading trajectory and its shadow intersect at $J_L=\Delta_L=0$ and transform in a log-multiplet. This corresponds to a double pole in the Lorentzian inversion formula. To see how a double pole in the Lorentzian inversion formula
yields $\log{r}$ scaling, for a moment we will consider a kinematically related observable, the Regge limit of a CFT four-point function, whose behavior is governed by the same spectrum of light-ray operators~\cite{Kravchuk:2018htv}. In the Regge limit, the leading
behavior of a four-point function is given by~\cite{Costa:2012cb,Caron-Huot:2017vep}
\begin{equation}
        \<\phi_1(x_1)\phi_2(x_2)\phi_3(x_3)\phi_4(x_4)\>\sim\int_{\frac{d}{2}-i\infty}^{\frac{d}{2}+i\infty}\frac{d\Delta}{2\pi i}\underset{J=j(\Delta)}{\text{Res}}\frac{C^t(\Delta,J)+e^{-i\pi J}C^u(\Delta,J)}{1-e^{2\pi iJ}}r^{1-J}
\end{equation}
where $C^t(\Delta,J)$ and $C^u(\Delta,J)$ are computed using the Lorentzian inversion formula with $C^u(\Delta,J)$ related to $C^t(\Delta,J)$ by swapping operators 1 and 2, and $r^{1-J}$ is the leading behavior of a conformal block in the Regge limit. If $C^t(\Delta,J)$ has a single pole at $j(\Delta)$, we
get the expected $r$ scaling of $r^{\Delta_L}$. However, a double pole at $\Delta_L=J_L=0$ yields $\log{r}$ scaling. It turns out that the Lorentzian inversion formula for $\<\s\s\s\s\>$ has a double pole at $\Delta_L=J_L=0$, while $\<\e\e\e\e\>$ has a single pole (details of this calculation can be found in appendix~\ref{sec:Lorentzian_Inversion_Formula}).

It is interesting to ask whether these Lorentzian inversion formula calculations can be considered a CFT prediction of which states have a thermodynamic interpretation under RG flow. The form factor expansion \eqref{eq:sigmaffexpasion} implies that the particles created by a local operator $\mathcal{O}$ are confined to a box
in rapidity space with volume $V\sim -\log{mr}$. A double pole then predicts that $N\sim V$ as $r\rightarrow 0$, and so the state has a thermodynamic interpretation, while a single pole predicts there is no such interpretation. It would be interesting to check if such a prediction holds in other 2d CFTs.

For higher-dimensional CFTs, the space of on-shell particles is given by hyperbolic space, and can be parametrized by $(\theta,\widehat{n})$ where $\theta$ is rapidity and $\widehat{n}\in S^{d-2}$. Again, the particles will be confined to a box with $\theta < -\log{mr}$, but now the volume goes as $V\sim r^{2-d}$ (with the appropriate
factors of $m$ for dimensional analysis). In order for a state to have a thermodynamic interpretation, the smallest value of $\Delta_L$ at $J_L=0$ (the spin of the number operator) must be $2-d$. We can then see that this condition will never hold for $d>2$ from the following argument: the Regge intercept is the largest value of $J$ at $\Delta=\frac{d}{2}$
and can be at most 1 \cite{Costa:2012cb,Caron-Huot:2017vep}. Then the smallest possible value of $\Delta_L(J_L=0)$ is $\frac{2-d}{2}$, meaning the number of particles in any state grows at most as $r^{\frac{2-d}{2}}$, which is slower than the growth of the volume. Thus, no state created by a local operator can have a thermodynamic interpretation.

\section{Evolution in the 2d Ising model in a magnetic field}
\label{sec:magnetic}

An important property of detector matching is its universality: we conjecture that the spectrum $\Delta_{L,i}(J_L)$ appearing in
 (\ref{eq:therexpansion}) corresponds to that of light-ray operators in the UV CFT, 
 and thus is universal for all RG flows stemming from a given fixed point --- even if these have completely different particle spectra, and thus different asymptotic detector operators $\mathcal{D}^{\mathrm{IR}}$.

To illustrate this idea, we now consider the 2d Ising model at critical temperature in the presence of a magnetic field,
\begin{equation}
        S = S_\text{Ising CFT} + h \int d^2x \sigma(x).
\end{equation}
In contrast to the thermal Ising model (\ref{eq:thermalising}), the magnetic theory is a highly nontrivial strongly interacting system. 
Thanks to the integrability of this theory, the particle spectrum is known exactly \cite{Zamolodchikov:1989hfa,Zamolodchikov:1989fp}. It consists of eight species with masses
\begin{align}
m_1&=m,&
m_2&=2m_1\cos\frac{\pi}{5},\\
m_3&=2m_1\cos\frac{\pi}{30},&
m_4&=2m_2\cos\frac{7\pi}{30},\\
m_5&=2m_2\cos\frac{2\pi}{15},&
m_6&=2m_2\cos\frac{\pi}{30},\\
m_7&=4m_2\cos\frac{\pi}{5}\cos\frac{7\pi}{30},&
m_8&=4m_2\cos\frac{\pi}{5}\cos\frac{2\pi}{15}.
\end{align}
Also due to integrability, their scattering amplitudes are purely elastic and known exactly, see appendix~\ref{sec:magneticapp}. 

\begin{figure}[p!]
  \thisfloatpagestyle{empty}
  \centering
  \vspace*{-1cm}

  \makebox[\textwidth][c]{%
    \includegraphics[
      width=190mm,
      height=1.05\textheight
      ,      keepaspectratio
    ]{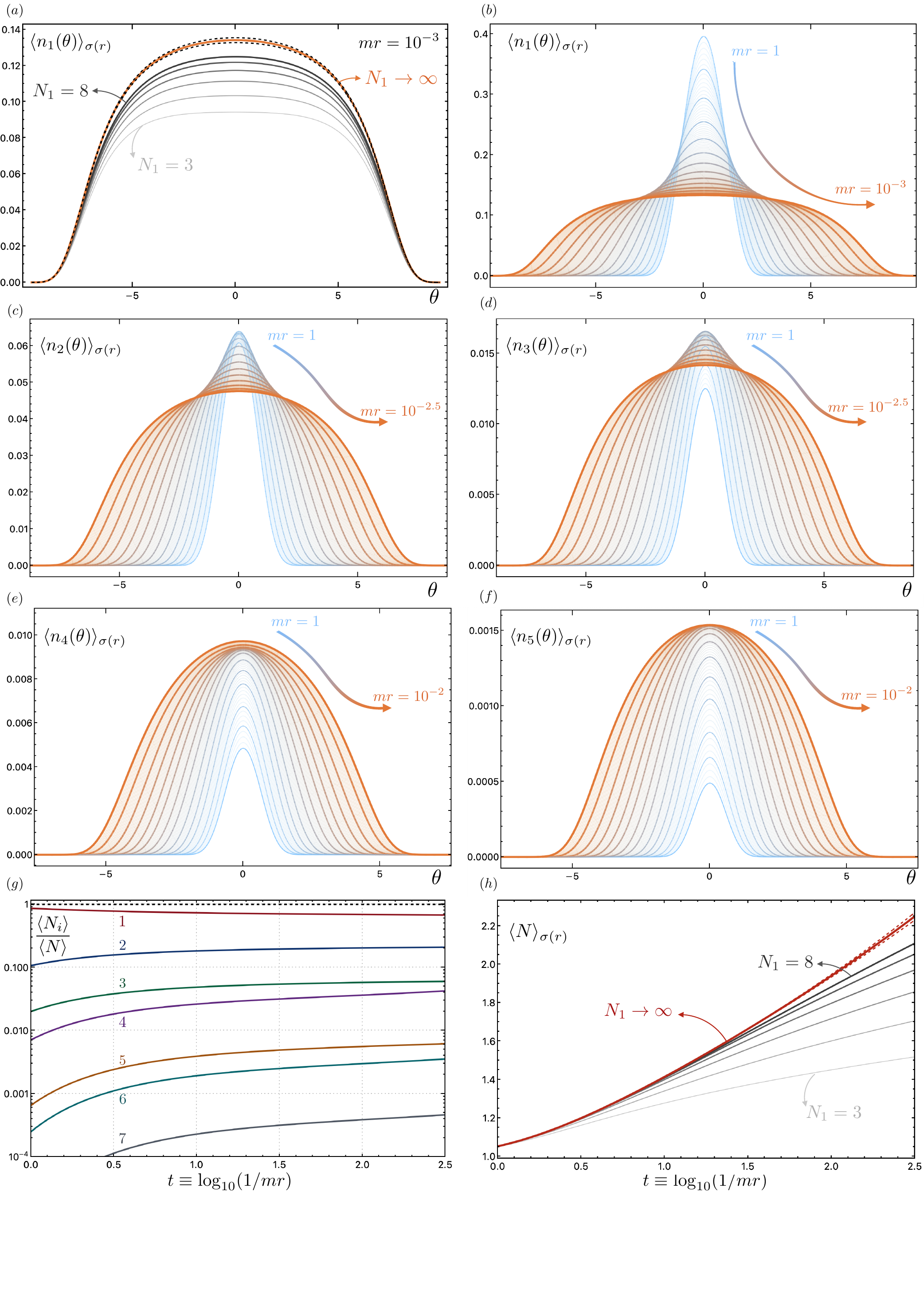}%
  }\vspace{-2cm}

  \caption{(a) We compute the densities using all form factors obtained via fusion from the $N_1$ type-1 particle form factor and extrapolate to $N_1 \to \infty$. This method allows us to extract the densities down to $m r \sim 10^{-3}$. Panels (b)--(f) show the density evolution for particles of types 1--5. Panel (g) computes the fraction of each particle type created by the $\sigma$ state at different scales. Panel (h) describes the expectation value of the number operator with energy. In panels (a) and (h), dashed lines estimate errors in the extrapolation procedure. }
  \label{fig:magneticdensities} 
\end{figure}

In the presence of multiple particle species, we may consider the exclusive number density operators $n_i(\theta) = a^\dagger_{i,\theta}a_{i,\theta}$, which count particles of type $i$ with rapidity $\theta$.
 The type-agnostic detector is then $n(\theta) = \sum_{i=1}^8 n_i(\theta)$. All of these operators are expected to match with detectors in the identity trajectory to leading order (with Wilson coefficients which depend on $i$). As in the thermal theory, this implies the various
densities evolve as massless waves in $(t,\theta)$ and
  \begin{equation}
      (\partial_t^2 - J_L^2) \langle \widehat{n}_i (J_L)\rangle_{\mathcal{O}(r)}  = 0 + \text{subleading} \qquad (r\to0, \forall i). \label{eq:magneticmatching}
 \end{equation}
Our goal will be to observe the evolution of $\langle n_i(\theta)\rangle$ and scaling of $\langle \widehat{n}_i (J_L)\rangle$ by evaluating them in the $\sigma$ state in this theory. 
 Note that this problem is akin to computing the ``distribution functions''\footnote{As opposed to parton distribution functions, here we are decomposing a UV state in terms of the well-defined stable IR particles.} describing how much of each particle type the $\sigma(r)$ state is composed of, as a function of the energy scale $1/r$, see figure~\ref{fig:magneticdensities}g.

 \begin{figure}[t!]
\vspace{-1cm}
  \centering
  \includegraphics[width=0.65\linewidth]{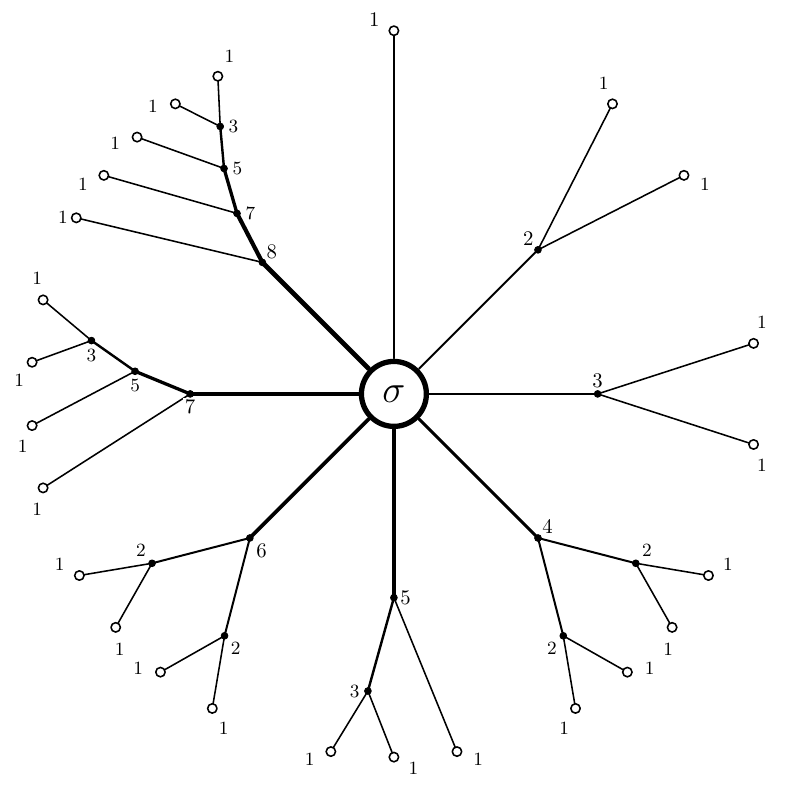}
  \caption{Form factors develop poles whenever intermediate on-shell
   particles can be formed. By tuning the external kinematics and computing 
   the appropriate residues, all form factors can be extracted from those 
   containing exclusively particles $m_1$ as external states.
  For example, in the picture we illustrate how the $F^\sigma_{12345678}$ 
  form factor can be obtained from the $N_1 = 25$ $m_1$-particle form factor. Each line corresponds to the particle labeled at its originating vertex.} \label{fig:fusion}
\end{figure}

We can once again think of this as a problem in statistical mechanics. The form factor expansion defines a grand canonical partition function
\begin{equation}
        \label{eq:magneticffexpansion}
Z[z_i]\equiv \langle\sigma(r) \sigma(0)\rangle = \sum_{N = 0}^\infty \sum_{a_i=1}^8 \int_{\theta_1<\cdots<\theta_N} \hspace{-0.5cm}{d}^N\theta |F^\sigma_{a_1,\dots,a_N}(\theta_1, \dots,\theta_N)|^2 \prod_{i=1}^N z_{a_i}(\theta_i),
\end{equation}
with the fugacities,
\begin{equation}
        z_{i}(\theta) = \frac{1}{2\pi} \exp(-m_i r \cosh\theta)
\end{equation}
and where
\begin{equation}
        F^\sigma_{a_1,\dots,a_N}(\theta_1, \dots,\theta_N) = \langle \sigma | (a_1, \theta_1), \dots,(a_N, \theta_N)\rangle
\end{equation}
are the form factors of $\sigma$ with $N$ particles of types $a_i$ and rapidities $\theta_i$. The density correlators are then computed by variations with respect to the fugacities,
\begin{equation}
       \langle n_i(\theta) \rangle_{\sigma(r)} = \frac{\delta \log Z}{\delta \log z_i(\theta)}. \label{eq:magneticdensity}
\end{equation}

The knowledge of the exact S-matrices allows us to, in principle, recursively compute all of the form factors in this theory. In practice,
this is a formidable task at even modest values of $N$. In particular, no closed-form expressions for the form factors are known, in contrast to the thermal theory.
Therefore, an exact resummation of the series (\ref{eq:magneticffexpansion}) akin to that of section~\ref{sec:fredholm} is beyond the scope of this work, even in the thermodynamic limit.

Nevertheless, we can rely on the convergence rate of the form factor expansion to compute a variety of detector correlators in 
the magnetic theory by truncating the series (\ref{eq:magneticffexpansion}) to sufficiently high $N$ at fixed $r$. If we can compute the correlator to sufficient 
accuracy at small enough $r$ in this way, we should be able to cleanly observe the density evolution and corresponding scaling behaviors predicted by detector matching,
as we did in figure~\ref{fig:sigmadensity} for the thermal theory.

An important aspect of the computation of the form factors is that it is sufficient to compute those involving
 exclusively the lightest particle of mass $m_1$. Once those are computed, all other form factors can be obtained via particle fusion.
 This is illustrated in figure~\ref{fig:fusion}.  In this work, we compute the form factor of $N_1=8$ particles of type 1, 
 as well as the form factors which can be obtained from it via fusion. This provides a total of 142 form factors. 
 Our implementation follows the recipe outlined in \cite{Delfino:2005bh,Delfino:1995zk}, which computed the $N_1=4$ type-1 form factor, 
 from which 18 form factors can be derived. We make the form factors available in an ancillary file attached to this submission. Both the computation of the form factors of particles $m_1$ 
 and the fusion procedure are described in detail in appendix~\ref{sec:magneticapp}.

  \begin{figure}[t!]
\vspace{-1cm}
  \centering
  \includegraphics[width=\linewidth]{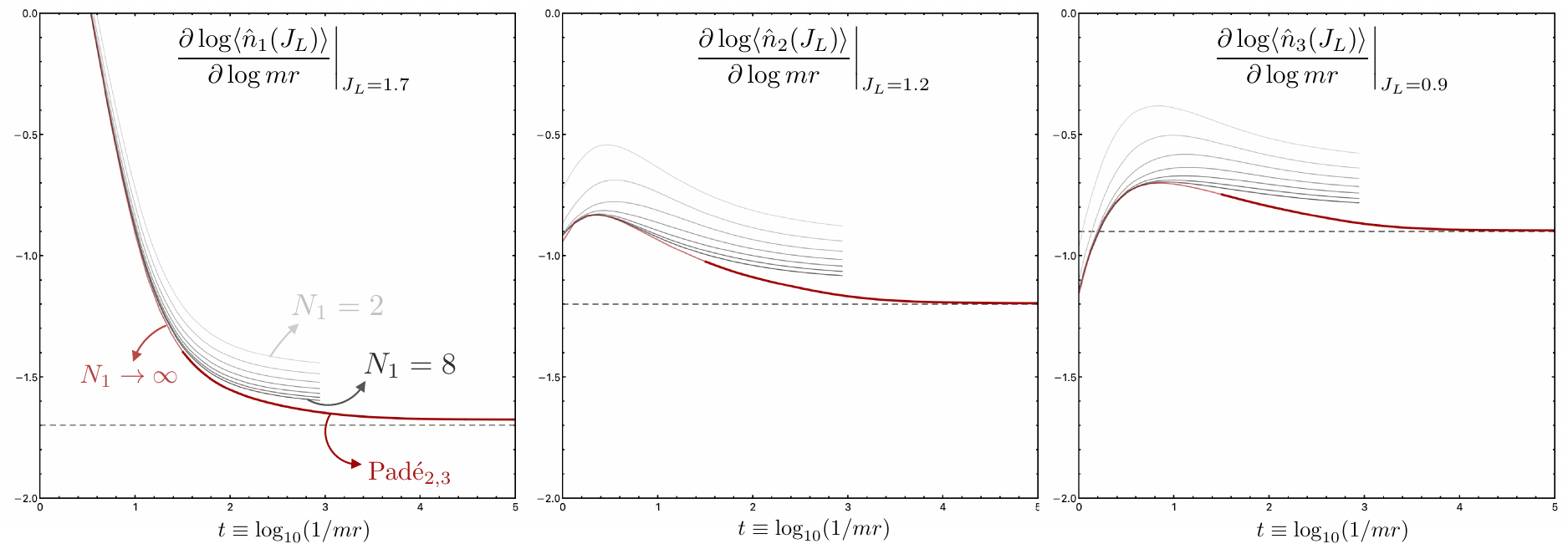}
  \caption{Detector matching predicts that all IR detector correlators
   $\langle \widehat{n}_i(J_L)\rangle_{\sigma(r)}$ scale as $r^{-J_L}$ as $r\to 0$. 
   We can extract the scaling exponent by computing $\partial \log \langle \widehat{n}_i \rangle/\partial\log mr$, 
   which for positive $J_L$ should approach $-J_L$  at small $r$. 
  In the figure we compute this observable for three particle species and values of $J_L$. We extract them from the densities obtained with finite $N_1$, figure~\ref{fig:magneticdensities}, and extrapolate to $N_1 \to \infty$. 
  This provides data for $ t\equiv -\log_{10} m r \leq 3$.
 We then fit a Pad\'e approximant (\ref{eq:pade}) which clearly converges to the anticipated plateau. We use data in a few different $t$ windows inside $t \in (1,3)$ to estimate the fit error, which corresponds to the thickness of the Pad\'e line. } 
   \label{fig:magneticmoments}
\end{figure}

 With the form factors computed, we then extract the densities $\langle n_i (\theta) \rangle_{\sigma(r)}$. We do so by 
 computing the partial sums in (\ref{eq:magneticffexpansion}), (\ref{eq:magneticdensity})
  using all form factors obtained by fusion of type-1 form factors, with $N_1$ playing the role of a truncation parameter\footnote{This is roughly equivalent to a truncation in the rest mass of the external state.}. We compute the partial sums for $N_1 \leq 8$ and then extrapolate
 the result to $N_1 \to \infty$.  The results are displayed and discussed in figure~\ref{fig:magneticdensities}.

From the extracted densities $\langle n_i (\theta) \rangle_{\sigma(r)}$ we can compute 
the harmonic moments $\langle \widehat{n}_i(J_L) \rangle_{\sigma(r)}$. Detector matching predicts 
these should scale as a power $r^{ \pm J_L}$ at small $r$, according to  (\ref{eq:magneticmatching}), the exponent sign being controlled by $\Re(J_L)$. To verify this prediction, 
in figure~\ref{fig:magneticmoments} we compute $ {{\partial \log \langle \widehat{n}_i(J_L) \rangle_{\sigma(r)}}/{\partial t}}$
 which should asymptote to $\pm J_L$ at large $t$. To estimate the large-$t$ asymptotics, we fit a Pad\'e approximant
 \begin{equation}
        \widehat{n}^{J_L}_\text{fit}(t) = \frac{a_0 + a_1 r + a_2 r^2}{b_0 + b_1 r + b_2 r^2 + b_3 r^3} \label{eq:pade}
 \end{equation}
 in a small window $t \in [2,3]$. 
  We observe that the resulting fit perfectly agrees with the expectation from detector matching for various values of $J_L$, see figure~\ref{fig:magneticmoments}.
  We have verified the result is stable under change in the fitting window used. In summary, we have confirmed, to leading order in $r$, the universality of detector matching and the evolution equations in this strongly interacting setup.
  
 {\footnotesize \paragraph*{Note}
  
  {
   When this work was nearly complete, GPT 6 Astra pointed out the reference \cite{Alekseev:2011my}.
    There, the author obtains explicit recursion relations expressing form factors of type-1 particles at generic
     kinematics in terms of lower-point form factors. This should be contrasted with the recursive method used in 
     this paper, in which equations (\ref{eq:Pn_kinematic_equation}), (\ref{eq:Pn_bound_state_equation}) and (\ref{eq:Pn_factorization_equation}) fix the form factor at particular kinematics. The form factors at generic kinematics are then obtained by constraining an ansatz and solving the resulting linear equations. The recursion relations of \cite{Alekseev:2011my} 
  trivialize the computation of the fundamental form factors.
  
  Using this method, we asked Opus 5.5 to verify the results in figure \ref{fig:magneticdensities}. In particular, it was asked to compute the bulk density ratios of various particle types $\rho_i \equiv \langle n_i\rangle/\langle n\rangle$ in a fugacity expansion for the pressure (i.e., the cluster expansion),
  \begin{equation}
  n_i = z_i\frac{\partial p}{\partial z_i}, \qquad p(z) = \sum_i c_i z_i + \sum_{i,j} c_{ij} z_i z_j + \dots.
  \end{equation}
  Opus performed the computation to third order in the fugacities and obtained, to this order, \begin{align}
        \rho_1 &= 65.62\%, & \rho_2 &= 22.25\%, & \rho_3 &= 6.78\%, & \rho_4 &= 4.17\%, \nonumber\\
        \rho_5 &= 0.74\%, & \rho_6 &=0.39\%, & \rho_7 &= 0.054\%, & \rho_8 &= 0.002\%, \label{eq:AIDoom} 
  \end{align}
  which are in agreement with the results shown in figure \ref{fig:magneticdensities}(g).
  }
 }

\section{High-energy evolution in Yang-Mills theory}\label{sec:higherDsec}

\subsection{Introduction: number density in higher dimensions}

\label{sec:introQCD}

In sections~\ref{sec:ising101} to~\ref{sec:magnetic}, we explored how light-ray matching for 2d RG flows leads to universal evolution equations for IR observables like the number density $n(\theta)$.  This idea works in higher dimensions as well --- the main new ingredient is that the Fourier transform with respect to rapidity gets replaced by harmonic analysis on the higher-dimensional Lorentz group $G=\SO(d-1,1)$.

As an example, consider again the number density of IR particles with mass $m$. Parametrizing the momentum as $p=m y$ with $y^2+1=0$, the number density $n(y)$ becomes a function of a point $y\in H^{d-1}$ on the hyperboloid. The first step in light-ray matching is to decompose $n(y)$ into Lorentz irreps.  The higher-dimensional analog of the harmonic decomposition (\ref{eq:fouriertransform}) is
\be
\widehat n_{J_L}(z) &= \int_{H^{d-1}} dy (-z\.y)^{J_L} n(y),
\ee
where $z \in \R^{d-1,1}$ is a null vector, and
\be
dy \equiv d^d y\, \de(y^2+1) \theta(y^0)
\ee
is the Lorentz-invariant measure on $H^{d-1}$. By construction $\widehat n_{J_L}(z)$ is a homogeneous function of the null vector $z$ with homogeneity $J_L$.\footnote{We can think of $z$ as an embedding-space coordinate for the celestial sphere, in which case $\widehat n_{J_L}(z)$ transforms like a conformal primary on the celestial sphere with dimension $-J_L$. It is obtained by integrating $n(y)$ against the bulk-to-boundary propagator $(-z\.y)^{J_L}$.} The inverse transformation is
\be
\label{eq:inversetransformation}
n(y) &= \int_0^\oo \frac{d\nu}{2\pi} \mu(\nu)\int \frac{D^{d-2} z}{\vol\,S^{d-2}} (-y\.z)^{2-d-J_L}\widehat n_{J_L}(z),
\ee
where $J_L=\frac{2-d}{2}+i\nu$ runs over principal series weights, $\frac{d\nu}{2\pi}\mu(\nu)$ is the Plancherel measure for scalar principal series representations $\cP_{J_L,0}$, and
\be
D^{d-2}z=\frac{1}{\vol\,\R^+}2d^d z\de(z^2) \theta(z^0)
\ee
is the Lorentz-invariant measure on the projective null cone. These formulas follow from the more general statement of the Plancherel theorem that we review in the next subsection.

By Lorentz invariance, we expect a matching equation
\be
\widehat n_{J_L}(z) &= \sum_i c_i(J_L,\mu) \cD^{\mathrm{UV}}_{J_L,i}(z,\mu),
\ee
where $\cD_{J_L,i}^{\mathrm{UV}}$ are UV detectors in the representation with weight $J_L$ (which can potentially depend on a renormalization scale $\mu$). For simplicity, let us assume that the UV theory is a CFT, so the detectors $\cD^{\mathrm{UV}}_{J_L,i}$ can be classified into dilatation eigenstates with mass dimensions $-\De_{L,i}(J_L)$. Let us furthermore assume that the right-hand side is dominated by a single trajectory, so we can drop the $i$ label. (We will relax both of these assumptions shortly.)

Taking expectation values of both sides, we find
\be
\<\widehat n_{J_L}(z)\>_{\cO(r)} &\sim  (z^0)^{J_L}r^{\De_{L}(J_L)},
\ee
where the matrix element $\<\cD_{J_L}^{\mathrm{UV}}(z)\>_{\cO(r)}$ on the right-hand side is fixed by Lorentz invariance and dimensional analysis, up to a constant. We see that in the single-trajectory approximation, the harmonic components $\<\widehat n_{J_L}\>$ satisfy the evolution equation
\be
(\ptl_t + \De_L(J_L))\<\widehat n_{J_L}(z)\>_{\cO(r)} &\sim 0,
\ee
where $t=-\log r$.

Transforming back to $y$-space via (\ref{eq:inversetransformation}), we find
\be
\ptl_t \<n(y)\>_{\cO(r)} + \int dy'  k(-y\.y') \<n(y')\>_{\cO(r)} &\sim 0,
\ee
where the Lorentz-invariant kernel $k$ is given by
\be
\label{eq:ourlorentzinvariantkernel}
k(\eta) &= \int_0^\oo \frac{d\nu}{2\pi} \mu(\nu) \De_L(J_L) P_{J_L}(\eta),
\ee
where $P_{J_L}$ is a Gegenbauer function
\be
\label{eq:thegegenbauerfunction}
P_{J_L}(\eta) &=  \int \frac{D^{d-2} z}{\vol\,S^{d-2}}\, \p{-y\.z}^{2-d-J_L}\p{-y'\.z}^{J_L} 
=
{}_2F_1\p{-J_L,d+J_L-2;\frac{d-1}{2};\frac{1-\eta}{2}}
\ee
with $\eta=-y\.y'$.

If we include more trajectories, we obtain a higher-degree approximate evolution equation
\be
Q(\ptl_t,J_L) \<\widehat n_{J_L}(z)\>_{\cO(r)} &\sim 0,
\ee
where
\be
Q(\ptl_t, J_L) &\equiv (\ptl_t + \De_{L,1}(J_L))(\ptl_t + \De_{L,2}(J_L)) \cdots (\ptl_t + \De_{L,n}(J_L))
\ee
is a polynomial in $\ptl_t$ with coefficients that are functions of $J_L$. We can perform a harmonic transform 
\be
\label{eq:harmonictransformagain}
Q(\ptl_t,J_L) &\mto Q(\ptl_t,\eta) = \int_0^\oo \frac{d\nu}{2\pi} \mu(\nu) Q(\ptl_t,J_L) P_{J_L}(\eta),
\ee
to obtain an evolution equation that is polynomial in $\ptl_t$ of degree equal to the number of trajectories, with each coefficient a Lorentz-invariant kernel. In $y$-space, the approximate evolution equation becomes
\be
\int dy' Q(\ptl_t, -y\.y') \<n(y')\>_{\cO(r)} &\sim 0.
\ee
We will see several examples below.

In the case where the UV theory is a perturbative non-CFT with a running coupling $\alpha(\mu)$, the above picture still holds, but now the $\De_{L,i}$ become nontrivial functions of $t$ through their dependence on the coupling $\alpha(1/r)$ evaluated at the hard scale $1/r\sim e^{t}$. In this case, the operator $Q(\ptl_t, J_L)$ has a potential ordering ambiguity, due to the dependence of the $\Delta_{L,i}$ on $t$. However, note that the ordering ambiguity is higher order in the coupling, so at leading order the above equations continue to hold, with the operator $Q(\ptl_t,\eta,t)$ now containing explicit $t$-dependence. To write an evolution equation that is valid at higher orders in the coupling, one must start from the matrix Callan-Symanzik equation satisfied by the perturbative detectors of the UV theory.

\subsection{Harmonic analysis generalities}

Before moving on to applications of these ideas, let us place the above technology in a more general context. An advantage of a more general treatment is that it will allow us to explore the evolution of more complicated observables, such as multi-point functions of the number density $\<n(p_1)\cdots n(p_n)\>$.

Our starting point is the Plancherel theorem, which gives an expression for a delta function $\de_e(g)$ on $G=\SO(d-1,1)$ centered at the identity element, expressed as an integral over the space $\widehat G$ of tempered representations of $G$:
\be
\de_e(g) &= \int_{\widehat G} d\mu(\pi) \Tr(\pi(g)).
\ee
Here, $\pi\in \widehat G$ runs over tempered representations, $d\mu(\pi)$ is the Plancherel measure on $\widehat G$, and $\Tr(\pi(g))$ is the trace of the representation map associated to $\pi$. The delta function $\de_e(g)$ is unit-normalized when integrated against the Haar measure $dg$. As an example, for $d=4$, the tempered representations will be principal series representations $\cP_{J_L,j}$ with transverse spin $j\in \frac 1 2 \Z$ and weight $J_L\in -1+i \R$.

Given a function $f(g)$ on the group $G$, we define its harmonic transform as 
\be
\widehat f(\pi) &= \int dg f(g) \pi(g)^{-1} \in \mathrm{End}(\pi).
\ee
The Plancherel theorem implies
\be
f(g) &= \int_{\widehat G} d\mu(\pi) \Tr(\pi(g) \widehat f(\pi)).
\ee

Consider an observable $N$, and let $N(g)=U(g)N U(g)^\dag$, so that $N(g)$ becomes a function on the Lorentz group. A $G$-invariant linear map $T:N\mto TN$ can be diagonalized in harmonic space with eigenvalue $\lambda(\pi)$:
\be
\widehat{T N}(\pi) &= \lambda(\pi) \widehat N(\pi).
\ee
As an example, we may be interested in the case where $T$ is evolution in dilatation time $\ptl_t$, so that $\lambda(\pi)=-\Delta_L(\pi)$ in the single-trajectory approximation. 
Performing the inverse harmonic transform of both sides, we have
\be
TN(g) &= \int d\mu(\pi) \lambda(\pi) \Tr(\pi(g) \widehat N(\pi)) = \int dg' K(gg'^{-1}) N(g'),
\ee
where $K$ is a kernel on the Lorentz group given by the inverse harmonic transform of $\lambda(\pi)$:
\be
K(g) &= \int d\mu(\pi) \lambda(\pi) \Tr(\pi(g)).
\ee
Note that $K(g)$ is a class function (as follows from $G$-invariance), so we can also write
\be
TN(g) &= \int dg' K(g'^{-1}g) N(g').
\ee

\subsubsection{Scalar observables on the hyperboloid}

An important special case occurs when $N$ is invariant under $H=\SO(d-1)$. In this case, $N(g)$ becomes a scalar function on the quotient space $G/H$, i.e.\ hyperbolic space $H^{d-1}$. An example is the one-point number density discussed in section~\ref{sec:introQCD}. Here, we set $N=n(e)$, where $e=(1,0,\dots,0)$ is a unit vector in the time direction. Then we have $N(g)=n(y)$ with $y=g e$, which is indeed a scalar function on the hyperboloid.

When $N$ is right-invariant under $H$, we can replace the group kernel $K$ with its average under $H$:
\be
\int dg' K(g'^{-1}g) N(g') &= \frac{1}{\vol H} \int dh \int dg' K(g'^{-1}g) N(g'h) 
= \int dg' \bar K( g'^{-1}g) N(g'),
\ee
where
\be
\bar K(g) &= \frac{1}{\vol H} \int dh K(hg) = \frac{1}{\vol H} \int dh \int d\mu(\pi) \lambda(\pi) \Tr( \pi(h)\pi(g)).
\ee
As we explain in appendix~\ref{eq:theappendixwhereweexplaingegenbauerfunctionthing}, averaging the trace $\Tr(\pi(h)\pi(g))$ over $H$ restricts $\pi$ to a scalar principal series representation $\pi=\cP_{J_L,0}$, and gives a Gegenbauer function:
\be
\frac{1}{\vol H}\int dh \Tr( \pi(h)\pi(g)) = \de_{\pi,\cP_{J_L,0}} P_{J_L}(\eta),\qquad \eta=-e\.ge.
\ee
Thus, writing $d\mu(\cP_{J_L,0}) = \frac{d\nu}{2\pi}\mu(\nu)$, we find
\be
\label{eq:singlyaveragedkernel}
\bar K(g) &= \int_0^\oo \frac{d\nu}{2\pi}\mu(\nu) \l(J_L) P_{J_L}(\eta) \equiv k(\eta),
\ee
reproducing (\ref{eq:ourlorentzinvariantkernel}) for $\l(J_L)=\De_L(J_L)$ and (\ref{eq:harmonictransformagain}) for $\l(J_L)=Q(\ptl_t,J_L)$.

Note that $\bar K(g)=k(\eta)$ is a function on the double coset $H\backslash G/H$, i.e.\ simply a function of a geodesic distance in hyperbolic space. The integral against $\bar K$ can be simplified as well. Let us write $N(g)=n(y)$, where $y=ge$ is a point in hyperbolic space, and also $y'=g'e$. Then we have
\be
\bar K(g'^{-1}g) = k(-e\.(g'^{-1} g e)) = k(-y'\.y) = k(\eta_{y,y'})
\ee
and
\be
\label{eq:oneHkernel}
TN(g) &= \int dg'\,\bar K(g'^{-1} g) N(g') = \vol H\int_{G/H} dy'\,k(\eta_{y,y'})\,n(y'),
\ee
where $G/H=H^{d-1}$.

\subsubsection{Rotationally symmetric scalar observables}

Another important special case is when an observable $N(g)$ is invariant under both left and right multiplication by $\SO(d-1)$. This applies, for example, to expectation values of the density $\<n(y)\>$ in rotationally invariant states, which are functions only of the distance between $y$ and the origin of hyperbolic space $\<n(y)\>=f(\eta)$ with $\eta=-y\.e$. In this case, we can simplify (\ref{eq:oneHkernel}) further:
\be
\vol H\int_{G/H} dy'\,k(\eta_{y,y'})\,\<n(y')\> &= \int_H dh \int_{G/H} dy' k(\eta_{y,y'}) \<n(hy')\> \nn\\
&= \vol\,H \int_{G/H} dy' \frac 1 { \vol\,H}\int_H dh\, k(\eta_{y,h^{-1}y'}) \<n(y')\>.
\ee
Let us define the ``doubly-averaged'' kernel
\be
k(\eta,\eta') &= \frac 1 { \vol\,H}\int_H dh\, k(\eta_{y,h^{-1}y'}).
\ee
Then our formula simplifies to
\be
\label{eq:radiallysymmetriccase}
TN(g) &= \vol\,H\,\vol\,S^{d-2} \int_0^\oo d\beta' \sinh^{d-2}\beta' k(\eta,\eta') n(\eta'),
\ee
where $\eta'=\cosh\beta'$. As we show in appendix~\ref{eq:theappendixwhereweexplaingegenbauerfunctionthing}, the doubly-averaged kernel is given by
\be
k(\eta,\eta')  %
&= \int \frac{d\nu}{2\pi}\mu(\nu)\l(J_L) P_{J_L}(\eta) P_{J_L}(\eta').
\ee
Note that the singly-averaged kernel $k(\eta)$ in (\ref{eq:singlyaveragedkernel}) is the $\eta'\to 1$ limit of the doubly-averaged kernel: $k(\eta)=k(\eta,1)$.

The formula (\ref{eq:radiallysymmetriccase}) reduces the action of $T$ on radially symmetric functions on $H^{d-1}$ to a 1-dimensional integral over the radial dependence alone.

\subsection{Evolution of rotationally symmetric observables in $d=4$}
Now we will focus on $d=4$, where we have the explicit formulas
\begin{equation}
        \begin{aligned}
                J_L&=-1+i\nu, \qquad \nu\geq 0 \\
                \mu(\nu)&=\frac{\nu^2}{8\pi^3} \\
                P_{J_L}(\eta)&=\frac{\sin{(\nu\b)}}{\nu\sinh\b}.
        \end{aligned}
\end{equation}

The evolution kernel of a physical theory is determined from its Chew-Frautschi plot, which in general
is given by the characteristic equation of the dilatation operator: 
\begin{equation}
        \widehat{D}(\nu)\equiv\det(\mathscr{D}+\De_L(\nu))=0
\end{equation}
where shadow symmetry dictates that $\widehat{D}(\nu)$ is even in $\nu$. Here we restrict to trajectories with zero transverse spin. 

Applying equation~\eqref{eq:radiallysymmetriccase},
we have the following explicit equation for the action of $\widehat{D}(\nu)$ on radially symmetric densities on $H^3$:
\begin{equation}
    \<{D},n\>(\b)\equiv\vol\,\SO(3)\,\vol\,S^2\int_0^{\infty}d\b'\sinh^2\b'\int_0^{\infty}\frac{d\nu}{2\pi}\mu(\nu)\widehat{D}(\nu)\frac{\sin{(\b\nu)}\sin{(\b'\nu)}}{\nu^2\sinh\b\sinh\b'}n(\b').
\end{equation}

We should think of $n(\b)$ as a tempered distribution on $[0,\infty)$. Because the rest of the integrand is even in $\b'$,
we can extend the $\b'$ integral to the whole real line, and impose that $n(\b')$ is an even tempered distribution. The integrand is also even in $\nu$, which allows us to also extend the $\nu$ integral.
We will also multiply both sides by $\sinh\b$. Putting this together gives
\begin{equation}
    (\sinh\b\<{D},n\>)(\b)=\frac{32\pi^3}{4}\int_{-\infty}^{\infty}\int_{-\infty}^{\infty}\frac{d\b'd\nu}{(2\pi)^4}\widehat{D}(\nu)\sin{(\b\nu)}\sin{(\b'\nu)}(\sinh\b'n(\b')).
\end{equation}

Now if we define
\begin{equation}
    u(\b)\equiv\sinh\b\,n(\b),
\end{equation}
we see that the problem reduces to Fourier theory on odd tempered distributions:
\begin{equation}
    \begin{aligned}
        (\sinh\b\<{D},n\>)(\b)\equiv\<{D},u\>(\b)&=8\pi^3\int_{-\infty}^{\infty}\int_{-\infty}^{\infty}\frac{d\b'd\nu}{(2\pi)^4}\widehat{D}(\nu)\sin{(\b\nu)}\sin{(\b'\nu)}u(\b') \\
        &=8\pi^3\int_{-\infty}^{\infty}\frac{d\b'}{(2\pi)^3}\frac{1}{4}(-D(-\b-\b')+D(\b-\b') \\
        &\qquad +D(-\b+\b')-D(\b+\b'))u(\b')
    \end{aligned}
\end{equation}
where
\begin{equation}
    D(\b)\equiv\int_{-\infty}^{\infty}\frac{d\nu}{2\pi}\widehat{D}(\nu)e^{i\b\nu}.
\end{equation}

To simplify further, we note that
\begin{equation}
    \int_{-\infty}^{\infty}d\b' D(\b+\b')u(\b')=-\int_{-\infty}^{\infty}d\b'\,D(\b-\b')u(\b'),
\end{equation}
due to the oddness of $u(\b')$. Additionally, $D(\b)$ is even. Thus, the action of $D$ simplifies to
\begin{equation}
    \<D,u\>(\b)=\int_{-\infty}^{\infty}d\b'\,D(\b-\b')u(\b').
\end{equation}

We see that the Chew-Frautschi plot determines a translation-invariant distribution acting on the sinh-weighted density, with the Chew-Frautschi plot being the diagonalization of this distribution in Fourier space.

\subsection{Pure Yang-Mills}\label{sec:PYM}
\subsubsection{Tree-level evolution}\label{sec:YM_tree_level}
Let us now apply these ideas to derive evolution equations in pure Yang-Mills. We truncate our detector matching to the leading twist-2 DGLAP trajectory, its shadow, and the BFKL trajectory. 

At tree level, 
the Chew-Frautschi plot is
\begin{equation}
        \De_L(\De_L-1+i\nu)(\De_L-1-i\nu)=0.
\end{equation}
The corresponding evolution equation is simple:
\begin{equation}
        \ptl_t((\ptl_t+1)^2-\ptl_{\b}^2)u(t,\b)=0. \label{eq:4dtree-evol}
\end{equation}
Solutions to this equation are a linear combination of a constant (the BFKL solution), and growing incoming waves and decaying outgoing waves (the DGLAP solutions).

\subsubsection{One-loop evolution}
At one loop, each trajectory gets modified by anomalous dimensions, which contain poles in $\nu$. Poles indicate mixing between different trajectories. Poles
corresponding to mixing between DGLAP and BFKL, or between DGLAP and its shadow, will cancel. However, there will also be mixing with subleading trajectories that we have omitted
in our detector matching, and so those poles will remain in our evolution kernel. A pole in the complex $\nu$ plane with imaginary part $\nu_0$ corresponds to an evolution
kernel in $\b$ space that decays as $e^{-\nu_0\b}$. To illustrate these concepts, we will first derive one-loop evolution equations for matching onto just BFKL, then just DGLAP
plus its shadow, before finally combining all three.

These calculations boil down to computing Fourier transforms of tempered distributions. One Fourier transform that we will be using repeatedly is the following:
\begin{equation}
        \Psi_{a,\pm b}(\b)\equiv\int_{-\infty}^{\infty}\frac{d\nu}{2\pi}\left(\psi(a\pm ib\nu)+\g_E\right)e^{i\b\nu}
\end{equation}
where $\psi(z)$ is the digamma function, and $a$ and $b$ are real with $b>0$.
To derive this Fourier transform, we use the following integral expressions for $\psi(z)$ and $\g_E$:
\begin{equation}
    \begin{aligned}
        \psi(z)&=\int_{0^+}^{\infty}dt\left(\frac{e^{-t}}{t}-\frac{e^{-zt}}{1-e^{-t}}\right) \\
        \g_E&=\int_{0^+}^1dx\frac{1-e^{-x}}{x}-\int_1^{\infty}dx\frac{e^{-x}}{x}.
    \end{aligned}
\end{equation}
Plugging these expressions into the Fourier transform yields
\begin{equation}
        \begin{aligned}
                \Psi_{a,\pm b}(\b)&=\de(\b)\int_{0^+}^1\frac{dx}{x}-\int_{-\infty}^{\infty}\frac{d\nu}{2\pi}\int_0^{\infty}dt\frac{e^{-(a\pm ib\nu)t}}{1-e^{-t}}e^{i\b\nu} \\
                &=\de(\b)\int_{0^+}^1\frac{dx}{x}-\frac{1}{b}\frac{\Theta(\pm \b)e^{\mp\frac{a}{b}\b}}{1-e^{\mp\frac{\b}{b}}}
        \end{aligned}
\end{equation}
where $\int_{0^+}$ denotes the regularization $\int_{\e}, \e>0$ and taking $\e\rightarrow 0$ at the end of the calculation.

We should interpret $\Psi_{a,\pm b}(\b)$ as a distribution. Let us integrate it against a test function in $\b$:
\begin{equation}
        \begin{aligned}
                \int_{-\infty}^{\infty}d\b\,\Psi_{a,\pm b}(\b)f(\b)&=\int_{0^+}^1dx\frac{f(0)}{x}-\int_{-\infty}^{\infty}\frac{d\b}{b}\frac{f(\b)\Theta(\pm \b)e^{\mp\frac{a}{b}\b}}{1-e^{\mp\frac{\b}{b}}} \\
                &=-\int_1^{\infty}\frac{d\b}{b}\frac{f(\pm\b)e^{-\frac{a}{b}\b}}{1-e^{-\frac{\b}{b}}}-\int_0^1 d\b\left(\frac{1}{b}\frac{f(\pm \b)e^{-\frac{a}{b}\b}}{1-e^{-\frac{\b}{b}}}-\frac{f(0)}{\b}\right).
        \end{aligned}
\end{equation}
We can write this as
\begin{equation}
        \Psi_{a,\pm b}(\b)=-\frac{1}{b}\frac{e^{\mp\frac{a}{b}\b}\Theta(\pm\b)}{1-e^{\mp\frac{\b}{b}}}\b^{\e}+\frac{1}{\e}\de(\b),
\end{equation}
where $\e\rightarrow 0$ after acting on test functions.

Let us now first consider just the BFKL trajectory. The BFKL eigenvalue is given by \cite{Lipatov:1985uk}
\begin{equation}
        \De_{\text{BFKL}}(\nu)=\frac{\a_S C_A}{\pi}\left(2\g_E+\psi\left(\frac{1+i\nu}{2}\right)+\psi\left(\frac{1-i\nu}{2}\right)\right).
\end{equation}
The Fourier transform is
\begin{equation}
        k_{\text{BFKL}}(\b)=-\frac{\a_S C_A}{\pi}\left(\frac{|\b|^{\e}}{\sinh|\b|}-\frac{2}{\e}\de(\b)\right)\equiv-2C_A\left[\frac{1}{\sinh|\b|}\right]_+,
\end{equation}
and the corresponding evolution equation is 
\begin{equation}
        \ptl_t u(t,\b)=-\int_{-\infty}^{\infty}d\b'\,k_{\text{BFKL}}(\b-\b')u(t,\b').
\end{equation}

For an even function $k(\b)$ whose behavior at the origin is given by
\begin{equation}
        k(\b)=\frac{k_0}{|\b|^p}+O(\b^0),
\end{equation}
we define its distributional action on test functions by
\begin{equation}
        \left[k(\b)\right]_+(f)\equiv\int_{-\infty}^{\infty}d\b\left(k(\b)f(\b)-\frac{2k_0f(0)\Theta(1-|\b|)}{|\b|^p}\right).
\end{equation}

Now let us consider the DGLAP trajectory plus its shadow at one loop. The characteristic equation is
\begin{equation}
        \begin{aligned}
                0&=\left(\De_L-1+i\nu-\De_{\text{DGLAP}}(-\nu)\right)\left(\De_L-1-i\nu-\De_{\text{DGLAP}}(\nu)\right) \\
                &=(\De_L-1)^2+\nu^2-\left((\De_L-1-i\nu)\De_{\text{DGLAP}}(-\nu)+(\nu\leftrightarrow -\nu)\right)+O(\alpha_S^2)
        \end{aligned}
\end{equation}
where \cite{Gross:1973zrg}
\begin{equation}
        \De_{\text{DGLAP}}(\nu)=\frac{\a_S}{2\pi}\left(4C_A\left(\psi(1-i\nu)+\g_E+\frac{i}{\nu}+\frac{1}{1+i\nu}+\frac{1}{-1+i\nu}-\frac{1}{-2+i\nu}\right)-\b_0\right).
\end{equation}
The Fourier transform gives the corresponding evolution equation
\begin{equation} \label{eq:DGLAP_radially_symmetric_evolution}
        0=\ptl_t^2u(t,\b)+\sum_{n=0}^1(-1)^n\ptl_t^n\int_{-\infty}^{\infty}d\b'\,k_{\text{DGLAP},n}(\b-\b')u(t,\b')
\end{equation}
where
\begin{equation}
        \begin{aligned}
                k_{\text{DGLAP},1}&=\left(-2+\frac{\alpha_S\b_0}{\pi}\right)\de(\b)+\frac{2\alpha_S C_A}{\pi}\left(-e^{-2|\b|}-\left[\frac{1}{1-e^{|\b|}}\right]_+\right) \\
                k_{\text{DGLAP},0}&=\left(1+\frac{\alpha_S}{\pi}(4C_A-\b_0)\right)\de(\b)-\de''(\b) \\
                &\quad +\frac{2\alpha_S C_A}{\pi}\left(2e^{-|\b|}-e^{-2|\b|}+\left[\frac{1}{1-e^{|\b|}}\right]_++\left[\frac{e^{|\b|}}{(1-e^{|\b|})^2}\right]_+\right)
        \end{aligned}
\end{equation}

The cancellation of the $\frac{1}{\nu}$ term, which in $\b$ space would correspond to a $\text{sgn}(\b)$ term in the kernel,
signals mixing between the DGLAP trajectory and its shadow at one loop. The poles closest to the origin are at $\pm i$,
corresponding to the fact that the kernel decays as $e^{-|\b|}$.

Now, we will consider the BFKL trajectory, the DGLAP trajectory, and its shadow at one loop. The characteristic polynomial is given by
\begin{equation}
    \left(\De_L-\De_{\text{BFKL}}(\nu)\right)\left(\De_L-1-i\nu-\De_{\text{DGLAP}}(\nu)\right)\left(\De_L-1+i\nu-\De_{\text{DGLAP}}(-\nu)\right).
\end{equation}
The corresponding evolution equation is
\begin{equation}
        0=-\ptl_t^3u(t,\b)+\sum_{n=0}^2(-1)^n\ptl_t^n\int_{-\infty}^{\infty}d\b'\,k_n(\b-\b')u(t,\b') \label{eq:full-evol-spin0}
\end{equation} 
where
\begin{equation} \label{eq:DGLAP-BFKL kernels}
        \begin{aligned}
                k_2(\b)&=\left(-2+\frac{\alpha_S\b_0}{\pi}\right)\de(\b)+\frac{2\alpha_S C_A}{\pi}\left(-e^{-2|\b|}+\frac{1}{2}\left[\frac{1}{\sinh|\b|}\right]_+-\left[\frac{1}{1-e^{|\b|}}\right]_+\right) \\
                k_1(\b)&=\left(1+\frac{\alpha_S}{\pi}(4C_A-\b_0)\right)\de(\b)-\de''(\b) \\
                & \quad +\frac{2\alpha_S C_A}{\pi}\left(2e^{-|\b|}-e^{-2|\b|}-\left[\frac{1}{\sinh|\b|}\right]_++\left[\frac{1}{1-e^{|\b|}}\right]_++\left[\frac{e^{|\b|}}{(1-e^{|\b|})^2}\right]_+\right) \\
                k_0(\b)&=-\frac{2\alpha_S C_A}{\pi}\left[\frac{1}{\sinh^3|\b|}\right]_+
        \end{aligned}
\end{equation}

In the characteristic polynomial for just DGLAP and its shadow, there are terms which have a pole at $\pm i$ in both the proportional to both $\De_L$ and constant in $\De_L$. In $\b$ space, this corresponds
to the fact that $k_{\text{DGLAP},1}$ and $k_{\text{DGLAP},0}$ decay as $e^{-|\b|}$. Once the BFKL trajectory is included, DGLAP/BFKL mixing causes the $\nu=\pm i$ pole proportional to $\De_L$ to cancel, which corresponds to the fact that $k_1(\b)$
decays as $e^{-2|\b|}$, not $e^{-|\b|}$. 

However, we see that $k_2(\b)$ still decays as $e^{-|\b|}$. This is because the DGLAP shadow trajectory has an additional pole at $\nu=i$, $\De_L=2$, which is only canceled by mixing with a different
subleading trajectory that we have not included. In \cite{Chang:2025zib}, this pole was dealt with by introducing a ``fake" horizontal trajectory. We employ this trick again in this work in solving the evolution equation numerically in section~\ref{sec:numerics4d}.\footnote{The results discussed in section~\ref{sec:numerics4d} are insensitive to the details of this auxiliary trajectory. In particular, the contribution from this trajectory is exponentially suppressed in $t$ relative to the leading trajectories, and are initiallized to be small.}

Introducing such trajectories to eliminate $e^{-|\b|}$ terms in the evolution kernel is required to obtain a regular evolution equation for total particle number. To see why, note that total particle number is given by
\begin{equation}
        N(t)=\int_{0}^{\infty}d\b\,\sinh\b\,u(t,\b).
\end{equation}
We can obtain an ordinary differential equation for $N(t)$ by integrating equation~\eqref{eq:full-evol-spin0} against $\sinh\b$, yielding
\begin{equation}
        0=-\ptl_t^3N(t)+\sum_{n=0}^2(-1)^n\ptl_t^n\int_{0}^{\infty}d\b\,\sinh\b\int_{-\infty}^{\infty}d\b'\,k_n(\b-\b')u(t,\b').
\end{equation}
The kernels $k_n(\b)$ in general decay as $e^{-j\b}$ for some $j>0$. To have regular evolution of $N(t)$, it must be that
\begin{equation}
        \int_{0}^{\infty}d\b\,\sinh\b\,e^{-j\b}
\end{equation}
converges, which requires $j>1$, and so the $\nu=\pm i$ poles must be canceled. 

This argument can be extended to the energy density and its moments. The $n$th moment of the energy density is given by
\begin{equation}
        E_n(t)=\int_{0}^{\infty}d\b\,(\cosh\b)^{n+1}\sinh\b\,u(t,\b). 
\end{equation}
Imposing regular evolution of $E_n(t)$ requires the cancelling of all poles in the $|\text{Im } \nu|\leq n+2$ strip.
Perturbatively, this is achieved through renormalization and mixing with subleading trajectories.  Nonperturbatively, we expect all singularities in the Chew-Frautschi plot to be resolved, and for $E_n$ to admit a regular evolution for all $n$.

\subsection{The hadron density contrail and hadron proliferation}\label{sec:numerics4d}

To illustrate the formalism above, let us consider a toy initial density, which we imagine being produced by some high-energy process, and evolve it to even higher energies using (\ref{eq:full-evol-spin0}). 

The first task is to determine initial conditions for the density. Consider the tree-level evolution (\ref{eq:4dtree-evol}), which we rewrite here:
\begin{equation}
      \ptl_t((\ptl_t+1)^2-\ptl_{\b}^2)u(t,\b)=0,\qquad u(t,\b) = \sinh(\b) n(t,\b).     \label{eq:4dtree-evol2}
\end{equation}
In analogy with the wave densities considered in section~\ref{sec:ising101},
 we consider a solution of  (\ref{eq:4dtree-evol2}) composed of an outgoing wave (the DGLAP solution) and 
 a background independent of the dilatation time $t$ (the BFKL solution),
 \begin{equation}
        n_\text{tree}(t,\beta) = \frac{B(\beta) + e^{-t}F(\beta - t)}{\sinh{\beta}}. \label{eq:initialdata}
 \end{equation}
For our numerical example we take $B(x) = C F(x)$ and
\begin{equation}
        F(x) = \mathcal{N} x^{a+1} e^{-a x}\Theta(x)
\end{equation}
with $C = 5 \times 10^{-4}$, $\mathcal{N}= 6^8/7!$ and $a=6$. This choice is natural in harmonic space where, at $t=0$, $\tilde{F}(\nu)$ is a simple rational factor.

Next, we consider the initial data $\partial_t^m n_\text{tree}(t,\beta)\vert_{t=0}$ and 
compute its one-loop evolution (\ref{eq:full-evol-spin0}). As explained above, at one loop we consider
 a fourth-order equation including an auxiliary trajectory so that the total particle number has a
 regular evolution. We therefore specify initial data for $m = 0,1,2,3$.
We fix $\alpha_S = 10^{-4}$ in the numerical example. Replacing the fixed coupling by a running coupling 
\begin{equation}
        \partial_t \alpha_S = -\beta_0 \alpha_S^2 + O(\alpha_S^3), \qquad \alpha_S(0) = 10^{-4}, \label{eq:betafunc}
\end{equation}
produces visually negligible differences over the time interval considered. In particular, the corrections are of the same magnitude as the ordering ambiguities associated with commuting $\partial_t$ and $\alpha_S$ discussed in section~\ref{sec:introQCD}.

\begin{figure}[t!]
  \centering
  \includegraphics[width=0.93\linewidth]{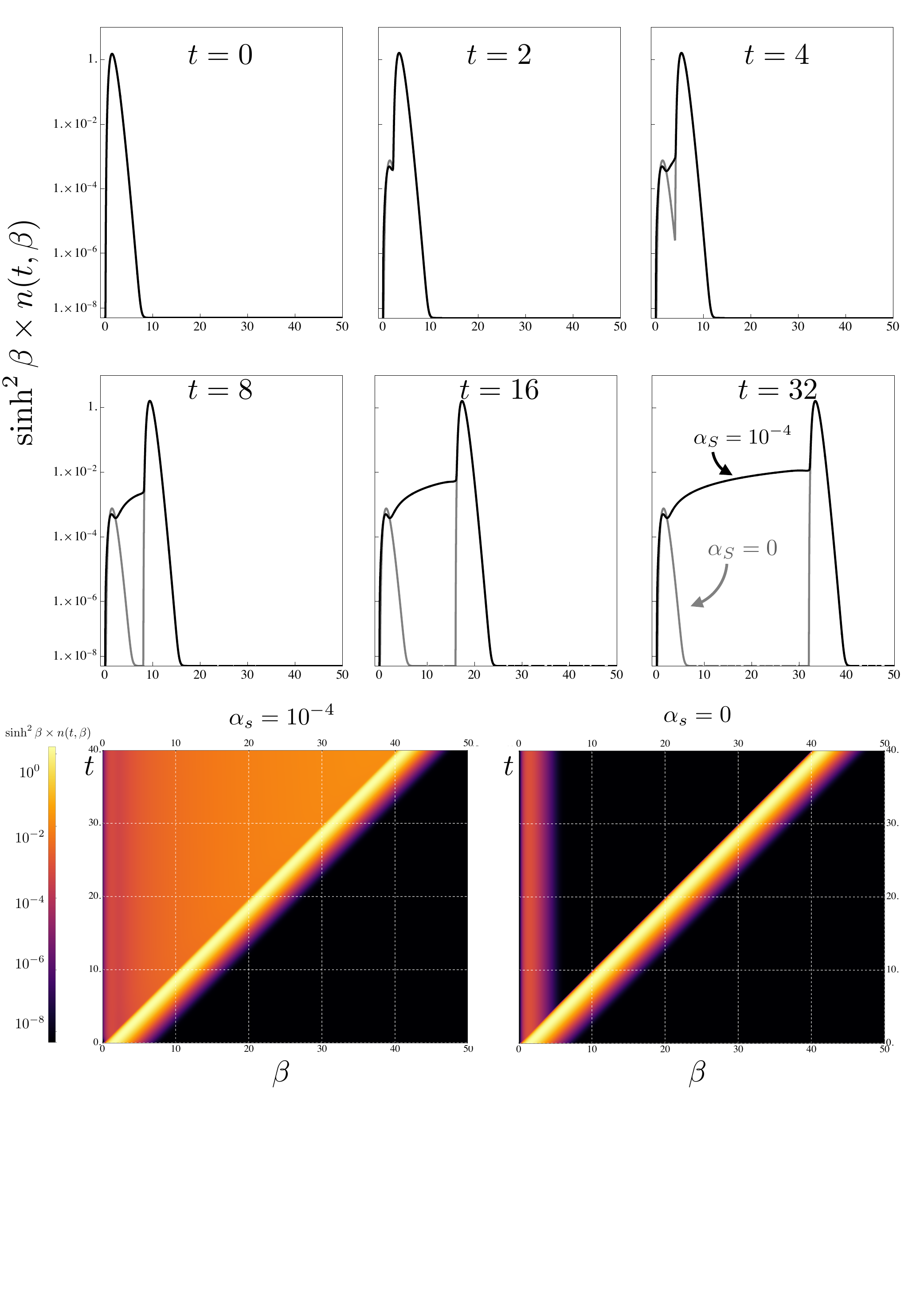} \vspace{-3.5cm}
  \caption{Evolution of the density (\ref{eq:initialdata}). We plot the radial density \mbox{$\rho(t,\beta) = \sinh^2\beta n(t,\beta)$}. Note that $\langle N \rangle  = \int d\beta \rho(t,\beta) $. The evolution is performed by first going to $\nu$ space, discretizing in $\nu$, evolving, and then performing the harmonic inverse. 
  \textbf{Top}: Six time slices comparing the radial density under tree-level evolution (gray) and one-loop evolution with $\alpha_S = 10^{-4}$ (black). Note that at one loop, as the hadron density propagates outward on the mass shell, it leaves a remnant \textit{contrail} which is approximately time independent. This is a signature of BFKL-DGLAP mixing. \textbf{Bottom}: A density plot of the same observable.} \label{fig:contrail}
\end{figure}

The numerical evolution, shown in figure~\ref{fig:contrail}, reveals a
remarkable difference compared to the tree-level evolution: the presence of a 
contrail. As the energy increases, our initial Gaussian-like hadron wave packet expands on the mass shell to higher rapidities, but leaves behind a persistent wake as it propagates. 

We can understand the presence of this BFKL contrail as coming from mixing between the DGLAP and BFKL trajectories in QCD at loop level. In rapidity space, this manifests in the $\a_S$ corrections in equation~\eqref{eq:full-evol-spin0} that causes the solutions to the tree-level wave equation to mix into each other. These terms are short-ranged. Locality in rapidity space
means the mixing of our initial outward moving wave into the BFKL contrail only occurs in the region where the wave is localized. Thus non-local effects
are neglible, which is what we see in figure~\ref{fig:contrail}.

Another prominent feature of figure~\ref{fig:contrail} is the growth of total particle number in 
dilatation time. Indeed, detector matching predicts that, in a high-energy state, 
hadron proliferation is controlled by the dimension $\Delta_0$ of the leading trajectory at $J_L =0$ \cite{Chang:2025zib}. This corresponds to the dominant branch of the resolved BFKL-shadow mixing, 
whose dimension is
 \begin{equation}\Delta_0(t) =  -\sqrt{2 C_A \alpha_S(t)/\pi} + O(\alpha_S).\end{equation}
The dominant asymptotics then follows from the RG equation
\begin{equation}
\partial_t \log  \langle  N\rangle_{\mathcal{O}(1/r)} = - \Delta_0(t)  + \texttt{subleading}
\end{equation}
combined with the running-coupling asymptotics $\alpha_S(t) \sim 1/\beta_0 t$ as $t\to \infty$, where $\beta_0 = 11 C_A/6\pi$ is the one-loop beta function coefficient. Integrating, we obtain
\begin{equation}
       \log \langle N\rangle_{\mathcal{O}(1/r)} = 2{ \sqrt{12 t/11}}  + \texttt{subleading} \qquad (t\to\infty)
\end{equation}
for the asymptotic growth of the hadron number in the high-energy state $\mathcal{O}(1/r)$. Note that the leading asymptotic is independent of $N_c$ in pure Yang-Mills! See \cite{Bassetto:1979nt,Amati:1980ch,Furmanski:1979jx,Konishi:1979ft,Mueller:1981ex} for the original perturbative computations of the multiplicity evolution from diagrammatic resummation.

Note that, given a positive initial density $n(t_0, \beta)$, it is not guaranteed that evolution will preserve positivity for all future values of $t$. This provides a nontrivial constraint relating the densities, determined by the family of states, and the evolution, determined by the Chew-Frautschi plot. This is akin to a bootstrap constraint, and it would be nice to explore it further.

\subsection{Multi-point detectors and general evolution}

We now discuss the one-loop evolution of a general function on the group. 
An example is the correlator of multiple rapidity detectors
 $N(g) = \<\psi_g | n(p_1)\cdots n(p_n) |\psi_g\>$ computed in a family of states $|\psi_g\rangle = U(g) |\psi\rangle$ parametrized by a Lorentz transformation $g$.

In such cases, 
the observable $N(g)$ has nontrivial transverse spin components, and so its evolution receives contributions from detectors of nontrivial $j$.
As above, we will restrict ourselves to the contribution from the BFKL,
 DGLAP and shadow trajectories in pure Yang-Mills. Their dimensions up to one loop are
  \cite{Gross:1973zrg,Lipatov:1985uk,Chen:2021gdk}
\begin{align}
        \Delta_\text{BFKL}(J_L,j) &=    \frac{\alpha_S C_A}{\pi}  \left(2 \gamma_E + \psi\left(\frac{2 + J_L + |j|}{2}\right)+ \psi\left(\frac{-J_L + |j|}{2}\right)\right), \\
        \Delta_\text{DGLAP}(J_L, j) &=  2 + J_L + \frac{\alpha_S C_A}{\pi} \left(2 \psi(-J_L) +  2 \gamma_E  -  \frac{11}{6} - \frac{4 (1 +J_L + J_L^2) \delta_{j,0}}{(J_L +1 ) (J_L + 2)(-J_L)(1-J_L)}\right), \nonumber
\end{align}
where the BFKL trajectory contributes for $j \in 2\mathbb{Z}$ while DGLAP and shadow contribute for $j \in \{-2,0,2\}$. As usual, $\Delta_\text{Shadow}(J_L,j) = \Delta_\text{DGLAP}(-2-J_L,-j)$. %

Define the characteristic polynomial
\begin{equation}
Q(T,\nu,j)=\begin{cases}
(T + \Delta_{\text{BFKL}})(T+\Delta_{\text{DGLAP}})(T+\Delta_{\text{Shadow}}) \label{eq:cpol} ,& j=0,\pm2,\\
(T+\Delta_{\mathrm{BFKL}})
, & j\in2\mathbb{Z},\quad |j|\ge4,
\end{cases}
\end{equation}
with $J_L = -1 + i \nu$. The corresponding harmonic-space evolution equations are\footnote{As discussed in section~\ref{sec:introQCD}, to one-loop order we can neglect running-coupling effects and treat $T$ and $\Delta_a(\alpha_S(t))$ as commuting variables.}
\begin{equation}
Q(\partial_t,\nu,j) \widehat{N}_t(\nu,j) = 0. \label{eq:hspaceevol}
\end{equation}

\label{sec:generalevol}
Let us derive the corresponding rapidity-space evolution equations. Performing the harmonic inverse of (\ref{eq:hspaceevol}) we obtain, for any $g$,
\begin{equation}
\int dg'\left({K(g g'^{-1}) \equiv {\sum_{j\in2\mathbb{Z}} \int_0^\infty \frac{d\nu}{2\pi} \mu_j(\nu) Q(\partial_t, \nu,j) \text{Tr}\left(\pi_{\nu,j}(g g'^{-1})\right)}} \right)N_t(g') = 0, \label{eq:generalevol}
\end{equation}
with $\mu_j(\nu) = (\nu^2 + j^2)/(8\pi^3)$.

Before an explicit computation of the kernel in (\ref{eq:generalevol}), two simplifications can be made. First, recall that the character~\cite{Dobrev:1977qv}
\begin{equation}
        \text{Tr}\left(\pi_{\nu,j}(g)\right) = \frac{\left(q^{\frac{-i \nu +j}{2}}\bar{q}^{\frac{-i \nu -j}{2}} + q^{\frac{i \nu -j}{2}}\bar{q}^{\frac{i \nu +j}{2}} \right)}{|q^{-1/2} - q^{1/2}|^2}, \qquad q=e^{\eta + i \theta}, \bar{q} = e^{\eta - i \theta}
\end{equation} 
is a class function and therefore depends only on the coordinates $\eta, \theta$ parametrizing the Cartan element $h_{\eta,\theta} = e^{\eta K_3 + \theta J_3} \in \text{SO}(1,1) \times \text{SO}(2)$. Defining first $\widehat{g} = g g'^{-1}$ and then $\widehat{g} = x h_{\eta,\theta} x^{-1}$, with $x$ parametrizing the conjugacy orbit of $h_{\eta,\theta}$, we have
\begin{equation}
0=\int_{\mathbb{R}^+\times S^1}d\eta d\theta \mu'(\eta,\theta) K(\eta,\theta) \int dx N_t(x h^{-1} x^{-1} g)  
\end{equation}
and the Jacobian $\mu'(\eta,\theta) = 4 (\cosh \eta - \cos \theta)^2 $ is computed in appendix~\ref{app:conjugacyorbit}. We thus see that the evolution kernel acts only on the orbit average of $N(g)$:
\begin{equation}
        \phi_{t,g}(\eta, \theta) = \int dx N_t(x h^{-1} x^{-1} g).
\end{equation}

The second simplification comes about from the Weyl reflection symmetry of the characteristic polynomial,
\begin{equation}
        Q(T,-\nu,-j) = Q(T,\nu ,j)
\end{equation}
which allows us to extend the $\nu$ contour in (\ref{eq:generalevol}) to the full real line and simplify the integrand
\begin{equation}
        K(\eta,\theta) = \frac{-1}{2(\cosh \eta - \cos \theta)}\frac{\partial_\eta^2 + \partial_\theta^2}{(2\pi)^4}
       \left(\widehat{Q} \equiv  \sum_{j\in2\mathbb{Z}} \int_\mathbb{R} d\nu Q(\partial_t,\nu,j) e^{i j \theta - i \nu \eta}\right).
\end{equation}
Above, we have replaced the Plancherel measure $\mu_j(\nu)$ by the differential operator outside the integral. Finally, note that due to the orbit averaging, the action of the distribution $\widehat{Q}$ only needs to be defined on functions
which extend evenly under Weyl reflection ${\phi(\eta, \theta) = \phi(-\eta, -\theta)}$.

As in the radially symmetric case, the computation has now been reduced to standard Fourier transforms, which can be performed by contour deformation. 
Consider first the tree-level term in  $Q(\partial_t,\nu,j)$, which results in\footnote{We define $\int_{\mathbb{R}^+} d\eta \delta^+(\eta) \phi(\eta) = \phi(0)$.}
\begin{align}
        \widehat{Q}_{\text{tree}} &=  \nonumber \sum_{j\in2\mathbb{Z}} \int_\mathbb{R} d\nu \left(\partial_t + \delta_{j,{0,\pm 2}} \partial_t(\partial_t^2 + 2 \partial_t + \nu^2)\right) e^{i j \theta - i \nu \eta}\\
        & = \pi^2 \partial_t \delta^+(\eta) (\delta(\theta) + \delta(\theta - \pi)) +  \pi (1 + 2 \cos2\theta) \partial_t \left((\partial_t^2 + 2 \partial_t)\delta^+(\eta) - \partial_\eta^2 \delta^+(\eta)\right).
\end{align}
Next, we consider the various one-loop terms. They are
\begin{equation}
    \label{eq:one-loopQ}    Q_\text{1-loop}      = \gamma_\text{BFKL} + \delta_{j,{0,\pm 2}} \left( (\partial_t^2 + 2 \partial_t + \nu^2)\gamma_\text{BFKL}(\nu,j) + \partial_t^2 \gamma_2(\nu,j) + \partial_t \gamma_1(\nu,j) \right) 
\end{equation}
where $\gamma_2  = \gamma_\text{DGLAP} + \gamma_\text{Shadow}$ and $\gamma_1 = (1- i \nu)\gamma_\text{DGLAP} + (1+ i\nu)\gamma_\text{Shadow} $. Here, $\gamma_a = \Delta_a - \Delta_a^\text{tree}$.

To illustrate the general method, let us consider the first term in detail. To compute
\begin{equation}
         \widehat{Q}_\text{BFKL} =  \sum_{j\in2\mathbb{Z}} \int_\mathbb{R} d\nu \gamma_\text{BFKL}(\nu,j) e^{i j \theta - i \nu \eta} \label{eq:hatBFKL}
\end{equation}
deform the $\nu$ contour into the lower half-plane for $\eta>0$, and then resum in $j$. The result is
\begin{equation}
         \widehat{Q}_{\text{BFKL}} = - 2 \alpha_S C_A\frac{\cosh{\eta}}{\cosh^2{\eta} - \cos^2\theta}, \qquad \eta>0 \label{eq:BFKLexample}
\end{equation}
which determines $\widehat{Q}_\text{BFKL} $ up to contact terms\footnote{Let us emphasize that ``up to contact terms'' is not a small detail here. In fact, the tree-level evolution is made entirely of contact terms! The same is true for the two-dimensional evolution (\ref{eq:thewaveequation}).}. 

Note that (\ref{eq:BFKLexample}) diverges logarithmically when integrated around $(\eta,\theta) = (0,0)$ and $(\eta,\theta) = (0,\pi)$. To extend (\ref{eq:BFKLexample}) to a distribution on $\mathbb{R}^+ \times S^1$, define for a test function $\phi(\eta,\theta)$,
\begin{align}
 \label{eq:2dplusdist}&\left\langle \left [ \frac{\cosh\eta}{\cosh^2\eta - \cos^2\theta}\right ]_{+},\phi(\eta,\theta)\right\rangle =\\& \int_{\mathbb{R}^+ \times S^1}  d\eta d\theta \left [ \left ( \frac{\cosh\eta}{\cosh^2\eta - \cos^2\theta}\right ) \phi(\eta, \theta) - \frac{\phi(0, 0)\Theta\left(1 - r_0\right)}{r_0^2}-\frac{\phi(0,\pi)\Theta\left(1 - r_\pi\right)}{r_\pi^2} \right ] \nonumber
\end{align}
where $r_{\theta^*}^2 \equiv (\theta-\theta^*)^2 + \eta^2$ and $\Theta$ is the step function. We then write the ansatz
\begin{equation}
        \widehat{Q}_{\text{BFKL}} = -2 \alpha_S C_A \left(\left [ \frac{\cosh\eta}{\cosh^2\eta - \cos^2\theta}\right ]_{+} + c \delta^+(\eta)(\delta(\theta)+ \delta(\theta - \pi) )\right),\label{eq:contactansatz}
\end{equation}
where the coefficient $c$ is the contact-term coefficient we aim to determine. In (\ref{eq:contactansatz}) both singular terms contribute symmetrically due to the $\pi$ periodicity of (\ref{eq:hatBFKL}).

To fix $c$, integrate (\ref{eq:contactansatz}) in a small half-disk of radius $x$ around $(\eta,\theta)=0$. We obtain
\begin{equation}
        \langle \widehat{Q}_{\text{BFKL}},  \Theta(x-r_0)\rangle = -2\alpha_S C_A (\pi \log x + c) + O(x).
\end{equation}
On the other hand, the same computation can be performed using (\ref{eq:hatBFKL}).  The small-$x$ behavior is then controlled by large $\nu$ and large $j$. A simple computation results in
\begin{equation}
        \langle \widehat{Q}_{\text{BFKL}},  \Theta(x-r_0)\rangle  = -2\alpha_S C_A \pi \log x  + O(x)
\end{equation}
from which we conclude $c=0$. This completes the computation of $\widehat{Q}_{\text{BFKL}}$ as a distribution.

The remaining terms in (\ref{eq:one-loopQ}) can be computed in complete analogy. We simply quote the results:
\begin{align}
          \widehat{Q}_2 &=  \sum_{j\in2\mathbb{Z}} \int_\mathbb{R} d\nu  \delta_{j,{0,\pm 2}} \gamma_2(\nu,j) e^{i j \theta - i \nu \eta} \\
          &=4 \alpha_S C_A \left(e^{-2\eta} - (1+ 2 \cos 2\theta) \left( \left[ \frac{1}{e^\eta-1}\right]_+ + \frac{11}{12}\delta^+(\eta)\right) \right)\nonumber,\\
          \widehat{Q}_1 &=  \sum_{j\in2\mathbb{Z}} \int_\mathbb{R} d\nu  \delta_{j,{0,\pm 2}} \gamma_1(\nu,j) e^{i j \theta - i \nu \eta} \\
          &=4 \alpha_S C_A \left(2e^{-\eta} -  e^{-2 \eta} - (1+ 2 \cos 2\theta) \left( (\partial_\eta + 1 ) \left[ \frac{1}{e^\eta-1}\right]_+ + \frac{11}{12}\delta^+(\eta)\right) \right)\nonumber,\\
   \intertext{and}
\widehat{Q}_{\text{BFKL},2}&=\sum_{j\in2\mathbb{Z}} \int_\mathbb{R} d\nu  \delta_{j,{0,\pm 2}} (\partial_t^2 + 2 \partial_t + \nu^2) \gamma_\text{BFKL}(\nu,j) e^{i j \theta - i \nu \eta} 
\\ &\hspace{-1.5em}=-2\alpha_S C_A (\partial_t^2 + 2 \partial_t - \partial_\eta^2)\left((1+ 2 \cos 2\theta) \left(\left[\frac{1}{\sinh\eta}\right]_{+} + \log (2) \delta^+(\eta) \right) - 4e^{-\eta}\cos 2\theta\right),\nonumber
\end{align}
where
\begin{equation}
\left\langle \left [f(\eta) \right ]_{+}, \phi(\eta,\theta)\right\rangle = \int_{\mathbb{R}^+ \times S^1} d\eta d\theta \left[ f(\eta){\phi(\eta, \theta)} - \frac{\phi(0,\theta) \Theta(1-\eta)}{\eta} \right].
\end{equation}
We conclude with the final one-loop evolution equation:
\begin{equation}
        0 = \langle \widehat{Q},  (\partial_\eta^2 + \partial_\theta^2) \left[(\cosh \eta - \cos \theta) \phi_{t,g}(\eta,\theta)\right]\rangle
\end{equation}
with 
\begin{equation}
        \widehat{Q} = \widehat{Q}_\text{tree} + \widehat{Q}_\text{BFKL}+ \widehat{Q}_{\text{BFKL},2} + \widehat{Q}_{2}\partial_t^2 +  \widehat{Q}_{1}\partial_t.
\end{equation}
It would be interesting to compare this evolution equation against data.

\section{Discussion}
\label{sec:discussion}

Our understanding of collider observables in QCD has been guided and shaped by decades of experiment. The theory of these observables grew organically alongside the data, so many of the theoretical tools (PDFs, DGLAP evolution, factorization theorems...) were developed with QCD primarily in mind. It is an important challenge to adapt these developments to improve our understanding of general quantum field theories. At the same time, generalizing theoretical tools can help us understand the tools themselves better, and in turn sharpen our understanding of QCD.

Unfortunately, we do not have much experimental data about collider-like processes in other QFTs to guide us. The integrable flows we study in this work have the virtue of being nonperturbative, but still under enough theoretical control that we can explore real-time physics. It is interesting to ask whether there are other theoretical laboratories where one could explore the idea of detector matching. One possibility is to study non-integrable Ising field theory, i.e.\ the 2d Ising model with both the magnetic and thermal deformations turned on. Scattering in this theory was recently studied using MPS methods in \cite{Jha:2024jan}. 
It would be interesting to explore event shapes and the high-energy evolution of particle distributions 
in this setup. One could also explore detector matching in {\it perturbative\/} 
RG flows, for example in theories that contain unstable particles. 
It would also be interesting to understand detector matching in 
holographic theories like $\cN=4$ SYM at strong coupling. 
The precise setup in this case is not immediately clear.
 What are the appropriate stringy detectors in the planar theory? In a confining deformation, what does semiclassical gravity mean for the evolution of particle density? What about stringy corrections?

In order for detector matching to be useful, we must understand what kind of UV light-ray operators appear on the right-hand side. One hypothesis is that the light-ray operators appearing in detector matching are the same as those appearing in the Regge limit of local correlation functions. These operators live on the Chew-Frautschi plot on Regge trajectories $\Delta_{L,i}(J_L)$. Some examples explored in this work support this hypothesis, but we also found some puzzling observables that seem to violate it: namely the ``multiscale detectors'' discussed in section~\ref{sec:multiscale}. In this case, an IR detector that was sufficiently spread out in rapidity space seems to match onto a ``composite'' detector built from products of leading-twist detectors. We do not know what the appropriate definition of a ``composite'' detector is in a general QFT. (In perturbation theory, a product of bare detectors generally requires renormalization. One potential working definition of a composite detector is one that is ``additively'' renormalized, as opposed to multiplicatively renormalized, see \cite{Caron-Huot:2022eqs}.) This is a huge gap in our understanding. Which IR observables match onto detectors on the Chew-Frautschi plot, and which do not? When they do not, what is the appropriate space of composite detectors? The recent work \cite{Gonzalez:2026aha} explores some versions of these questions in QCD. They propose that IR detectors that are spaced out at angles larger than $\Lambda_\mathrm{QCD}/Q$ (where $Q$ is the energy of the state) on the celestial sphere should be matched individually to UV detectors, resulting in composites of UV detectors.

Another important question is what happens to detector matching at locations on the Chew-Frautschi plot where there are an infinite number of trajectories. In gauge theories, there are an infinite number of horizontal trajectories near $\Delta_L=0$. There are also an infinite number of nearly-degenerate trajectories for higher-twist operators, see e.g.\ \cite{Homrich:2022cfq,Henriksson:2023cnh,Homrich:2024nwc}. It is an interesting challenge to understand these objects, and in particular their mixing, in perturbation theory, in order to build a complete picture of the Chew-Frautschi plot of perturbative gauge theories.  Naively, when an infinite number of terms contribute nearly equally, it makes it hard to make predictions for relationships between different observables. But perhaps there is structure in the matching coefficients and matrix elements of these infinite trajectories. What do their relative contributions look like in simple states created by local operators? What do they look like in more complicated states like the quark-gluon plasma? We expect that the $1/Q$ expansion obtained from detector matching is likely asymptotic, due to the presence of infinite nearly-degenerate trajectories.

In this work, we considered theories that are massive in the IR. What happens when the IR theory is instead a CFT? There are no massive asymptotic states, and so instead the detector operators are light ray operators, which would match onto light ray operators of the UV theory. What do the analogous evolution equations look like? Instead of particle density, the natural observable to consider are matrix elements of the (weighted) energy flux operators $\mathcal{E}_{J_L}(z)$ where $z\in S^{d-2}$ is a point on the celestial sphere and $J_L$ is a boost weight. Perhaps we will have evolution on a null cone. A simple setup to explore some of these questions could be RG flows between minimal models. What other differences in detector matching and evolution are there when the IR theory is a CFT? 

\looseness=-1
All in all, detector matching provides a nonperturbative organizing principle for how infrared observables retain imprints of the ultraviolet dynamics, encoded in the spectrum~of light-ray operators. We hope that understanding the full scope of this connection may enrich our perspective on real-time physics in quantum field theory. We look forward to exploring it. 

\section*{Acknowledgements}

We thank Simon Caron-Huot, Hao Chen, David Gross, Zohar Komargodski, Yue-Zhou Li, Ian Moult, Douglas Stanford, Pedro Vieira and Hao-Lan Xu for discussions. We especially thank Ian Moult for early discussions of detector matching in 2d that helped motivate this project. This material is
based upon work supported by the U.S. Department of Energy, Office of Science, Office of High Energy Physics, under Award Number DE-SC0011632. This research was supported in part by grant NSF PHY-2309135 to the Kavli Institute for Theoretical Physics (KITP). AH thanks the KITP for generous hospitality while this work was carried out.

\section*{AI/Human statement}

GPT 5.6 Sol was used to write and run code for Monte Carlo integration of terms in the form factor expansion for the thermally deformed and magnetically deformed 2d Ising CFT, and to prepare the corresponding figures. GPT-6 Astra was used to simplify the results of section~\ref{sec:generalevol}, to find the references and obtain the numerical results for magnetic Ising number densities described in (\ref{eq:AIDoom}), and for minor edits to the manuscript. Claude Opus 5.5 was used to create figures~\ref{fig:penrose} and \ref{fig:chewfrautschi}, and minor edits to the manuscript.

All other ideas and results in this paper were the work of the human authors.

\appendix

\section{Conformal Regge theory, chaos, and the flow equations}
\label{app:conformalreggedetails}

In this appendix, we give more technical details on the relation between the flow equations for IR detectors and the spread of chaos in hyperbolic space.

We first recall that Minkowski space in Rindler quantization is Weyl-equivalent to $H^{d-1} \x S^1$. Start with the flat metric on $\R^d$ and use polar coordinates for two of the directions:
\be
ds^2_{\R^d} = dr^2 + r^2 d\theta^2 + d\vec x^2,
\ee
where $\vec x\in \R^{d-2}$ is a transverse direction. The angular variable $\theta\in [0,2\pi)$ plays the role of a periodic Euclidean time in Rindler quantization. Performing a Weyl transformation, we obtain
\be
ds^2_{\R^d} \to ds_{H^{d-1} \x S^1}^2 = d\theta^2 + \frac{dr^2 + d\vec x^2}{r^2}.
\ee

Consider now a Euclidean four-point function on $\R^d$. Using conformal symmetry, we can place the operators in the following configuration on $H^{d-1}\x S^1$:
\be
\<V(e,\pi)W(y,\pi+\theta)V(e,0) W(y,\theta)\>_{H^{d-1} \x S^1}.
\ee
Here, $e$ and $y$ are points in hyperbolic space. We take $e$ to be the ``center'' of $H^{d-1}$, which is fixed by an $\SO(d-1)$ action. Note that here we write $V(y,\tau)$ and $W(y,\tau)$ as operators in Euclidean time, as opposed to the notation in the introduction where we write them in Lorentzian time. The conformal cross-ratios associated to this configuration in the $VV\to WW$ channel are
\be
z = \frac{1}{\cosh^2 \frac{\b-i \th}{2}},\qquad\bar z = \frac{1}{\cosh^2 \frac{\b+i \th}{2}},
\ee
where $\cosh \b=-e\.y$. The OTOC is obtained by setting $\theta\to -\e + it$. As $t$ increases, $z$ moves toward the origin, while $\bar z$ goes around $1$ before approaching the origin. In the long-time limit, we have
\be
\label{eq:thereggelimit}
z \sim e^{-t-\b}, \quad \bar z \sim e^{-t+\b}.
\ee
This is precisely the continuation of cross-ratios that corresponds to a time-ordered correlator in Minkowski space in the Regge limit \cite{Costa:2012cb}. We emphasize that the $i\e$ prescription for the OTOC in $H^{d-1}\x S^1$ becomes the $i\e$ prescription for a time-ordered correlator in Minkowski space.

Conformal Regge theory \cite{Costa:2012cb} gives a prediction for the behavior of a four-point function in the Regge limit:
\be
\<V(e,\pi)W(y,\pi+\theta)V(e,0) W(y,\theta)\>_{H^{d-1} \x S^1} \sim \sum_i \int_{-\oo}^\oo d\nu\, \alpha_i(\nu) e^{(J_i(\nu)-1)t} P_{\frac{2-d}{2}+i\nu}(\cosh \b),
\label{eq:leadingreggelimit}
\ee
where $P_{\frac{2-d}{2}+i\nu}(\cosh \b)$ is a Gegenbauer function with $J_L=\frac{2-d}{2}+i\nu$. The index $i$ runs over Regge trajectories with spins $J_i(\nu)$. The coefficient $\alpha_i(\nu)$ is related to the analytic continuation of OPE coefficients in spin. Applying the dictionary $\De_{L,i} = 1-J_i(\nu)$, we see that time evolution $\ptl_t$ evolves each harmonic component with eigenvalue $-\De_{L,i}(\nu)$, just as for expectation values of rotationally symmetric scalar observables like the number density $\<n(y)\>$ discussed in section~\ref{sec:introQCD}.

This gives a vivid physical picture to understand the evolution of the number density: it evolves on the mass shell like the spread of chaos in hyperbolic space. This relationship is only approximate --- there are corrections on both sides of the correspondence. In the detector picture for a deformed CFT, there are corrections from conformal perturbation theory in the deformation. In the Regge limit, there are subleading corrections to the formula (\ref{eq:leadingreggelimit}) described, e.g.\ in \cite{Caron-Huot:2020nem}.

Ultimately, the arguments above can be made without cross-ratios --- the correspondence stems from the idea that the same set of light-ray operators controls the Regge limit and detector matching at high energies. In particular, we expect similar correspondences between higher-point observables in the Regge limit and higher-point collider observables.

\section{Lorentzian inversion formula calculations}\label{sec:Lorentzian_Inversion_Formula}
Here, we show the details for Lorentzian inversion formula calculations of 2d Ising CFT correlators. The inversion formula in the $t$ channel reads
\begin{equation}
        c^t(J,\Delta)\propto\int_0^1 dz d\bar{z}\mu(z,\bar{z})G_{J+d-1,\Delta+1-d}(z,\bar{z})\text{dDisc}(\mathcal{G}(z,\bar{z}))
\end{equation}
where
\begin{equation}
        \text{dDisc}(\mathcal{G}(z,\bar{z}))=\mathcal{G}(z,\bar{z})-\frac{1}{2}(\mathcal{G}^{\circlearrowleft}(z,\bar{z})+\mathcal{G}^{\circlearrowright}(z,\bar{z}))
\end{equation}
where $\mathcal{G}^{\circlearrowleft,\circlearrowright}(z,\bar{z})$ denote the continuations around $\bar{z}=1$, and in two dimensions,
\begin{equation}
        \begin{aligned}
                G_{\Delta,J}(z,\bar{z})&=\frac{k_{\Delta-J}(z)k_{\Delta+J}(\bar{z})+k_{\Delta+J}(z)k_{\Delta-J}(\bar{z})}{1+\delta_{J,0}} \\
                k_{\beta}(z)&=z^{\beta/2} {}_{2}F_1\left(\frac{\beta}{2}+a,\frac{\beta}{2}+b,\beta,z\right) \\
                a&=\frac{\Delta_2-\Delta_1}{2} \quad b=\frac{\Delta_3-\Delta_4}{2}.
        \end{aligned}
\end{equation}

The inversion formula for $\<\s\s\s\s\>$ was already computed in Appendix B of~\cite{Caron-Huot:2017vep}, which is given by
\begin{equation}
        \begin{aligned}
                c^t(J,\Delta)\propto I_{-1/4}^0(J+\Delta)I_{-1/4}^0(J-\Delta+2)+I_{-1/4}^{1/2}(J+\Delta)I_{-1/4}^{1/2}(J-\Delta+2) \\
                -\frac{1}{\sqrt{2}}I_{-1/4}^{1/2}(J+\Delta)I_{-1/4}^0(J-\Delta+2)-\frac{1}{\sqrt{2}}I_{-1/4}^0(J+\Delta)I_{-1/4}^{1/2}(J-\Delta+2)
        \end{aligned}
\end{equation}
where
\begin{equation}
        I_{p_1}^{p_0}(\beta)\equiv\int_0^1\frac{d\rho (1-\rho^2)}{4\rho^2}k_{\beta}(\rho)\rho^{p_0}(1-\rho^2)^{p_1}.
\end{equation}
Importantly, the integrals $I$ each have a pole at $\beta=2$, and so $c^t(J,\Delta)$ will have a double pole when $J=\Delta=1$ (which corresponds to $J_L=\Delta_L=0$).

For $\<\e\e\e\e\>$ in the 2d Ising model, this integral becomes
\begin{equation}
        \int_0^1\frac{dz}{z^2}k_{J-\Delta+2}(z)\frac{1-z+z^2}{1-z}\int_0^1\frac{d\bar{z}}{\bar{z}^2}k_{J+\Delta}(\bar{z})\text{dDisc}\left(\frac{1-\bar{z}+\bar{z}^2}{1-\bar{z}}\right)+(\Delta\leftrightarrow 2-\Delta).
\end{equation}
Denote the $z$ integral as $I_1(\beta)$ and the $\bar{z}$ integral as $I_2(\bar{\beta},a)$, where $\beta=J-\Delta+2$ and $\bar{\beta}=J+\Delta$. Let us work out $I_2$ first. To do so, we first write
\begin{equation}
        \text{dDisc}\left(\frac{1-\bar{z}+\bar{z}^2}{1-\bar{z}}\right)=\text{dDisc}\left(\frac{\bar{z}^2}{1-\bar{z}}\right).
\end{equation}
Naively, this double discontinuity evaluates to zero. However, this does not take into account the singular behavior at $\bar{z}=1$,  where we should interpret this double discontinuity to have $\delta$-distribution-like terms.
To perform the calculation, we will first work with a regulated version of the double discontinuity:
\begin{equation}
        \text{dDisc}\left(\frac{\bar{z}^2}{1-\bar{z}}\right)^a=2\sin^2{(\pi a)}\left(\frac{\bar{z}^2}{1-\bar{z}}\right)^a,
\end{equation}
then send $a\rightarrow 1$. We get
\begin{equation}
        I_2(\bar{\beta},a)=2\sin^2{(\pi a)}\frac{\Gamma(1-a)\Gamma\left(-1+2a+\frac{\bar{\beta}}{2}\right) {}_3F_2\left(\frac{\bar{\beta}}{2},\frac{\bar{\beta}}{2},-1+2a+\frac{\bar{\beta}}{2}; \bar{\beta},a+\frac{\bar{\beta}}{2}; 1\right)}{\Gamma\left(a+\frac{\bar{\beta}}{2}\right)}
\end{equation}
${}_3F_2$ with these arguments has the following analytic continuation:
\begin{equation}
        {}_3F_2\left(\frac{\bar{\beta}}{2},\frac{\bar{\beta}}{2},-1+2a+\frac{\bar{\beta}}{2}; \bar{\beta},a+\frac{\bar{\beta}}{2}; 1\right)=\frac{\sqrt{\pi}\Gamma(1-a)\Gamma\left(\frac{\bar{\beta}+1}{2}\right)\Gamma\left(a+\frac{\bar{\beta}}{2}\right)}{\Gamma\left(1-a+\frac{\bar{\beta}}{4}\right)\Gamma\left(a+\frac{\bar{\beta}}{4}\right)\Gamma^2{\left(\frac{2+\bar{\beta}}{4}\right)}}.
\end{equation}
We can now take the $a\rightarrow 1$ limit, yielding
\begin{equation}
        I_2(\bar{\beta},1)=\frac{2^{1+\bar{\beta}/2}\pi^2\Gamma\left(\frac{\bar{\beta}+1}{2}\right)}{\Gamma\left(\frac{\bar{\beta}}{4}\right)\Gamma\left(\frac{2+\bar{\beta}}{4}\right)}.
\end{equation}
Notice that there is no pole at $\bar{\beta}=2$.

Now let us turn to $I_1$. In performing the integral, we find that it diverges for any $\beta$. The nature of this divergence comes from the $z\rightarrow 1$ limit, but as argued in \cite{Caron-Huot:2017vep}, this type of divergence does not affect the pole structure in $(\Delta,J)$ that is relevant for light-ray operators. To separate this
divergence from our analysis, we first break up the integral into two pieces by writing
\begin{equation}
        \frac{1-z+z^2}{1-z}=1+\frac{z^2}{1-z}.
\end{equation}
The $z\rightarrow 1$ divergence comes from the second term. To regulate this divergence, we replace
\begin{equation}
        \frac{z^2}{1-z}\rightarrow\left(\frac{z^2}{1-z}\right)^a.
\end{equation}
The integral then becomes
\begin{equation}
        I_1(\beta)=\frac{2^{\beta-1}\Gamma\left(\frac{\beta+1}{2}\right)}{\left(\frac{\beta}{2}-1\right)\sqrt{\pi}\Gamma\left(1+\frac{\beta}{2}\right)}+\frac{\sqrt{\pi}\Gamma^2(1-a)\Gamma\left(-1+2a+\frac{\beta}{2}\right)\Gamma\left(\frac{\beta+1}{2}\right)\Gamma\left(a+\frac{\beta}{2}\right)}{\Gamma\left(a+\frac{\beta}{2}\right)\Gamma\left(1-a+\frac{\beta}{4}\right)\Gamma\left(a+\frac{\beta}{4}\right)\Gamma^2\left(\frac{\beta+2}{4}\right)}.
\end{equation}
The $z\rightarrow 1$ divergence is now entirely captured by the $a\rightarrow 1$ divergence in the second term, but has no pole at $\beta=2$. Meanwhile the first term is finite, and has a single pole at $\beta=2$. Thus, we see that the Lorentzian inversion formula for $\<\e\e\e\e\>$ only has a single pole at $\Delta=J=1$.

\section{Details on the 2d Ising model in a magnetic field}
\label{sec:magneticapp}
\subsection{Scattering theory}
Here, we give a quick review of the scattering theory necessary for the computation of the form factors of the Ising model in a magnetic field. We restrict to the scattering of neutral, scalar, non-degenerate particles. In two dimensions,
momenta of on-shell asymptotic particles can be parametrized by a single variable, the rapidity $\th$. Asymptotic states are denoted by
\begin{equation}
        |(a_1,\th_1),\dots,(a_n,\th_n)\>
\end{equation}
where $a_i$ labels particle species.

The $S$-matrix obeys the usual unitarity and crossing-symmetric conditions, and, in the case of integrable models, is elastic. A result of these conditions is that
the two-particle $S$-matrix is a meromorphic function of $\th$ obeying the equations
\begin{equation}
        \begin{aligned}
                S_{ab}(\th)S_{ab}(-\th)=1 \\
                S_{ab}(\th)=S_{ab}(i\pi-\th)
        \end{aligned}
\end{equation}
with the bound states of the theory corresponding to simple poles along the imaginary axis between
$0$ and $i\pi$. That is, we have
\begin{equation}
        \text{Res}_{\th=iu_{ab}^c}S_{ab}(\th)=i\left(\Gamma_{ab}^c\right)^2
\end{equation}
where $a,b,c$ label asymptotic states of the theory, $\Gamma_{ab}^c$ is the three-particle coupling between $a$, $b$, and $c$, and $u_{ab}^c$ are the numbers in parentheses in Table~\ref{table:magnetic_ising_s_matrix}.
These equations for the $S$-matrix and $\Gamma_{ab}^c$ can be solved exactly. For the Ising model in a magnetic field, the solution is~\cite{Zamolodchikov:1989fp}
\begin{equation}
        S_{ab}(\th)=\prod_{\a\in\cA_{ab}}\left(f_{\a}(\th)\right)^{p_{\a}}
\end{equation}
where
\begin{equation}
        \begin{aligned}
                f_{\a}(\th)&=\frac{\tanh\frac{1}{2}(\th+i\pi\a)}{\tanh\frac{1}{2}(\th-i\pi\a)} \\
                f_{\a}(\th)&=f_{1-\a}(\th) \\
                f_{\a}(\th)&=f_{\a}(i\pi-\th)
        \end{aligned}
\end{equation}
and $\cA_{ab}$ and $p_{\a}$ can be found in Table~\ref{table:magnetic_ising_s_matrix}.
\begin{table}
\centering
\begin{tabular}{||c | c ||}
        \hline
        $a$ $b$ & $S_{ab}$ \\ [0.5ex]
        \hline\hline
        1 1 & $ \begin{matrix}
                        \textbf{1} \\
                        (20)
                \end{matrix} \,
                \begin{matrix}
                        \textbf{2} \\
                        (12)
                \end{matrix} \,
                \begin{matrix}
                        \textbf{3} \\
                        (2)
                \end{matrix}$ \\
        \hline
        1 2 & $\begin{matrix}
                        \textbf{1} \\
                        (24)
                \end{matrix} \,
                \begin{matrix}
                        \textbf{2} \\
                        (18)
                \end{matrix} \,
                \begin{matrix}
                        \textbf{3} \\
                        (14)
                \end{matrix} \,
                \begin{matrix}
                        \textbf{4} \\
                        (8)
                \end{matrix} \,$ \\
        \hline
        1 3 & $\begin{matrix}
                        \textbf{1} \\
                        (29)
                \end{matrix} \,
                \begin{matrix}
                        \textbf{2} \\
                        (21)
                \end{matrix} \,
                \begin{matrix}
                        \textbf{4} \\
                        (13)
                \end{matrix} \,
                \begin{matrix}
                        \textbf{5} \\
                        (3)
                \end{matrix} \,
                \begin{matrix}
                         \\
                        (11)^2
                \end{matrix} \,$ \\
        \hline
        1 4 & $\begin{matrix}
                        \textbf{2} \\
                        (25)
                \end{matrix} \,
                \begin{matrix}
                        \textbf{3} \\
                        (21)
                \end{matrix} \,
                \begin{matrix}
                        \textbf{4} \\
                        (17)
                \end{matrix} \,
                \begin{matrix}
                        \textbf{5} \\
                        (11)
                \end{matrix} \,
                \begin{matrix}
                        \textbf{6} \\
                        (7)
                \end{matrix} \,
                \begin{matrix}
                        \\
                        (15)
                \end{matrix} \,$ \\
        \hline
        1 5 & $\begin{matrix}
                        \textbf{3} \\
                        (28)
                \end{matrix} \,
                \begin{matrix}
                        \textbf{4} \\
                        (22)
                \end{matrix} \,
                \begin{matrix}
                        \textbf{6} \\
                        (14)
                \end{matrix} \,
                \begin{matrix}
                        \textbf{7} \\
                        (4)
                \end{matrix} \,
                \begin{matrix}
                        \\
                        (10)^2
                \end{matrix} \,
                \begin{matrix}
                        \\
                        (12)^2
                \end{matrix} \,$ \\
        \hline
        1 6 & $\begin{matrix}
                        \textbf{4} \\
                        (25)
                \end{matrix} \,
                \begin{matrix}
                        \textbf{5} \\
                        (19)
                \end{matrix} \,
                \begin{matrix}
                        \textbf{7} \\
                        (9)
                \end{matrix} \,
                \begin{matrix}
                        \\
                        (7)^2
                \end{matrix} \,
                \begin{matrix}
                        \\
                        (13)^2
                \end{matrix} \,
                \begin{matrix}
                        \\
                        (15)
                \end{matrix} \,$ \\
        \hline
        1 7 & $\begin{matrix}
                        \textbf{5} \\
                        (27)
                \end{matrix} \,
                \begin{matrix}
                        \textbf{6} \\
                        (23)
                \end{matrix} \,
                \begin{matrix}
                        \textbf{8} \\
                        (5)
                \end{matrix} \,
                \begin{matrix}
                        \\
                        (9)^2
                \end{matrix} \,
                \begin{matrix}
                        \\
                        (11)^2
                \end{matrix} \,
                \begin{matrix}
                        \\
                        (13)^2
                \end{matrix} \,
                \begin{matrix}
                        \\
                        (15)
                \end{matrix} \,$ \\
        \hline
        1 8 & $\begin{matrix}
                        \textbf{7} \\
                        (26)
                \end{matrix} \,
                \begin{matrix}
                        \textbf{8} \\
                        (16)^3
                \end{matrix} \,
                \begin{matrix}
                        \\
                        (6)^2
                \end{matrix} \,
                \begin{matrix}
                        \\
                        (8)^2
                \end{matrix} \,
                \begin{matrix}
                        \\
                        (10)^2
                \end{matrix} \,
                \begin{matrix}
                        \\
                        (12)^2
                \end{matrix} \,$ \\
        \hline
        2 2 & $\begin{matrix}
                        \textbf{1} \\
                        (24)
                \end{matrix} \,
                \begin{matrix}
                        \textbf{2} \\
                        (20)
                \end{matrix} \,
                \begin{matrix}
                        \textbf{4} \\
                        (14)
                \end{matrix} \,
                \begin{matrix}
                        \textbf{5} \\
                        (8)
                \end{matrix} \,
                \begin{matrix}
                        \textbf{6} \\
                        (2)
                \end{matrix} \,
                \begin{matrix}
                        \\
                        (12)^2
                \end{matrix} \,$ \\
        \hline
        2 3 & $\begin{matrix}
                        \textbf{1} \\
                        (25)
                \end{matrix} \,
                \begin{matrix}
                        \textbf{3} \\
                        (19)
                \end{matrix} \,
                \begin{matrix}
                        \textbf{6} \\
                        (9)
                \end{matrix} \,
                \begin{matrix}
                        \\
                        (7)^2
                \end{matrix} \,
                \begin{matrix}
                        \\
                        (13)^2
                \end{matrix} \,
                \begin{matrix}
                        \\
                        (15)
                \end{matrix} \,$ \\
        \hline
        2 4 & $\begin{matrix}
                        \textbf{1} \\
                        (27)
                \end{matrix} \,
                \begin{matrix}
                        \textbf{2} \\
                        (23)
                \end{matrix} \,
                \begin{matrix}
                        \textbf{7} \\
                        (5)
                \end{matrix} \,
                \begin{matrix}
                        \\
                        (9)^2
                \end{matrix} \,
                \begin{matrix}
                        \\
                        (11)^2
                \end{matrix} \,
                \begin{matrix}
                        \\
                        (13)^2
                \end{matrix} \,
                \begin{matrix}
                        \\
                        (15)
                \end{matrix} \,$ \\
        \hline
        2 5 & $\begin{matrix}
                        \textbf{2} \\
                        (26)
                \end{matrix} \,
                \begin{matrix}
                        \textbf{6} \\
                        (16)^3
                \end{matrix} \,
                \begin{matrix}
                        \\
                        (6)^2
                \end{matrix} \,
                \begin{matrix}
                        \\
                        (8)^2
                \end{matrix} \,
                \begin{matrix}
                        \\
                        (10)^2
                \end{matrix} \,
                \begin{matrix}
                        \\
                        (12)^2
                \end{matrix} \,$ \\
        \hline
        2 6 & $\begin{matrix}
                        \textbf{2} \\
                        (29)
                \end{matrix} \,
                \begin{matrix}
                        \textbf{3} \\
                        (25)
                \end{matrix} \,
                \begin{matrix}
                        \textbf{5} \\
                        (19)^3
                \end{matrix} \,
                \begin{matrix}
                        \textbf{7} \\
                        (13)^3
                \end{matrix} \,
                \begin{matrix}
                        \textbf{8} \\
                        (3)
                \end{matrix} \,
                \begin{matrix}
                        \\
                        (7)^2
                \end{matrix} \,
                \begin{matrix}
                        \\
                        (9)^2
                \end{matrix} \,
                \begin{matrix}
                        \\
                        (15)
                \end{matrix} \,$ \\
        \hline
        2 7 & $\begin{matrix}
                        \textbf{4} \\
                        (27)
                \end{matrix} \,
                \begin{matrix}
                        \textbf{6} \\
                        (21)^3
                \end{matrix} \,
                \begin{matrix}
                        \textbf{7} \\
                        (17)^3
                \end{matrix} \,
                \begin{matrix}
                        \textbf{8} \\
                        (11)^3
                \end{matrix} \,
                \begin{matrix}
                        \\
                        (5)^2
                \end{matrix} \,
                \begin{matrix}
                        \\
                        (7)^2
                \end{matrix} \,
                \begin{matrix}
                        \\
                        (15)^2
                \end{matrix} \,$ \\
        \hline
        2 8 & $\begin{matrix}
                        \textbf{6} \\
                        (28)
                \end{matrix} \,
                \begin{matrix}
                        \textbf{7} \\
                        (22)^3
                \end{matrix} \,
                \begin{matrix}
                        \\
                        (4)^2
                \end{matrix} \,
                \begin{matrix}
                        \\
                        (6)^2
                \end{matrix} \,
                \begin{matrix}
                        \\
                        (10)^4
                \end{matrix} \,
                \begin{matrix}
                        \\
                        (12)^4
                \end{matrix} \,
                \begin{matrix}
                        \\
                        (16)^4
                \end{matrix} \,$ \\
        \hline
        3 3 & $\begin{matrix}
                        \textbf{2} \\
                        (22)
                \end{matrix} \,
                \begin{matrix}
                        \textbf{3} \\
                        (20)^3
                \end{matrix} \,
                \begin{matrix}
                        \textbf{5} \\
                        (14)
                \end{matrix} \,
                \begin{matrix}
                        \textbf{6} \\
                        (12)^3
                \end{matrix} \,
                \begin{matrix}
                        \textbf{7} \\
                        (4)
                \end{matrix} \,
                \begin{matrix}
                        \\
                        (2)^2
                \end{matrix} \,$ \\
        \hline
        3 4 & $\begin{matrix}
                        \textbf{1} \\
                        (26)
                \end{matrix} \,
                \begin{matrix}
                        \textbf{5} \\
                        (16)^3
                \end{matrix} \,
                \begin{matrix}
                        \\
                        (6)^2
                \end{matrix} \,
                \begin{matrix}
                        \\
                        (8)^2
                \end{matrix} \,
                \begin{matrix}
                        \\
                        (10)^2
                \end{matrix} \,
                \begin{matrix}
                        \\
                        (12)^2
                \end{matrix} \,$ \\
        \hline
        3 5 & $\begin{matrix}
                        \textbf{1} \\
                        (29)
                \end{matrix} \,
                \begin{matrix}
                        \textbf{3} \\
                        (23)
                \end{matrix} \,
                \begin{matrix}
                        \textbf{4} \\
                        (21)^3
                \end{matrix} \,
                \begin{matrix}
                        \textbf{7} \\
                        (13)^3
                \end{matrix} \,
                \begin{matrix}
                        \textbf{8} \\
                        (5)
                \end{matrix} \,
                \begin{matrix}
                        \\
                        (3)^2
                \end{matrix} \,
                \begin{matrix}
                        \\
                        (11)^4
                \end{matrix} \,
                \begin{matrix}
                        \\
                        (15)
                \end{matrix} \,$ \\
        \hline
\end{tabular}
\caption{The bolded number labels particle species.
The number in parentheses is the location of the pole in units of $\frac{i\pi}{30}$. The superscript number is the exponent $p_\a$ of the corresponding factor $f_\a$, which is the degree of the pole except for $(15)$, where each factor $f_{1/2}$ contributes a double pole.}
\label{table:magnetic_ising_s_matrix}
\end{table}
\begin{table}
\centering
\begin{tabular}{||c | c ||}
        \hline
        $a$ $b$ & $S_{ab}$ \\ [0.5ex]
        \hline\hline
        3 6 & $\begin{matrix}
                        \textbf{2} \\
                        (26)
                \end{matrix} \,
                \begin{matrix}
                        \textbf{3} \\
                        (24)^3
                \end{matrix} \,
                \begin{matrix}
                        \textbf{6} \\
                        (18)^3
                \end{matrix} \,
                \begin{matrix}
                        \textbf{8} \\
                        (8)^3
                \end{matrix} \,
                \begin{matrix}
                        \\
                        (10)^2
                \end{matrix} \,
                \begin{matrix}
                        \\
                        (16)^4
                \end{matrix} \,$ \\
        \hline
        3 7 & $\begin{matrix}
                        \textbf{3} \\
                        (28)
                \end{matrix} \,
                \begin{matrix}
                        \textbf{5} \\
                        (22)^3
                \end{matrix} \,
                \begin{matrix}
                        \\
                        (4)^2
                \end{matrix} \,
                \begin{matrix}
                        \\
                        (6)^2
                \end{matrix} \,
                \begin{matrix}
                        \\
                        (10)^4
                \end{matrix} \,
                \begin{matrix}
                        \\
                        (12)^4
                \end{matrix} \,
                \begin{matrix}
                        \\
                        (16)^4
                \end{matrix} \,$ \\
        \hline
        3 8 & $\begin{matrix}
                        \textbf{5} \\
                        (27)
                \end{matrix} \,
                \begin{matrix}
                        \textbf{6} \\
                        (25)^3
                \end{matrix} \,
                \begin{matrix}
                        \textbf{8} \\
                        (17)^5
                \end{matrix} \,
                \begin{matrix}
                        \\
                        (7)^4
                \end{matrix} \,
                \begin{matrix}
                        \\
                        (9)^4
                \end{matrix} \,
                \begin{matrix}
                        \\
                        (11)^4
                \end{matrix} \,
                \begin{matrix}
                        \\
                        (15)^3
                \end{matrix} \,$ \\
        \hline
        4 4 & $\begin{matrix}
                        \textbf{1} \\
                        (26)
                \end{matrix} \,
                \begin{matrix}
                        \textbf{4} \\
                        (20)^3
                \end{matrix} \,
                \begin{matrix}
                        \textbf{6} \\
                        (16)^3
                \end{matrix} \,
                \begin{matrix}
                        \textbf{7} \\
                        (12)^3
                \end{matrix} \,
                \begin{matrix}
                        \textbf{8} \\
                        (2)
                \end{matrix} \,
                \begin{matrix}
                        \\
                        (6)^2
                \end{matrix} \,
                \begin{matrix}
                        \\
                        (8)^2
                \end{matrix} \,$ \\
        \hline
        4 5 & $\begin{matrix}
                        \textbf{1} \\
                        (27)
                \end{matrix} \,
                \begin{matrix}
                        \textbf{3} \\
                        (23)^3
                \end{matrix} \,
                \begin{matrix}
                        \textbf{5} \\
                        (19)^3
                \end{matrix} \,
                \begin{matrix}
                        \textbf{8} \\
                        (9)^3
                \end{matrix} \,
                \begin{matrix}
                        \\
                        (5)^2
                \end{matrix} \,
                \begin{matrix}
                        \\
                        (13)^4
                \end{matrix} \,
                \begin{matrix}
                        \\
                        (15)^2
                \end{matrix} \,$ \\
        \hline
        4 6 & $\begin{matrix}
                        \textbf{1} \\
                        (28)
                \end{matrix} \,
                \begin{matrix}
                        \textbf{4} \\
                        (22)^3
                \end{matrix} \,
                \begin{matrix}
                        \\
                        (4)^2
                \end{matrix} \,
                \begin{matrix}
                        \\
                        (6)^2
                \end{matrix} \,
                \begin{matrix}
                        \\
                        (10)^4
                \end{matrix} \,
                \begin{matrix}
                        \\
                        (12)^4
                \end{matrix} \,
                \begin{matrix}
                        \\
                        (16)^4
                \end{matrix} \,$ \\
        \hline
        4 7 & $\begin{matrix}
                        \textbf{2} \\
                        (28)
                \end{matrix} \,
                \begin{matrix}
                        \textbf{4} \\
                        (24)^3
                \end{matrix} \,
                \begin{matrix}
                        \textbf{7} \\
                        (18)^5
                \end{matrix} \,
                \begin{matrix}
                        \textbf{8} \\
                        (14)^5
                \end{matrix} \,
                \begin{matrix}
                        \\
                        (4)^2
                \end{matrix} \,
                \begin{matrix}
                        \\
                        (8)^4
                \end{matrix} \,
                \begin{matrix}
                        \\
                        (10)^4
                \end{matrix} \,$ \\
        \hline
        4 8 & $\begin{matrix}
                        \textbf{4} \\
                        (29)
                \end{matrix} \,
                \begin{matrix}
                        \textbf{5} \\
                        (25)^3
                \end{matrix} \,
                \begin{matrix}
                        \textbf{7} \\
                        (21)^5
                \end{matrix} \,
                \begin{matrix}
                        \\
                        (3)^2
                \end{matrix} \,
                \begin{matrix}
                        \\
                        (7)^4
                \end{matrix} \,
                \begin{matrix}
                        \\
                        (11)^6
                \end{matrix} \,
                \begin{matrix}
                        \\
                        (13)^6
                \end{matrix} \,
                \begin{matrix}
                        \\
                        (15)^3
                \end{matrix} \,$ \\
        \hline
        5 5 & $\begin{matrix}
                        \textbf{4} \\
                        (22)^3
                \end{matrix} \,
                \begin{matrix}
                        \textbf{5} \\
                        (20)^5
                \end{matrix} \,
                \begin{matrix}
                        \textbf{8} \\
                        (12)^5
                \end{matrix} \,
                \begin{matrix}
                        \\
                        (2)^2
                \end{matrix} \,
                \begin{matrix}
                        \\
                        (4)^2
                \end{matrix} \,
                \begin{matrix}
                        \\
                        (6)^2
                \end{matrix} \,
                \begin{matrix}
                        \\
                        (16)^4
                \end{matrix} \,$ \\
        \hline
        5 6 & $\begin{matrix}
                        \textbf{1} \\
                        (27)
                \end{matrix} \,
                \begin{matrix}
                        \textbf{2} \\
                        (25)^3
                \end{matrix} \,
                \begin{matrix}
                        \textbf{7} \\
                        (17)^5
                \end{matrix} \,
                \begin{matrix}
                        \\
                        (7)^4
                \end{matrix} \,
                \begin{matrix}
                        \\
                        (9)^4
                \end{matrix} \,
                \begin{matrix}
                        \\
                        (11)^4
                \end{matrix} \,
                \begin{matrix}
                        \\
                        (15)^3
                \end{matrix} \,$ \\
        \hline
        5 7 & $\begin{matrix}
                        \textbf{1} \\
                        (29)
                \end{matrix} \,
                \begin{matrix}
                        \textbf{3} \\
                        (25)^3
                \end{matrix} \,
                \begin{matrix}
                        \textbf{6} \\
                        (21)^5
                \end{matrix} \,
                \begin{matrix}
                        \\
                        (3)^2
                \end{matrix} \,
                \begin{matrix}
                        \\
                        (7)^4
                \end{matrix} \,
                \begin{matrix}
                        \\
                        (11)^6
                \end{matrix} \,
                \begin{matrix}
                        \\
                        (13)^6
                \end{matrix} \,
                \begin{matrix}
                        \\
                        (15)^3
                \end{matrix} \,$ \\
        \hline
        5 8 & $\begin{matrix}
                        \textbf{3} \\
                        (28)
                \end{matrix} \,
                \begin{matrix}
                        \textbf{4} \\
                        (26)^3
                \end{matrix} \,
                \begin{matrix}
                        \textbf{5} \\
                        (24)^5
                \end{matrix} \,
                \begin{matrix}
                        \textbf{8} \\
                        (18)^7
                \end{matrix} \,
                \begin{matrix}
                        \\
                        (8)^6
                \end{matrix} \,
                \begin{matrix}
                        \\
                        (10)^6
                \end{matrix} \,
                \begin{matrix}
                        \\
                        (16)^8
                \end{matrix} \,$ \\
        \hline
        6 6 & $\begin{matrix}
                        \textbf{3} \\
                        (24)^3
                \end{matrix} \,
                \begin{matrix}
                        \textbf{6} \\
                        (20)^5
                \end{matrix} \,
                \begin{matrix}
                        \textbf{8} \\
                        (14)^5
                \end{matrix} \,
                \begin{matrix}
                        \\
                        (2)^2
                \end{matrix} \,
                \begin{matrix}
                        \\
                        (4)^2
                \end{matrix} \,
                \begin{matrix}
                        \\
                        (8)^4
                \end{matrix} \,
                \begin{matrix}
                        \\
                        (12)^6
                \end{matrix} \,$ \\
        \hline
        6 7 & $\begin{matrix}
                        \textbf{1} \\
                        (28)
                \end{matrix} \,
                \begin{matrix}
                        \textbf{2} \\
                        (26)^3
                \end{matrix} \,
                \begin{matrix}
                        \textbf{5} \\
                        (22)^5
                \end{matrix} \,
                \begin{matrix}
                        \textbf{8} \\
                        (16)^7
                \end{matrix} \,
                \begin{matrix}
                        \\
                        (6)^4
                \end{matrix} \,
                \begin{matrix}
                        \\
                        (10)^6
                \end{matrix} \,
                \begin{matrix}
                        \\
                        (12)^6
                \end{matrix} \,$ \\
        \hline
        6 8 & $\begin{matrix}
                        \textbf{2} \\
                        (29)
                \end{matrix} \,
                \begin{matrix}
                        \textbf{3} \\
                        (27)^3
                \end{matrix} \,
                \begin{matrix}
                        \textbf{6} \\
                        (23)^5
                \end{matrix} \,
                \begin{matrix}
                        \textbf{7} \\
                        (21)^7
                \end{matrix} \,
                \begin{matrix}
                        \\
                        (5)^4
                \end{matrix} \,
                \begin{matrix}
                        \\
                        (11)^8
                \end{matrix} \,
                \begin{matrix}
                        \\
                        (13)^8
                \end{matrix} \,
                \begin{matrix}
                        \\
                        (15)^4
                \end{matrix} \,$ \\
        \hline
        7 7 & $\begin{matrix}
                        \textbf{2} \\
                        (26)^3
                \end{matrix} \,
                \begin{matrix}
                        \textbf{4} \\
                        (24)^5
                \end{matrix} \,
                \begin{matrix}
                        \textbf{7} \\
                        (20)^7
                \end{matrix} \,
                \begin{matrix}
                        \\
                        (2)^2
                \end{matrix} \,
                \begin{matrix}
                        \\
                        (8)^6
                \end{matrix} \,
                \begin{matrix}
                        \\
                        (12)^8
                \end{matrix} \,
                \begin{matrix}
                        \\
                        (16)^8
                \end{matrix} \,$ \\
        \hline
        7 8 & $\begin{matrix}
                        \textbf{1} \\
                        (29)
                \end{matrix} \,
                \begin{matrix}
                        \textbf{2} \\
                        (27)^3
                \end{matrix} \,
                \begin{matrix}
                        \textbf{4} \\
                        (25)^5
                \end{matrix} \,
                \begin{matrix}
                        \textbf{6} \\
                        (23)^7
                \end{matrix} \,
                \begin{matrix}
                        \textbf{8} \\
                        (19)^9
                \end{matrix} \,
                \begin{matrix}
                        \\
                        (9)^8
                \end{matrix} \,
                \begin{matrix}
                        \\
                        (13)^{10}
                \end{matrix} \,
                \begin{matrix}
                        \\
                        (15)^5
                \end{matrix} \,$ \\
        \hline
        8 8 & $\begin{matrix}
                        \textbf{1} \\
                        (28)^3
                \end{matrix} \,
                \begin{matrix}
                        \textbf{3} \\
                        (26)^5
                \end{matrix} \,
                \begin{matrix}
                        \textbf{5} \\
                        (24)^7
                \end{matrix} \,
                \begin{matrix}
                        \textbf{7} \\
                        (22)^9
                \end{matrix} \,
                \begin{matrix}
                        \textbf{8} \\
                        (20)^{11}
                \end{matrix} \,
                \begin{matrix}
                        \\
                        (12)^{12}
                \end{matrix} \,
                \begin{matrix}
                        \\
                        (16)^{12}
                \end{matrix} \,$ \\
        \hline
\end{tabular}
\caption{Continuation of the magnetic Ising $S$-matrix.}
\label{table:magnetic_ising_s_matrix_continued}
\end{table}

\subsection{Form factors}

Here, we describe the computation of the form factors for the Ising model in a magnetic field, in which we follow~\cite{Delfino:1995zk}. Reference~\cite{Delfino:1995zk} describes two
different methods for computing the form factors. We follow their second method, described in Appendix A of the cited paper. However, we note that several
of the formulas written there have typos. We will provide the corrected versions of their formulas in this section.

The possibility of computing the form factors of the Ising model in a magnetic field comes from its integrability. In integrable theories, Lorentz invariance, elasticity,
and crossing symmetry imply that the form factors of scalar Hermitian operators obey the following equations, known as Watson's equations:
\begin{equation} \label{eq:Waton's_equations}
        \begin{aligned}
                F_{a_1\dots a_i a_{i+1}\dots a_n}^{\cO}(\th_1,\dots,\th_i,\th_{i+1},\dots,\th_n)&=S_{a_ia_{i+1}}(\th_i-\th_{i+1})F_{a_1\dots a_{i+1}a_i \dots a_n}^{\cO}(\th_1,\dots,\th_{i+1},\th_i,\dots,\th_n) \\
                F_{a_1 a_2 \dots a_n}^{\cO}(\th_1+2\pi i,\dots,\th_{n-1},\th_n)&=F_{a_2 a_3 \dots a_n a_1}^{\cO}(\th_2,\dots,\th_n,\th_1).
        \end{aligned}
\end{equation}

Additionally, the poles of the $S$-matrix induce poles in the form factors and give the following residue relations between $n$-particle form factors and $(n-1)$- and $(n-2)$-particle form factors:
\begin{equation} \label{eq:FF_pole_equations}
        \begin{aligned}
                &\text{Res}_{\th_1=\th_2+i\pi}F_{a_1a_2a_3\dots a_n}^{\cO}(\th_1,\th_2,\th_3,\dots,\th_n)=i\de_{a_1a_2}\left(1-\prod_{j=1}^nS_{a_ja_1}(\th_2-\th_j)\right)F_{a_3 \dots a_n}^{\cO}(\th_3,\dots,\th_n) \\
                &\text{Res}_{\th_1=\th_2+iu_{a_1a_2}^c}F_{a_1a_2a_3\dots a_n}^{\cO}(\th_1,\th_2,\th_3,\dots,\th_n)=i\left(\Gamma_{a_1a_2}^c\right)F_{ca_3 \dots a_n}^{\cO}(\th_c,\th_3,\dots,\th_n),
        \end{aligned}
\end{equation}
where $\th_1=\th_c+i (\pi - u_{a_1 c}^{a_2})$, $\th_2=\th_c - i (\pi - u_{a_2 c}^{a_1})$. 

An ansatz that solves these equations is
\begin{equation}
        F_{a_1\dots a_n}^{\cO}(\th_1,\dots,\th_n)=Q_{a_1\dots a_n}^{\cO}(x_1,\dots,x_n)\prod_{i<j}\frac{F_{a_ia_j}^{\text{min}}(\th_i-\th_j)}{(x_i+x_j)^{\de_{a_ia_j}}D_{a_ia_j}(\th_i-\th_j)}
\end{equation}
where $x_i\equiv e^{\th_i}$ and the dynamical poles are contained in $D$:
\begin{equation}
        \begin{aligned} 
                &D_{ab}(\th)=\prod_{\a\in\cA_{ab}}\left(P_{\a}(\th)\right)^{i_{\a}}\left(P_{1-\a}(\th)\right)^{j_{\a}} \\
                &P_{\a}(\th)=\frac{\cos\pi\a-\cosh\th}{2\cos^2\frac{\pi\a}{2}}
        \end{aligned}
\end{equation}
where
\begin{equation}
        \begin{aligned}
                i_{\a}&=n+1, && j_{\a}=n \quad \text{if} \quad p_{\a}=2n+1 \\
                i_{\a}&=n, && j_{\a}=n \quad \text{if} \quad p_{\a}=2n,
        \end{aligned}
\end{equation}
and
\begin{equation}
        \begin{aligned}
                F_{ab}^{\text{min}}(\th)=\left(-i\sinh\frac{\th}{2}\right)^{\de_{ab}}\prod_{\a\in\cA_{ab}}\left(G_{\a}(\th)\right)^{p_{\a}} \\
                G_{\a}(\th)=\exp\left\{2\int_0^{\infty}\frac{dt}{t}\frac{\cosh\left(\a-\frac{1}{2}\right)t}{\cosh\frac{t}{2}\sinh t}\sin^2{\frac{(i\pi-\th)t}{2\pi}}\right\}.
        \end{aligned}
\end{equation}
A useful expression for numerical evaluations of $G_{\a}(\th)$ is
\begin{equation}
        \begin{aligned}
                G_{\a}(\th)&=\prod_{k=0}^{N-1}\left[\frac{\left[1+\left(\frac{\widehat{\th}/2\pi}{k+1-\frac{\a}{2}}\right)^2\right]\left[1+\left(\frac{\widehat{\th}/2\pi}{k+\frac{1}{2}+\frac{\a}{2}}\right)^2\right]}{\left[1+\left(\frac{\widehat{\th}/2\pi}{k+1+\frac{\a}{2}}\right)^2\right]\left[1+\left(\frac{\widehat{\th}/2\pi}{k+\frac{3}{2}-\frac{\a}{2}}\right)^2\right]}\right]^{k+1} \\
                &\times \exp\left[2\int_0^{\infty}\frac{dt}{t}\frac{\cosh\left[\frac{t}{2}(1-2\a)\right]}{\cosh\frac{t}{2}\sinh t}(N+1-Ne^{-2t})e^{-2Nt}\sin^2\frac{\widehat{\th}t}{2\pi}\right]
        \end{aligned}
\end{equation}
where $\widehat{\th}\equiv i\pi-\th$. This expression is true for any positive integer $N$.

Some properties of $G_{\a}(\th)$ that are useful in deriving the later equations are 
\begin{equation} \label{eq:Gfunc_identity_1}
        G_{\a}(\th)=G_{1-\a}(\th)
\end{equation}
\begin{equation}
        G_{\a}(\th+i\pi)G_{\a}(\th)=-i\frac{G_{\a}(0)}{\sin\pi\a}(\sinh\th+i\sin\pi\a)
\end{equation}
\begin{equation}
        \frac{G_{\a}(\th+i\pi\g)G_{\a}(\th-i\pi\g)}{G_{\a+\g}(\th)G_{\a-\g}(\th)}=\frac{G_{\a}(i\pi\g)G_{\a}(-i\pi\g)}{G_{\a+\g}(0)G_{\a-\g}(0)}
\end{equation}
\begin{equation} \label{eq:Gfunc_identity_5}
        G_{1-\a}(\th)G_{\a-1}(\th)=\frac{\sinh\frac{1}{2}(\th-i(\a-1)\pi)\sinh\frac{1}{2}(\th+i(\a+1)\pi)}{\sin^2\frac{\pi\a}{2}}
\end{equation}
\begin{equation}
        \lim_{\th\rightarrow\infty}G_{\a}(\th)=-i\sqrt{\frac{G_{\a}(0)}{2\sin\pi\a}}e^{\frac{\th}{2}}.
\end{equation}

The operator dependence is contained in $Q^{\cO}$, which is what we will need to solve for. $Q^{\cO}$ is a rational function, whose degree is constrained by
the asymptotic behavior of the form factors~\cite{Delfino:1995zk}:
\begin{equation}
        \lim_{|\th_i|\rightarrow\infty}F^{\cO}_{a_1\dots a_n}(\th_1,\dots,\th_n)\sim \exp{(Y_{\cO}|\th_i|)}
\end{equation}
where
\begin{equation} \label{eq:FF_asymptotic_bound}
        Y_{\cO}\leq \De_{\cO}/2
\end{equation}

As discussed in section~\ref{sec:magnetic}, we will need to solve for $Q_{1\dots 1}^{\s}(x_1,\dots,x_n)\equiv Q_n^{\s}(x_1,\dots,x_n)$.
The bound in equation~\eqref{eq:FF_asymptotic_bound} implies that $Q_n^{\s}$ has minimal degree $1-n$ in $x_i$. Thus it is convenient to define the polynomial
\begin{equation}
        P_n^{\s}(x_1,\dots,x_n)\equiv Q_n^{\s}(x_1,\dots,x_n)\prod_{i=1}^n x_i^{n-1}
\end{equation}
where generically we write
\begin{equation}
        P_n^{\s}(x_1,\dots,x_n)=\sum_{\a_1,\dots,\a_n}c_{\a_1\dots \a_n}\prod_{i=1}^n x_i^{\a_i}.
\end{equation}

Additional constraints come from particle exchange symmetry and Lorentz invariance. Symmetry under particle exchange implies that $P_n^{\s}$ is a symmetric polynomial in $x_i$, while Lorentz invariance of the form factors implies that $P_n^{\s}$ only contains
terms such that
\begin{equation}
        \sum_{i=1}^n \a_i= \frac{n(n-1)}{2}+n(n-1).
\end{equation}

Now we apply the residue equations given in equation~\eqref{eq:FF_pole_equations} to $P_n^{\s}$. They give the equations\footnote{Equations~\eqref{eq:Pn_kinematic_equation} and~\eqref{eq:Pn_bound_state_equation} are the correct versions of Equations A.10 and A.6 in~\cite{Delfino:1995zk}.
Our form factors are still consistent with theirs, since they use a different method described in their main text.}
\begin{equation} \label{eq:Pn_kinematic_equation}
    \begin{aligned}
        \frac{P_{n}(-x,x,\dots)}{x^3P_{n-2}(x_3,\dots,x_n)}&=iD^{n-2}(\prod_{j=3}^n\prod_{a\in\cA_{11}}\left(x_j^2+2i\sin{(\pi a)}xx_j-x^2\right) \\ &\quad +(-1)^{n+1}\prod_{j=3}^n\prod_{a\in\cA_{11}}\left(x^2+2i\sin{(\pi a)}xx_j-x_j^2\right)) \\
        D&=\frac{1}{16}\prod_{a\in\cA_{11}}\frac{\sin{\frac{\pi a}{2}}}{G_a(0)\cos^3{\frac{a\pi}{2}}}
    \end{aligned}
\end{equation}
and
\begin{equation} \label{eq:Pn_bound_state_equation}
        \begin{aligned}
                \frac{P_n^{\s} (x e^{i\pi/3}, x e^{-i\pi/3}, x_3, \dots, x_n)}{x^3 P_{n-1} (x , x_3, \dots, x_n) } &=A B^{n-2}  \prod_{j=3}^n (x + x_j) \left(x^2 + 2 \sin\left(\frac{7 \pi}{30}\right) x x_j + x_j^2\right) \\
                A&=\frac{2^{3/4}(15-3\sqrt{5})^{1/4}(1+2\cos{\frac{\pi}{15}})\sqrt{\cot{\frac{\pi}{30}}}\cot{\frac{2\pi}{15}}\sec^2{\frac{\pi}{30}}}{(1+\sqrt{5})^{3/2}G_{\frac{1}{15}}G_{\frac{2}{5}}G_{\frac{2}{3}}\left(\frac{2i\pi}{3}\right)} \\
                B&=\frac{2(-3+\sqrt{5})\cos^2{\frac{2\pi}{15}}\sec^2{\frac{\pi}{30}}}{\Tilde{G}_{\frac{1}{15}}\Tilde{G}_{\frac{2}{5}}\Tilde{G}_{\frac{2}{3}}\left(\frac{1}{3}\right)} \\
                \Tilde{G}_{\lambda}(\gamma)&=\frac{G_{\lambda}(i\pi\gamma)G_{\lambda}(-i\pi\gamma)}{G_{\lambda+\gamma}(0)G_{\lambda-\gamma}(0)}.
        \end{aligned}
\end{equation}

One final condition on the form factors is the asymptotic factorization equation which holds for relevant scalars with no internal symmetries~\cite{Delfino:1996nf}:
\begin{equation}
        \lim_{\L\rightarrow\infty}F_{a_1\dots a_r b_1\dots b_l}^{\s}(\th_1+\L,\dots,\th_r+\L,\th_1',\dots,\th_l')=\frac{1}{\<\s\>}F_{a_1\dots a_r}^{\s}(\th_1,\dots,\th_r)F_{b_1\dots b_l}^{\s}(\th_1',\dots,\th_l').
\end{equation}
Applied to $P_n^{\s}$, this gives
\begin{equation} \label{eq:Pn_factorization_equation}
    \begin{aligned}
        \lim_{\Lambda\rightarrow\infty}\frac{P_n^{\s}(x_1\Lambda,\dots,x_r\Lambda,x_{r+1},\dots,x_n)}{\Lambda^{r(n-1)+\frac{r(r-1)}{2}+2r(n-r)}}&=\frac{1}{K^{r(n-r)}\langle\sigma\rangle}\prod_{j=1}^r x_j^{3n-3r}P_r{(x_1,\dots,x_r)}P_{n-r}(x_{r+1},\dots,x_n) \\
        K&=-32\prod_{a\in\cA_{11}}\cos^2\frac{\pi a}{2}\sqrt{\frac{G_{a}(0)}{2\sin a\pi}}
    \end{aligned}
\end{equation}

Using equations~\eqref{eq:Pn_kinematic_equation},~\eqref{eq:Pn_bound_state_equation}, and~\eqref{eq:Pn_factorization_equation}, we can recursively compute $P_n^{\s}$. The problem boils down to an overconstrained linear system.
The total number of free parameters that one needs to solve for in $P_n^{\s}$ in this linear system is given in Table~\ref{tb:Pansatzparameters}.
\begin{table}[ht]
\centering
\begin{tabular}{r r}
\hline
$n$ & parameters \\
\hline
2 & 2 \\
3 & 8 \\
4 & 43 \\
5 & 252 \\
6 & 1\,636 \\
7 & 11\,322\\
8 & 81\,805\\
9 & 610\,358\\
10 & 4\,672\,836\\
\hline
\end{tabular}
\caption{Number of degrees of freedom for $P_n^{\s}$.}
\label{tb:Pansatzparameters}
\end{table}
For our numerical purposes, rationalizing the coefficients to high precision and using row-reduction algorithms over the rational numbers sufficed; for this we used the program SparseRREF~\cite{SparseRREF}. 

Note that this procedure also works for computing form factors of $\e$.
To distinguish the $\s$ solution from the $\e$ solution, we can use the fact that the trace of the stress tensor $\Theta(x)$ is proportional to $\s$ in the magnetic Ising model:
\begin{equation}
        \Theta(x)=2\pi h(2-\De_{\s})\s(x)
\end{equation}
where $h$ is related to the mass of the lightest particle by~\cite{Fateev:1993av}
\begin{equation}
        \begin{aligned}
                m_1&=Ch^{8/15} \\
                C&=\frac{4\sin\frac{\pi}{5}\Gamma\left(\frac{1}{5}\right)}{\Gamma\left(\frac{2}{3}\right)\Gamma\left(\frac{8}{15}\right)}\left(\frac{4\pi^2\Gamma\left(\frac{3}{4}\right)\Gamma^2\left(\frac{13}{16}\right)}{\Gamma\left(\frac{1}{4}\right)\Gamma^2\left(\frac{3}{16}\right)}\right)^{4/15}.
        \end{aligned}
\end{equation}
The mass of the lightest particle sets the scale of the problem, which in this work we set equal to 1.

The VEV of $\Theta(x)$ can be computed separately (for example, with the thermodynamic Bethe ansatz) and is equal to~\cite{Fateev:1993av}
\begin{equation}
        \<0|\Theta(x)|0\>=\frac{\pi m_1^2}{2\left(\sin\frac{2\pi}{3}+\sin\frac{2\pi}{5}+\sin\frac{\pi}{15}\right)}.
\end{equation}

Additionally, note that the described method of solving for the form factors is ambiguous up to an overall rescaling of the operators. In our calculations, we fix this overall normalization by rescaling $\Theta(x)$ so that its VEV is equal to 1.

Putting everything together, one can compute $P_n^{\s}$ up to arbitrarily high $n$. In this work, we compute up to $n=8$. We note that the coefficients of $P_n^{\s}$ computed using this method are real. 

Once the appropriate $P_n$ is computed, any form factor can be computed by repeated use of the second equation in~\eqref{eq:FF_pole_equations} and simplifications using the identities in equations~\eqref{eq:Gfunc_identity_1}--\eqref{eq:Gfunc_identity_5}. Performing these particle fusions starting with $P_n^{\s}$ with real coefficients will yield
subsequent form factors with real coefficients, except for form factors involving particles of either type 4 or type 6, in which there will be an overall complex phase which complicates numerical computations. Luckily, it turns out that one particular way of obtaining type 4 or type 6 particles, that is, by fusing $1111\rightarrow 22 \rightarrow 4(6)$, yields real coefficients. Thus, we are able to
obtain every form factor with real coefficients.

\section{Some harmonic analysis results}

In this appendix, we describe some results from harmonic analysis on $\SO(d-1,1)$ that are needed in the main text.

\subsection{Plancherel measure}

The support of the Plancherel measure for $\SO(d-1,1)$ includes principal series representations $\cP_{J_L,\l}$, where $J_L\in \frac{2-d}{2}+i \R_{\geq 0}$ and $\l$ is a finite-dimensional irrep of $\SO(d-2)$. When $d$ is odd, there are also discrete series contributions that we will not describe. We will be interested in the case where $\l$ is a rank-$j$ traceless symmetric tensor, and we refer to $j$ as ``transverse spin''. In generic $d$, the Plancherel measure is (see e.g.\ \cite{Dobrev:1977qv,Karateev:2018oml})
\be
d\mu(\cP_{J_L,j}) &= \frac{d\nu}{2\pi} \mu_j(\nu),
\ee
where $J_L=\frac{2-d}{2}+i\nu$ and 
\be
\label{eq:formulaformuj}
\mu_j(\nu) &= \frac{1}{\cN(-J_L,j)} \frac{\dim \rho_j}{2^{d-2} \vol\,\SO(d-2)}, \nn\\
\cN(-J_L,j) &= \frac{\pi^{d-2}\G(-J_L-\frac{d-2}{2})\G(\frac{d-2}{2}+J_L)}{\G(-J_L-1)\G(d+J_L-3)(-J_L+j-1)(d+J_L+j-3)}.
\ee
Here, 
\be
\dim \rho_j &= \frac{\G(j+d-4)(2j+d-4)}{\G(j+1)\G(d-3)},
\ee
is the dimension of the spin-$j$ traceless symmetric tensor representation of $\SO(d-2)$. We normalize $\vol\,\SO(n)$ so that
\be
\frac{\vol\,\SO(n)}{\vol\, \SO(n-1)} &= \vol\,S^{n-1},
\ee
and $\vol\,\SO(2) = 2\pi$.

In $d=4$, the spin-$j$ representation of $\SO(d-2)$ is reducible, so we instead let $j\in \Z$ run over both positive and negative values, representing the 1-dimensional representations of $\SO(2)$. In this case, we have $\dim\rho_j=1$, and the Plancherel measure becomes
\be
\mu_j(\nu) &= \frac{\nu^2 + j^2}{8\pi^3}\qquad (d=4).
\ee

\subsection{Scalar principal series representations}
\label{eq:theappendixwhereweexplaingegenbauerfunctionthing}

The scalar principal series representations $\cP_{J_L,0}$ of $G=\SO(d-1,1)$ play an important role in describing scalar observables in hyperbolic space like the number density $n(y)$. They consist of functions $f(z)$ on the projective null cone with homogeneity $J_L=\frac{2-d}{2}+i\nu$. The inner product is
\be
\<f_1|f_2\> &= \int D^{d-2} z\,f_1(z)^* f_2(z),
\ee
where $D^{d-2}z$ is the $\SO(d-1,1)$-invariant projective measure
\be
D^{d-2} z &= \frac{2}{\vol\,\R^+} d^d z\, \de(z^2) \th(z^0).
\ee
Note that we can gauge-fix an integral over the projective null cone to an integral over the sphere:
\be
\int D^{d-2} z\,f(z) &= \int_{S^{d-2}} d^{d-2} \Omega_{\vec n}\,f(z=(1,\vec n)).
\ee

Formally, we can write $f(z)=\<z|f\>$, where $\<z|$ is a dual basis vector that is homogeneous of degree $J_L$ in $z$. These basis vectors are dual to elements $|z\>$ that are homogeneous of degree $J_L^*$ and satisfy $\<z|z'\>=\de(z,z')$. The inner product gives a completeness relation
\be
1 &= \int D^{d-2}z |z\>\<z|.
\ee
We have $\pi(g)|z\>=|g z\>$ and $\<z|\pi(g)=(\pi(g^{-1})|z\>)^\dag = \<g^{-1} z|$, where we used unitarity $\pi(g)^\dag = \pi(g^{-1})$.

Each scalar principal series representation contains a unique (up to a phase) unit-normalized vector $|e\>$ that is invariant under $\SO(d-1)$. The corresponding function is 
\be
\<z|e\>=\frac{1}{\sqrt{\vol\,{S^{d-2}}}}(-e\.z)^{J_L},
\ee
where on the right-hand side, $e=(1,0,\dots,0)$ denotes a unit vector in the time direction.
By acting with $G$, we generate states $|ge\>=\pi(g)|e\>$, labeled by elements $v=g e$ of the hyperboloid $v^2=-1$. Note that we abuse notation and use the same symbols for null-cone states $|z\>$ and hyperboloid states $|v\>$, disambiguating them by the name of the variable. The inner product between hyperboloid states is a Gegenbauer function (\ref{eq:thegegenbauerfunction})
\be
\<v_1|v_2\> &= P_{J_L}(-v_1\.v_2).
\ee

Let $H=\SO(d-1)$. An important lemma is that the projector onto $H$-invariant states in $\cP_{J_L,0}$ is given by
\be
\Pi_H &= \frac{1}{\vol H} \int_{H} dh\, \pi(h) = |e\>\<e|.
\ee
This follows because $|e\>$ is the unique $H$-invariant state. However, we can also see it by direct computation
\be
\frac{1}{\vol\,H} \int_{H} dh \<z'|\pi(h)|z\> &= \frac{1}{\vol\,H} \int_{H} dh\, \<z'|hz\>\nn\\
&= \frac{1}{\vol\,S^{d-2}} \int d\Omega_{\vec n}\, \<z'|z^0(1,\vec n)\> \nn\\
&= \frac{1}{\vol\,S^{d-2}} \int D^{d-2} w\, \<z'|w\> (w^0)^{J_L} (z^0)^{J_L^*} \nn\\
&= \frac{(z'^0)^{J_L}(z^0)^{J_L^*}}{\vol\,S^{d-2}} = \<z'|e\>\<e|z\>.
\ee
This implies in particular the following identities needed in the main text:
\be
\Tr(\Pi_H \pi(g)) &= \<e|\pi(g)|e\> = P_{J_L}(\eta),\nn\\
\Tr(\Pi_H \pi(g) \Pi_H \pi(g'^{-1})) &= \<e|\pi(g)|e\>\<e|\pi(g'^{-1})|e\> = P_{J_L}(\eta) P_{J_L}(\eta')
\ee
where $\eta=-e\.g e$ and $\eta'=-e\.g' e$.

\section{Haar measure and conjugacy orbit}
\label{app:conjugacyorbit}

In this appendix, we compute the Haar measure in the coordinates $(\eta, \theta, x)$, where $(\eta,\theta)$ parametrize the Cartan subgroup as  $h = e^{\theta J_3 + \eta K_3} \in H$ and $x$ parametrizes the conjugacy orbit of $h$, $g = x h x^{-1}$.

The invariant metric on $\SO(3,1)$ is given by
\begin{equation}
        \langle U, V \rangle_g = -\frac{1}{2}\text{Tr}\left(g^{-1} U g^{-1} V\right) \label{eq:metric}
\end{equation}
where $U, V \in T_g G$ and $g \in \SO(3,1)$. Note that it satisfies left and right invariance under group multiplication:
\begin{equation}
 \langle g' U, g' V \rangle_{g'g} =  \langle U g', V g' \rangle_{gg'} = \langle U, V \rangle_{g}.
\end{equation}
Its volume density therefore defines a Haar measure.
 
The Haar measure used in the main text is normalized by requiring that $\vol\,\mathrm{U}(1) = 2\pi$. Let us check that this is the case for the measure derived from (\ref{eq:metric}).
At the identity, the metric has components  $\text{diag}(-1,-1,-1,1,1,1)$ in the basis $(K_1,K_2,K_3,J_1,J_2,J_3)$ of $T_e$. The subgroup
\begin{equation}
        g(\theta) = e^{\theta J_3}
\end{equation}
has the induced line element
\begin{equation}
ds = \sqrt{|\langle \partial_\theta g,\partial_\theta g \rangle_{ g}|} d\theta = d\theta
\end{equation}
and so
\begin{equation}
        \vol\,\mathrm{U}(1) = \int_{0}^{2\pi} d\theta  = 2\pi
\end{equation}
as required.

We can now compute the Haar measure in the desired coordinates. Note that, due to left and right invariance, it is sufficient to compute the measure at $x=e$. 
The conjugacy orbit is generated by $X_a = (J_1,K_1, J_2, K_2)$. The tangent vectors at $h$ are $K_3h$, $J_3h$, and 
\begin{equation}
        \partial_s \left(e^{s X_a} h e^{-s X_a}\right)\vert_{s=0} = X_a h - h X_a.
\end{equation}
The metric elements at $h$ can be computed by left translation by $h^{-1}$:
\begin{align}
    \langle K_3 h, K_3 h \rangle_h &=  -1\\
        \langle J_3 h, J_3 h \rangle_h &=  1\\
            \langle X_a h - h X_a, X_b h - h X_b \rangle_h &=
            \left(
\begin{array}{cccc}
\alpha & \beta & 0 & 0 \\
\beta & -\alpha & 0 & 0 \\
0 & 0 & \alpha & \beta \\
0 & 0 & \beta & -\alpha
\end{array}
\right)_{ab},
\end{align}
where 
$\alpha = 2(1-\cosh\eta\cos\theta)$ and $\beta = -2\sinh\eta\sin\theta$. All other elements are zero. 

Finally, computing the volume density from the metric determinant, we find
\begin{equation}
        dg = 4(\cosh \eta - \cos\theta)^2 d\eta d\theta dx
\end{equation}
with $dx$ the invariant measure on $G/H$ induced by (\ref{eq:metric}). This is the result quoted in section~\ref{sec:generalevol}.
\bibliographystyle{JHEP}
\bibliography{refs}

\providecommand{\href}[2]{#2}\begingroup\raggedright\begin{thebibliography}{10}

\bibitem{Chang:2025zib}
C.-H. Chang, H.~Chen, D.~Simmons-Duffin and H.~X. Zhu, \emph{{Seeing through the confinement screen: DGLAP/BFKL mixing and light-ray matching in QCD}}, \href{http://dx.doi.org/10.1007/JHEP02(2026)251}{\emph{JHEP} {\bf 02} (2026) 251}, [\href{https://arxiv.org/abs/2506.06431}{{\tt 2506.06431}}].

\bibitem{Gonzalez:2026aha}
M.~Gonzalez, K.~Lee and I.~Moult, \emph{{Generalized Detectors at Colliders}},  \href{https://arxiv.org/abs/2609.16114}{{\tt 2609.16114}}.

\bibitem{Chen:2026charge}
H.~Chen, P.~F. Monni, G.~Vita and H.~X. Zhu, ``{Taming Charge-Charge Correlation with Celestial Harmonic Analysis}.'' To appear.

\bibitem{Chang:2025kgq}
C.-H. Chang, H.~Chen, X.~Liu, D.~Simmons-Duffin, F.~Yuan and H.~X. Zhu, \emph{{Quantum Scaling in Energy Correlators beyond the Confinement Transition}}, \href{http://dx.doi.org/10.1103/9ml8-xkfc}{\emph{Phys. Rev. Lett.} {\bf 136} (2026) 081903}, [\href{https://arxiv.org/abs/2507.15923}{{\tt 2507.15923}}].

\bibitem{Hofman:2008ar}
D.~M. Hofman and J.~Maldacena, \emph{{Conformal collider physics: Energy and charge correlations}}, \href{http://dx.doi.org/10.1088/1126-6708/2008/05/012}{\emph{JHEP} {\bf 05} (2008) 012}, [\href{https://arxiv.org/abs/0803.1467}{{\tt 0803.1467}}].

\bibitem{Kravchuk:2018htv}
P.~Kravchuk and D.~Simmons-Duffin, \emph{{Light-ray operators in conformal field theory}}, \href{http://dx.doi.org/10.1007/JHEP11(2018)102}{\emph{JHEP} {\bf 11} (2018) 102}, [\href{https://arxiv.org/abs/1805.00098}{{\tt 1805.00098}}].

\bibitem{Caron-Huot:2017vep}
S.~Caron-Huot, \emph{{Analyticity in Spin in Conformal Theories}}, \href{http://dx.doi.org/10.1007/JHEP09(2017)078}{\emph{JHEP} {\bf 09} (2017) 078}, [\href{https://arxiv.org/abs/1703.00278}{{\tt 1703.00278}}].

\bibitem{Mukhanov:1981xt}
V.~F. Mukhanov and G.~V. Chibisov, \emph{{Quantum Fluctuations and a Nonsingular Universe}}, {\emph{JETP Lett.} {\bf 33} (1981) 532--535}.

\bibitem{Hawking:1982cz}
S.~W. Hawking, \emph{{The Development of Irregularities in a Single Bubble Inflationary Universe}}, \href{http://dx.doi.org/10.1016/0370-2693(82)90373-2}{\emph{Phys. Lett. B} {\bf 115} (1982) 295}.

\bibitem{Guth:1982ec}
A.~H. Guth and S.~Y. Pi, \emph{{Fluctuations in the New Inflationary Universe}}, \href{http://dx.doi.org/10.1103/PhysRevLett.49.1110}{\emph{Phys. Rev. Lett.} {\bf 49} (1982) 1110--1113}.

\bibitem{Zamolodchikov:1978xm}
A.~B. Zamolodchikov and A.~B. Zamolodchikov, \emph{{Factorized s Matrices in Two-Dimensions as the Exact Solutions of Certain Relativistic Quantum Field Models}}, \href{http://dx.doi.org/10.1016/0003-4916(79)90391-9}{\emph{Annals Phys.} {\bf 120} (1979) 253--291}.

\bibitem{Zamolodchikov:1989hfa}
A.~B. Zamolodchikov, \emph{{Integrable field theory from conformal field theory}}, {\emph{Adv. Stud. Pure Math.} {\bf 19} (1989) 641--674}.

\bibitem{Liu:2026asy}
Y.~Liu, \emph{Asymptotics of spin-spin correlators weighted by fermion number measurements with low rapidity threshold in the 2d ising free-fermion qft}, \href{http://dx.doi.org/10.1007/JHEP01(2026)064}{\emph{JHEP} {\bf 01} (2026) 064}, [\href{https://arxiv.org/abs/2507.13093}{{\tt 2507.13093}}].

\bibitem{Feynman:1969PRL}
R.~P. Feynman, \emph{Very high-energy collisions of hadrons}, \href{http://dx.doi.org/10.1103/PhysRevLett.23.1415}{\emph{Phys. Rev. Lett.} {\bf 23} (Dec, 1969) 1415--1417}.

\bibitem{Wilson:1970zzb}
K.~G. Wilson, \emph{{Some experiments on multiple production}},  Preprint CLNS-131, Cornell University, 1970.

\bibitem{Cardy:1990pc}
J.~L. Cardy and G.~Mussardo, \emph{{Form-factors of Descendent Operators in Perturbed Conformal Field Theories}}, \href{http://dx.doi.org/10.1016/0550-3213(90)90452-J}{\emph{Nucl. Phys. B} {\bf 340} (1990) 387--402}.

\bibitem{Zamolodchikov:1989fp}
A.~B. Zamolodchikov, \emph{{Integrals of Motion and S Matrix of the (Scaled) T=T(c) Ising Model with Magnetic Field}}, \href{http://dx.doi.org/10.1142/S0217751X8900176X}{\emph{Int. J. Mod. Phys. A} {\bf 4} (1989) 4235}.

\bibitem{Caron-Huot:2013fea}
S.~Caron-Huot, \emph{{When does the gluon reggeize?}}, \href{http://dx.doi.org/10.1007/JHEP05(2015)093}{\emph{JHEP} {\bf 05} (2015) 093}, [\href{https://arxiv.org/abs/1309.6521}{{\tt 1309.6521}}].

\bibitem{Caron-Huot:2015bja}
S.~Caron-Huot, \emph{{Resummation of non-global logarithms and the BFKL equation}}, \href{http://dx.doi.org/10.1007/JHEP03(2018)036}{\emph{JHEP} {\bf 03} (2018) 036}, [\href{https://arxiv.org/abs/1501.03754}{{\tt 1501.03754}}].

\bibitem{Maldacena:2015waa}
J.~Maldacena, S.~H. Shenker and D.~Stanford, \emph{{A bound on chaos}}, \href{http://dx.doi.org/10.1007/JHEP08(2016)106}{\emph{JHEP} {\bf 08} (2016) 106}, [\href{https://arxiv.org/abs/1503.01409}{{\tt 1503.01409}}].

\bibitem{McCoy:1977er}
B.~M. McCoy, C.~A. Tracy and T.~T. Wu, \emph{{Two-Dimensional Ising Model as an Exactly Solvable Relativistic Quantum Field Theory: Explicit Formulas for n Point Functions}}, \href{http://dx.doi.org/10.1103/PhysRevLett.38.793}{\emph{Phys. Rev. Lett.} {\bf 38} (1977) 793--796}.

\bibitem{Berg:1978sw}
B.~Berg, M.~Karowski and P.~Weisz, \emph{{Construction of Green Functions from an Exact S Matrix}}, \href{http://dx.doi.org/10.1103/PhysRevD.19.2477}{\emph{Phys. Rev. D} {\bf 19} (1979) 2477}.

\bibitem{2020sfta.book.....M}
G.~{Mussardo}, \emph{{Statistical Field Theory: An Introduction to Exactly Solved Models in Statistical Physics (2nd edn)}}.
\newblock 2020.

\bibitem{Delfino:1996nf}
G.~Delfino, P.~Simonetti and J.~L. Cardy, \emph{{Asymptotic factorization of form-factors in two-dimensional quantum field theory}}, \href{http://dx.doi.org/10.1016/0370-2693(96)01035-0}{\emph{Phys. Lett. B} {\bf 387} (1996) 327--333}, [\href{https://arxiv.org/abs/hep-th/9607046}{{\tt hep-th/9607046}}].

\bibitem{Jha:2024jan}
R.~G. Jha, A.~Milsted, D.~Neuenfeld, J.~Preskill and P.~Vieira, \emph{{Real-time scattering in Ising field theory using matrix product states}}, \href{http://dx.doi.org/10.1103/9dxz-k5wb}{\emph{Phys. Rev. Res.} {\bf 7} (2025) 023266}, [\href{https://arxiv.org/abs/2411.13645}{{\tt 2411.13645}}].

\bibitem{Wu:1975mw}
T.~T. Wu, B.~M. McCoy, C.~A. Tracy and E.~Barouch, \emph{{Spin spin correlation functions for the two-dimensional Ising model: Exact theory in the scaling region}}, \href{http://dx.doi.org/10.1103/PhysRevB.13.316}{\emph{Phys. Rev. B} {\bf 13} (1976) 316--374}.

\bibitem{Tracy:1990tn}
C.~A. Tracy, \emph{{Asymptotics of a tau function arising in the two-dimensional Ising model}}, \href{http://dx.doi.org/10.1007/BF02102065}{\emph{Commun. Math. Phys.} {\bf 142} (1991) 297--312}.

\bibitem{Babelon:1992sn}
O.~Babelon and D.~Bernard, \emph{{From form-factors to correlation functions: The Ising model}}, \href{http://dx.doi.org/10.1016/0370-2693(92)91964-B}{\emph{Phys. Lett. B} {\bf 288} (1992) 113--120}, [\href{https://arxiv.org/abs/hep-th/9206003}{{\tt hep-th/9206003}}].

\bibitem{Bernard:1994re}
D.~Bernard and A.~LeClair, \emph{{Differential equations for Sine-Gordon correlation functions at the free fermion point}}, \href{http://dx.doi.org/10.1016/0550-3213(94)90020-5}{\emph{Nucl. Phys. B} {\bf 426} (1994) 534--558}, [\href{https://arxiv.org/abs/hep-th/9402144}{{\tt hep-th/9402144}}].

\bibitem{Caron-Huot:2022eqs}
S.~Caron-Huot, M.~Kologlu, P.~Kravchuk, D.~Meltzer and D.~Simmons-Duffin, \emph{{Detectors in weakly-coupled field theories}}, \href{http://dx.doi.org/10.1007/JHEP04(2023)014}{\emph{JHEP} {\bf 04} (2023) 014}, [\href{https://arxiv.org/abs/2209.00008}{{\tt 2209.00008}}].

\bibitem{Costa:2012cb}
M.~S. Costa, V.~Goncalves and J.~Penedones, \emph{{Conformal Regge theory}}, \href{http://dx.doi.org/10.1007/JHEP12(2012)091}{\emph{JHEP} {\bf 12} (2012) 091}, [\href{https://arxiv.org/abs/1209.4355}{{\tt 1209.4355}}].

\bibitem{Delfino:2005bh}
G.~Delfino, P.~Grinza and G.~Mussardo, \emph{{Decay of particles above threshold in the Ising field theory with magnetic field}}, \href{http://dx.doi.org/10.1016/j.nuclphysb.2005.12.024}{\emph{Nucl. Phys. B} {\bf 737} (2006) 291--303}, [\href{https://arxiv.org/abs/hep-th/0507133}{{\tt hep-th/0507133}}].

\bibitem{Delfino:1995zk}
G.~Delfino and G.~Mussardo, \emph{{The Spin spin correlation function in the two-dimensional Ising model in a magnetic field at T = T(c)}}, \href{http://dx.doi.org/10.1016/0550-3213(95)00464-4}{\emph{Nucl. Phys. B} {\bf 455} (1995) 724--758}, [\href{https://arxiv.org/abs/hep-th/9507010}{{\tt hep-th/9507010}}].

\bibitem{Alekseev:2011my}
O.~Alekseev, \emph{{Form factors in the Bullough-Dodd related models: The Ising model in a magnetic field}}, \href{http://dx.doi.org/10.1134/S0021364012040030}{\emph{JETP Lett.} {\bf 95} (2012) 201--205}, [\href{https://arxiv.org/abs/1106.4758}{{\tt 1106.4758}}].

\bibitem{Lipatov:1985uk}
L.~N. Lipatov, \emph{{The Bare Pomeron in Quantum Chromodynamics}}, {\emph{Sov. Phys. JETP} {\bf 63} (1986) 904--912}.

\bibitem{Gross:1973zrg}
D.~J. Gross and F.~Wilczek, \emph{{Asymptotically Free Gauge Theories~2}}, \href{http://dx.doi.org/10.1103/PhysRevD.9.980}{\emph{Phys. Rev. D} {\bf 9} (1974) 980--993}.

\bibitem{Bassetto:1979nt}
A.~Bassetto, M.~Ciafaloni and G.~Marchesini, \emph{{Inelastic Distributions and Color Structure in Perturbative QCD}}, \href{http://dx.doi.org/10.1016/0550-3213(80)90413-7}{\emph{Nucl. Phys. B} {\bf 163} (1980) 477--518}.

\bibitem{Amati:1980ch}
D.~Amati, A.~Bassetto, M.~Ciafaloni, G.~Marchesini and G.~Veneziano, \emph{{A Treatment of Hard Processes Sensitive to the Infrared Structure of QCD}}, \href{http://dx.doi.org/10.1016/0550-3213(80)90012-7}{\emph{Nucl. Phys. B} {\bf 173} (1980) 429--455}.

\bibitem{Furmanski:1979jx}
W.~Furmanski, R.~Petronzio and S.~Pokorski, \emph{{Heavy Flavor Multiplicities at Very High-Energies}}, \href{http://dx.doi.org/10.1016/0550-3213(79)90365-1}{\emph{Nucl. Phys. B} {\bf 155} (1979) 253--268}.

\bibitem{Konishi:1979ft}
K.~Konishi, \emph{{Multiplicity distributions in QCD}},  Preprint RL-79-035, Rutherford Laboratory, 1979.

\bibitem{Mueller:1981ex}
A.~H. Mueller, \emph{{On the Multiplicity of Hadrons in QCD Jets}}, \href{http://dx.doi.org/10.1016/0370-2693(81)90581-5}{\emph{Phys. Lett. B} {\bf 104} (1981) 161--164}.

\bibitem{Chen:2021gdk}
H.~Chen, I.~Moult and H.~X. Zhu, \emph{{Spinning gluons from the QCD light-ray OPE}}, \href{http://dx.doi.org/10.1007/JHEP08(2022)233}{\emph{JHEP} {\bf 08} (2022) 233}, [\href{https://arxiv.org/abs/2104.00009}{{\tt 2104.00009}}].

\bibitem{Dobrev:1977qv}
V.~K. Dobrev, G.~Mack, V.~B. Petkova, S.~G. Petrova and I.~T. Todorov, \emph{{Harmonic Analysis on the n-Dimensional Lorentz Group and Its Application to Conformal Quantum Field Theory}}, vol.~63.
\newblock 1977, \href{http://dx.doi.org/10.1007/BFb0009678}{10.1007/BFb0009678}.

\bibitem{Homrich:2022cfq}
A.~Homrich, D.~Simmons-Duffin and P.~Vieira, \emph{{Multitwist Trajectories and Decoupling Zeros in Conformal Field Theory}}, \href{http://dx.doi.org/10.1103/PhysRevLett.134.011602}{\emph{Phys. Rev. Lett.} {\bf 134} (2025) 011602}, [\href{https://arxiv.org/abs/2211.13754}{{\tt 2211.13754}}].

\bibitem{Henriksson:2023cnh}
J.~Henriksson, P.~Kravchuk and B.~Oertel, \emph{{Missing local operators, zeros, and twist-4 trajectories}}, \href{http://dx.doi.org/10.1007/JHEP07(2024)248}{\emph{JHEP} {\bf 07} (2024) 248}, [\href{https://arxiv.org/abs/2312.09283}{{\tt 2312.09283}}].

\bibitem{Homrich:2024nwc}
A.~Homrich, D.~Simmons-Duffin and P.~Vieira, \emph{{Light-ray wave functions and integrability}}, \href{http://dx.doi.org/10.1007/JHEP10(2024)125}{\emph{JHEP} {\bf 10} (2024) 125}, [\href{https://arxiv.org/abs/2409.02160}{{\tt 2409.02160}}].

\bibitem{Caron-Huot:2020nem}
S.~Caron-Huot and J.~Sandor, \emph{{Conformal Regge Theory at Finite Boost}}, \href{http://dx.doi.org/10.1007/JHEP05(2021)059}{\emph{JHEP} {\bf 05} (2021) 059}, [\href{https://arxiv.org/abs/2008.11759}{{\tt 2008.11759}}].

\bibitem{SparseRREF}
Z.~Li, ``{SparseRREF: Exact Sparse Reduced Row Echelon Form in C++}.'' Version 0.4.0, \href{https://github.com/munuxi/SparseRREF}{{\tt github.com/munuxi/SparseRREF}}, \href{https://doi.org/10.5281/zenodo.21278863}{doi:10.5281/zenodo.21278863}.

\bibitem{Fateev:1993av}
V.~A. Fateev, \emph{{The Exact relations between the coupling constants and the masses of particles for the integrable perturbed conformal field theories}}, \href{http://dx.doi.org/10.1016/0370-2693(94)00078-6}{\emph{Phys. Lett. B} {\bf 324} (1994) 45--51}.

\bibitem{Karateev:2018oml}
D.~Karateev, P.~Kravchuk and D.~Simmons-Duffin, \emph{{Harmonic Analysis and Mean Field Theory}}, \href{http://dx.doi.org/10.1007/JHEP10(2019)217}{\emph{JHEP} {\bf 10} (2019) 217}, [\href{https://arxiv.org/abs/1809.05111}{{\tt 1809.05111}}].

\end{thebibliography}\endgroup

\end{document}